\documentclass{article}

\usepackage{geometry}
\usepackage{amsmath}
\usepackage[textwidth=8em,textsize=small]{todonotes}
\usepackage{caption}
\usepackage{nicefrac}
\usepackage{subcaption}
\usepackage{color}
\usepackage{tikz}
\usepackage{float}
\usepackage{setspace}
\usepackage{enumitem}
\usepackage{mathtools, nccmath}
\usepackage{threeparttable}
\usepackage{setspace}
\usepackage{url}
\usepackage[colorlinks=true, allcolors=blue]{hyperref}
\usepackage[notquote]{hanging}
\usepackage{lettrine}
\usepackage{titlesec}
\usepackage{multicol}
\usepackage[superscript,biblabel]{cite}

\usepackage{fancyhdr}
\titleformat{\section}[block]{\large\bfseries\filcenter\sffamily}{}{1em}{}
\titleformat{\subsection}[runin]{\small\bfseries\filcenter\sffamily}{}{0em}{}

\title{\sffamily{\textbf{Causal Selection of Covariates in Regression Calibration for Mismeasured Continuous Exposure}}}
\author{Wenze Tang \textsuperscript{a}, Donna Spiegelman \textsuperscript{b, c}, Xiaomei Liao \textsuperscript{a,d,e}, Molin Wang \textsuperscript{a, d}}

\begin{document}
\maketitle
\pagestyle{plain}
\fancyhf{} 
\fancyhead[R]{ \sffamily Epidemiology \bullet \text{Volume 0, Number 0, XXX 0000}}
\fancyhead[L]{\textit{Tang et al.}}

\begin{multicols}{2}
\noindent\rule{8.95cm}{1pt}\\
\textbf{Abstract: } \normalsize{Regression calibration as developed by Rosner, Spiegelman and Willet is used to correct the bias in effect estimates due to measurement error in continuous exposures. The method involves two models: a measurement error model (MEM) relating the mismeasured exposure to the true exposure and an outcome model relating the mismeasured exposure to outcome. However, no comprehensive guidance exists for determining which covariates should be included in each model. In this paper, we investigate the selection of the minimal and most efficient covariate adjustment sets under a causal inference framework. We show that in order to correct for the measurement error, researchers must adjust for, in both MEM and outcome model, any common causes (1) of true exposure and the outcome and (2) of measurement error and the outcome. When such variable(s) are only available in the main study, researchers should still adjust for them in the outcome model to reduce bias, provided that these covariates are at most weakly associated with measurement error. We also show that adjusting for so called prognostic variables that are independent of true exposure and measurement error in outcome model, may increase efficiency, while adjusting for any covariates that are associated only with true exposure generally results in efficiency loss in realistic settings. We apply the proposed covariate selection approach to the Health Professional Follow-up Study dataset to study the effect of fiber intake on cardiovascular disease. Finally, we extend the originally proposed estimators to a non-parametric setting where effect modification by covariates is allowed. (250 words).}
\newline
\\
\noindent\text{(Epidemiology 0000; 00: 0 - 0)}
\newline
\\

\footnotesize{
\noindent\rule{8.95cm}{0.2pt}
\begin{hangparas}{.25in}{1}
Submitted Oct, 2022; accepted 00, 0000.

From the \textsuperscript{a} Department of Epidemiology, Harvard School of Public Health, Boston, MA; \textsuperscript{b} Department of Biostatistics, Yale School of Public Health, New Haven, CT;   \textsuperscript{c} Center on Methods for Implementation and Prevention Science, Yale School of Public Health, New Haven, CT;   \textsuperscript{d} Department of Biostatistics, Harvard School of Public Health, Boston, MA; \textsuperscript{e} Department of Nutrition, Harvard School of Public Health, Boston, MA. 

Supported in part through grants 5R01ES026246 from National Institute of Envirnomental Health Sciences(NIEHS) and grant R01 DC017717 from National Institute of Health(NIH). All statements in this article, including its findings and conclusions, are solely those of the authors and do not necessarily represent the views of the NIEHS or the NIH. 

The authors report no conflicts of interest.

Supplemental digital content is avaialble through the URL citations in the HTML and PDF versions of this article (www.epidem.com). 

Correspondence: Molin Wang, Department of Biostatistics, Harvard School of Public Health, Boston, MA. Email: stmow@channing.harvard.edu.
\end{hangparas}
}

\normalsize
\section{INTRODUCTION}\label{sec1}

\pagenumbering{arabic} 

\lettrine{R}egression calibration is a popular tool for correcting bias due to non-differential measurement error in continuous exposures.\cite{RSWapp1,RSWapp2,RSWapp3,CRSapp1,CRSapp2,CRSapp3}. There are two versions of regression calibration (RC) methods: one developed by Rosner, Spiegelman and Willet\cite{RSW1,RSW2,liao2011survival,white2006commentary} (hereinafter RSW) and the other developed by Carroll, Ruppert and Stefanski \cite{CRS_book,CRS1}(hereinafter CRS). Both versions involve two models: a measurement error model (also known as calibration model, hereafter MEM) linking the surrogate or measured exposure to the true exposure and an outcome model linking either the surrogate exposure or imputed true exposure to the error-free outcome. For both methods, a set of pre-treatment covariates are typically adjusted for in the MEM and the outcome model but what should be included in this set of covariates has not been systematically discussed.   

For example, in assessing the effect of fiber intake on the risk of cardiovascular disease (CVD), age and smoking are two important confounding variables for the true exposure-outcome relationship, with age potentially also influencing measurement error by affecting memory and cognitive function\cite{ambrose2004pathophysiology,mandolesi2018effects,mcdermott2009meta,ma2020association}. Other risk factors for CVD such as marital status, in contrast, may not be associated with either fiber intake or measurement error\cite{cahill2013prospective}.

In this paper, we aim to answer the optimal covariate selection question in RSW RC method under a counterfactual framework. With the aid of directed acyclic graph (DAG), we investigate (1) what is the minimal covariate adjustment set that would guarantee validity and (2) whether adjusting for additional covariates outside of the minimal covariate adjustment set in the MEM and/or outcome model increases efficiency, including the question of whether different sets of covariates can be adjusted for in the two models \cite{RSW1,rosner1990correction,LoganSpiegelman2012blinplus}. 

In section 2, we define the causal contrast of interest, the statistical models and the resulting estimators, followed by an introduction to the relevant DAGs and assumptions. In section 3, we provide the identification formula for our causal estimand. We then present the results regarding the validity and efficiency of RSW estimators under each DAG structure. In section 4, we present results from a series of Monte Carlo simulations evaluating the finite sample performance of each estimator. In section 5, we apply the proposed covariate selection strategy to the study of the effect of fiber intake on risk of CVD within the Health Professional Follow-up Study (HPFS). We conclude this paper with a summary and further discussions. 

\section{REVIEW} \label{sec2}

\subsection{2.1 Causal Estimand of Interest}
We denote $X$, $Z$, $V$, $Y$ respectively as true exposure, surrogate/mismeasured exposure, covariate(s) and outcome, where $X$, $Z$ are assumed to be scalar and $V$ can be either a scalar or a set of covariates. We focus on the main study/external validation study design; that is, independent samples $(Z,V,Y)$ and $(X,Z,V)$ are available from the main study and validation study respectively. We assume that the parameters in the MEM estimated using validation study population would be the same as those estimated from the main study population had we been able to run the MEM in both populations (i.e. transportability condition). Following the standard causal inference literature, we define the potential outcome $Y^x$ as the value of $Y$ that each participant would have experienced had their exposure value been fixed at $X=x$. 

We are primarily interested in the conditional average treatment effect (CATE) on the additive scale denoted by $E[Y^x - Y^{x^{\prime}} |V]$, i.e. the mean difference in the potential outcome had everyone been assigned exposure level $x$ versus exposure level $x'$ within levels of the covariates $V$. Parametric regression models typically estimate CATEs. We may also be interested in the average treatment effect (ATE) $E[Y^x - Y^{x^{\prime}}]$. One can always obtain ATE from CATE by standardizing over the distribution of the covariates $V$.

\subsection{2.2 Generalized RSW Regression Calibration Estimators} \label{parametricRSWEstimators}
In the standard RSW method, for a given covariate $V$, analyst typically choose one of the two linear MEMs (adjusted model \eqref{eq1} versus unadjusted model \eqref{eq2}) and one of the two linear outcome models (adjusted model \eqref{eq3} versus unadjusted model \eqref{eq4}):
\begin{align}
    E[X|Z,V] & = \alpha_0 + \alpha_1 Z + \alpha_2 V \label{eq1}, \\
    E[X|Z] & = \alpha_0^* + \alpha_1^* Z \label{eq2}, \\
    E[Y|Z,V] & = \gamma_0 + \gamma_1 Z + \gamma_2 V  \label{eq3}, \\
    E[Y|Z] & = \gamma_0^* + \gamma_1^* Z \label{eq4}, 
\end{align} where \eqref{eq1} and \eqref{eq3} imply \eqref{eq2} and \eqref{eq4} respectively. 

We denote by subscript a RSW estimator as $\hat{\beta}_{(OM)}$ if it includes a given $V$ in both MEM and outcome model and as $\hat{\beta}_{(--)}$ if the covariate is included in neither model:
\begin{equation} \label{eq5}
\hat{\beta}_{(OM)}=\frac{\hat{\gamma}_1}{\hat{\alpha}_1}, 
\end{equation}
\begin{equation} \label{eq6}
\hat{\beta}_{(--)}=\frac{\hat{\gamma}_1^*}{\hat{\alpha}_1^*}.
\end{equation}

In model \eqref{eq5} for example, $\alpha_1$ is also known as attenuation factor and by dividing $\gamma_1$, the biased effect estimate subject to measurement error, by $\alpha_1$ we can recover the true effect \cite{RSW1,RSW2,CRS_book,vanderweele2012results,weinberg1994will}. In this paper, we also investigate the following extended RSW estimators:
\begin{equation} \label{eq7}
\hat{\beta}_{(-M)}=\frac{\hat{\gamma}_1^*}{\hat{\alpha_1}},
\end{equation}
\begin{equation} \label{eq8}
\hat{\beta}_{(O-)}=\frac{\hat{\gamma}_1}{\hat{\alpha_1^*}}.
\end{equation}

Thus, these estimators respectively represent covariate adjustment strategies: (i)$\hat{\beta}_{(OM)}$: $V$ is included in both models, (ii)$\hat{\beta}_{(--)}$: $V$ is included in neither model, (iii) $\hat{\beta}_{(-M)}$: $V$ is only included in MEM and (iv) $\hat{\beta}_{(O-)}$: $V$ is only included in outcome model.  

\subsection{2.3 Directed Acyclic Graphs for Measurement Error Structures}
\begin{figure*}[!t]
     \caption{\label{fig1}Directed Acyclic Graphs Representing Common Measurement Error Structures}
     \begin{subfigure}[b]{0.23\textwidth}
         \begin{tikzpicture}[>= stealth, shorten >= 1pt, auto, node distance = 0.8 cm, semithick]
            \node[text centered] (v) {$V_{1(-{}-Y)}$};
            \node[right= of v] (x) {$X$};
            \node[right= of x] (y) {$Y$};
            \node[below= of x] (z) {$Z$};
            \path[->] (x) edge node {} (z);
            \path[->] (v) edge [out=45, in=135] node {} (y);
            \path[->] (x) edge node {} (y);
        \end{tikzpicture}
        \caption{DAG 1}
     \end{subfigure}
     \begin{subfigure}[b]{0.23\textwidth}
         \begin{tikzpicture}[>= stealth, shorten >= 1pt, auto, node distance = 0.8 cm, semithick]
            \node[text centered] (v) {$V_{2(-ZY)}$};
            \node[right= of v] (x) {$X$};
            \node[right= of x] (y) {$Y$};
            \node[below= of x] (z) {$Z$};
            \path[->] (x) edge node {} (z);
            \path[->] (v) edge [out=45, in=135] node {} (y);
            \path[->] (x) edge node {} (y);
            \path[->] (v) edge node {} (z);
        \end{tikzpicture}
        \caption{DAG 2}
     \end{subfigure}
     \begin{subfigure}[b]{0.23\textwidth}
         \begin{tikzpicture}[>= stealth, shorten >= 1pt, auto, node distance = 0.8 cm, semithick]
            \node[text centered] (v) {$V_{3(X-Y)}$};
            \node[right= of v] (x) {$X$};
            \node[right= of x] (y) {$Y$};
            \node[below= of x] (z) {$Z$};
            \path[->] (x) edge node {} (z);
            \path[->] (v) edge [out=45, in=135] node {} (y);
            \path[->] (x) edge node {} (y);
            \path[->] (v) edge node {} (x);
        \end{tikzpicture}
        \caption{DAG 3}
     \end{subfigure}
     \begin{subfigure}[b]{0.23\textwidth}
         \begin{tikzpicture}[>= stealth, shorten >= 1pt, auto, node distance = 0.8 cm, semithick]
            \node[text centered] (v) {$V_{4(XZY)}$};
            \node[right= of v] (x) {$X$};
            \node[right= of x] (y) {$Y$};
            \node[below= of x] (z) {$Z$};
            \path[->] (x) edge node {} (z);
            \path[->] (v) edge [out=45, in=135] node {} (y);
            \path[->] (x) edge node {} (y);
            \path[->] (v) edge node {} (x);
            \path[->] (v) edge node {} (z);
        \end{tikzpicture}
        \caption{DAG 4}
     \end{subfigure}
     \begin{subfigure}[b]{0.24\textwidth}
         \begin{tikzpicture}[>= stealth, shorten >= 1pt, auto, node distance = 0.8 cm, semithick]
            \node[text centered] (v) {$V_{5(-{}-{}-)}$};
            \node[right= of v] (x) {$X$};
            \node[right= of x] (y) {$Y$};
            \node[below= of x] (z) {$Z$};
            \path[->] (x) edge node {} (z);
            \path[->] (x) edge node {} (y);
        \end{tikzpicture}
        \caption{DAG 5}
     \end{subfigure}
     \begin{subfigure}[b]{0.24\textwidth}
         \begin{tikzpicture}[>= stealth, shorten >= 1pt, auto, node distance = 0.8 cm, semithick]
            \node[text centered] (v) {$V_{6(-{}Z{}-)}$};
            \node[right= of v] (x) {$X$};
            \node[right= of x] (y) {$Y$};
            \node[below= of x] (z) {$Z$};
            \path[->] (x) edge node {} (z);
            \path[->] (x) edge node {} (y);
            \path[->] (v) edge node {} (z);
        \end{tikzpicture}
        \caption{DAG 6}
     \end{subfigure}
     \begin{subfigure}[b]{0.24\textwidth}
         \begin{tikzpicture}[>= stealth, shorten >= 1pt, auto, node distance = 0.8 cm, semithick]
            \node[text centered] (v) {$V_{7(X{}-{}-)}$};
            \node[right= of v] (x) {$X$};
            \node[right= of x] (y) {$Y$};
            \node[below= of x] (z) {$Z$};
            \path[->] (x) edge node {} (z);
            \path[->] (x) edge node {} (y);
            \path[->] (v) edge node {} (x);
        \end{tikzpicture}
        \caption{DAG 7}
     \end{subfigure}
     \begin{subfigure}[b]{0.24\textwidth}
         \begin{tikzpicture}[>= stealth, shorten >= 1pt, auto, node distance = 0.8 cm, semithick]
            \node[text centered] (v) {$V_{8(X{}Z{}-)}$};
            \node[right= of v] (x) {$X$};
            \node[right= of x] (y) {$Y$};
            \node[below= of x] (z) {$Z$};
            \path[->] (x) edge node {} (z);
            \path[->] (x) edge node {} (y);
            \path[->] (v) edge node {} (z);
            \path[->] (v) edge node {} (x);
        \end{tikzpicture}
        \caption{DAG 8}
     \end{subfigure}
\end{figure*}
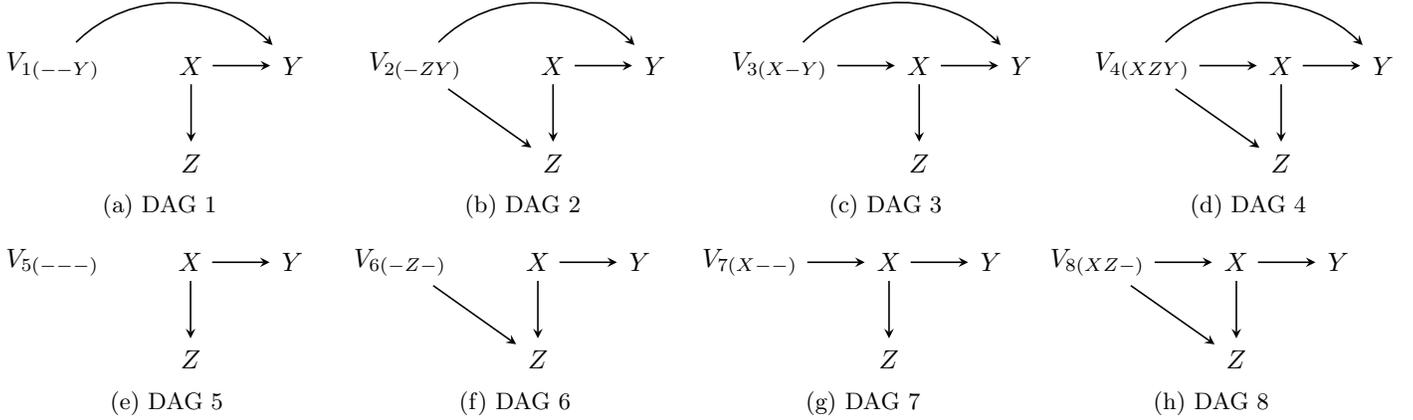

In this paper, we focus on eight variations of the non-differential exposure measurement error structure \cite{HernanCole2009,VanderweeleHernan2012,weisskopf2017trade}, represented in the eight causal DAGs in Figure \ref{fig1}. The non-differentiality assumption requires the independence between $Z$ and $Y$ conditional on $X$ in the absence of covariate \cite{CRS_book}. 

We partition the covariate set V into eight mutually exclusive subsets of covariate(s) $V_j$ where $j =1,\dots,8$ , with each DAG reflecting how $V_j$ relates to $X$, $Z$ and $Y$ (Figure \ref{fig1})\cite{zivichuse}. For example, DAG 4 describes the scenario where covariate $V_{4(XZY)}$ is both a confounder and directly contributes to measurement error. To facilitate reading, we write the additional subscript $(XZY)$ to emphasize the fact that the covariate has an arrow to $X, Z$ and $Y$ under DAG 4. DAG 5 through 8 are replicates of DAG 1 through 4 except the covariate $V_j$ under consideration is no longer a risk factor for the outcome. 

We can establish the following non-differential measurement error conditions (i.e. surrogacy assumptions): (i) $Y \perp Z |X$ under DAG 1, 3, 5, 6, 7 and 8, and (ii) $Y \perp Z| X, V_j$ under all DAGs, where $j=1,\dots,8$\cite{HernanRobinsWhatIf,Richardson2013SWIG}. We can also establish no confounding assumptions (i.e. exchangeability assumptions) (i) $Y^x \perp X$ for DAG 1, 2, 5, 6, 7 and 8 and (ii) $Y^x \perp X|V_j$ for all DAGs, where $j=1,\dots,8$\cite{Richardson2013SWIG}.  

For simplicity but with no loss of generality, we consider covariates of each type in the absence of the other types. In realistic settings, all eight types would likely appear on the same DAG. We demonstrate how the proof for validity can be generalized to such complex settings via one particular example in section 1.4 of supplemental material (SM), where we assume that both $(V_{3(X-Y)}, V_{4(XZY)})$ are present. 

\section{VALIDITY AND EFFICIENCY OF REGRESSION CALIBRATION ESTIMATORS}

\subsection{3.1 Identification of Causal Effect of Interest}
We show in section 1.2 of the online SM that under the following sufficient conditions:  
\begin{itemize}
    \item[C.1.1] Exchangeability: $Y^x \perp X |V$ (i.e. no confounding for the effect of $X$ on $Y$ given $V$),
    \item[C.1.2] Non-differential measurement error (surrogacy): $Y \perp Z| (X, V)$, 
    \item[C.1.3] Consistency: $Y^x$ and $Z^x$ take observed values $Y$ and $Z$ when $X=x$, and
    \item[C.1.4] Linearity: CATE follows the form $ E[Y^x - Y^{x^{\prime}}|V] = \beta (V) (x-x')$, where $\beta (V)$ is the constant effect associated with one unit increase in $X$ given $V$,    
    \end{itemize} 
we can non-parametrically identify $\beta(V)$ and thus CATE defined in 2.1 for all DAGs 1 through 8 as: 

\begin{equation} 
\label{identify_formula_V}
\beta (V) = \frac{E[Y|Z=z, V] - E[Y|Z=z^{'},V]}{E[X|Z=z, V] - E[X|Z=z^{'},V]}.
\end{equation}

Furthermore, if we additionally make the following modeling assumption: 
\begin{itemize}
\item[C.1.5] The models for $E[X|Z=z,V=v]$ and $E[Y|Z=z,V=v]$ as in (\ref{eq1}) and (\ref{eq3}) are correctly specified and that $E[V|Z=z]$ is a linear function of $Z$,
\end{itemize} $\beta(V)$ reduces to $\beta$, where $\beta$ is a constant effect associated with one unit increase in $X$, and is identified as:
\begin{equation} \label{eq10}
\beta=\frac{\gamma_1}{\alpha_1},
\end{equation} which has the same form as RSW estimator \eqref{eq5}. 

Similarly, we show in the SM that under the following sufficient conditions:  
\begin{itemize}
    \item[C.2.1] Exchangeability: $Y^x \perp X$ (i.e. no confounding for $X$'s effect on $Y$),
    \item[C.2.2] Non-differential measurement error (surrogacy): $Y \perp Z| X$, 
    \item[C.2.3] Consistency: $Y^x, Z^x$ take observed value $Y$, $Z$ when $X=x$, and
    \item[C.2.4] Linearity: ATE follows the form $ E[Y^x - Y^{x^{\prime}}] = \beta (x-x')$, where $\beta$ is the constant effect associated with one unit increase in $X$,   

\end{itemize} we can non-parametrically identify $\beta$ and thus ATE defined in 2.1 for DAGs 1, 5, 6, 7 and 8 as: 
\begin{equation}  \label{identify_formula_noV}
\beta = \frac{E[Y|Z=z] - E[Y|Z=z^{'}]}{E[X|Z=z] - E[X|Z=z^{'}]}.
\end{equation}

Under further modeling assumptions:
\begin{itemize}
    \item[C.2.5] The models for $E[X|Z=z]$ and $E[Y|Z=z]$ as in (\ref{eq2}) and (\ref{eq4}) are correct and that $E[V|Z=z]$ is a linear function of $Z$, 
\end{itemize} $\beta$ is identified as: 
\begin{equation} \label{eq9}
\beta=\frac{\gamma_1^*}{\alpha_1^*},
\end{equation} which has the same form as RSW estimator \eqref{eq6}. 

\subsection{3.2 Validity and Efficiency of Regression Calibration Estimators}
We use results from section 2 and 3.1 to evaluate whether each RC estimator, $\hat{\beta}_{(OM)}$, $\hat{\beta}_{(--)}$, $\hat{\beta}_{(-M)}$, and $\hat{\beta}_{(O-)}$, correctly estimates the causal effect. We also analytically evaluate the asymptotic efficiency of valid RSW estimators for a given DAG for a continuous outcome under linear models \eqref{eq1} through \eqref{eq4}\cite{RSW2}. See sections 1 and 2 of the SM for proofs. 

\subsection{3.3 Summary of Analytical Results for RSW Estimators} \label{theoreticalResults}

\begin{table*}[!t]  
\centering 
  \begin{threeparttable} 
  \caption{Validity and Efficiency of RSW Estimators for Continuous Outcomes with Linear Models \label{tab:validity}} 
    \begin{tabular}{|c | c  c  c  c|} 
        \hline
        $V_j$ as in \tnote{a} & $\hat{\beta}_{(OM)}$ & $\hat{\beta}_{(--)}$ & $\hat{\beta}_{(-M)}$ & $\hat{\beta}_{(O-)}$ \\[1ex] 
        \hline
        DAG 1, $V_{1(-{}-{}Y)}$  & valid (efficient) & valid & valid \tnote{b} & valid \tnote{c} (efficient)\\ 
        DAG 2, $V_{2(-{}Z{}Y)}$ & valid & biased & biased & biased \\
        DAG 3, $V_{3(X{}-{}Y)}$ & valid & biased & biased & biased \\
        DAG 4, $V_{4(X{}Z{}Y)}$ & valid & biased & biased & biased \\
        DAG 5, $V_{5(-{}-{}-)}$ & valid & valid & valid \tnote{b} & valid \tnote{c} \\ 
        DAG 6, $V_{6(-{}Z{}-)}$ & valid (efficient) & valid & biased & biased\\ 
        DAG 7, $V_{7(X{}-{}-)}$ & valid & valid (efficient) & biased & biased\\ 
        DAG 8, $V_{8(X{}Z{}-)}$ & valid \tnote{d} & valid \tnote{d} & biased & biased\\ 
        \hline
    \end{tabular}
    
    \begin{tablenotes}
        \item[a] The subscript such as $(-{}-{}Y)$ emphasizes how the given covariate relate to $X, Z$ and $Y$. For example, DAG 2 describes a situation where covariate $V_{2(-ZY)}$ systematically affects measurement error and is a risk factor for the outcome.
        \item[b] The estimator is valid due to independence $Y \perp V |Z$ implied by the DAG. 
        \item[c] The estimator is valid due to independence $X \perp V |Z$ implied by the DAG. 
        \item[d] Relative efficiency depends on strength and direction of $\rho_{_{V,X}}$, $\rho_{_{X,Z|V}}$, $\rho_{_{V,Z|X}}$ and $\rho_{_{X,Y|V}}$, as well as the sample size of main study and validation study.  
    \end{tablenotes}
 \end{threeparttable}
\end{table*}

Table \ref{tab:validity} summarizes the validity of different RSW estimators under DAGs 1 through 8, with the most efficient estimators indicated. Taking DAG 7 as an example, our result suggests that while it would be valid to adjust for $V_{7(X{}-{}-)}$ in either both outcome model and MEM (i.e. $\hat{\beta}_{(OM)}$) or neither of the two models(i.e. $\hat{\beta}_{(- -)}$), which is, in this case, the most efficient option, it is invalid to adjust for this type of variable in either MEM alone (i.e. $\hat{\beta}_{(-M)}$) or outcome model alone (i.e. $\hat{\beta}_{(O-)}$). 

Our result suggests that $V_{2(-ZY)}, V_{3(X-Y)}, V_{4(XZY)}$ are the minimal covariate adjustment set and need to be included in both MEM and outcome model. With the exception of $V_{1(-{}-Y)}$ and trivially $V_{5(-{}-{}-)}$, including any other covariates in either MEM alone or outcome model alone would result in bias. Including $V_{1(-{}-Y)}$ in either outcome model alone or both MEM and outcome model would increase efficiency, so does including $V_{6(-Z-)}$ in both MEM and outcome model. Under DAG 8, the relative efficiency of estimators $\hat{\beta}_{_{(OM)}}$ and $\hat{\beta}_{_{(--)}}$ depend on the strength and direction of the correlation of $V-X$ ($\rho_{_{V,X}}$), of $X-Z$ conditional on $V$ ($\rho_{_{X,Z|V}}$), of $V$ and the measurement error (i.e. $\rho_{_{V,Z|X}}$) and of $X-Y$ conditional on $V$ ($\rho_{_{X,Y|V}}$). To evaluate to what extent the relative efficiency is determined by $\rho_{_{V,X}}$, $\rho_{_{X,Z|V}}$ and $\rho_{_{V,Z|X}}$, we plot the analytical asymptotic relative efficiency ($\widehat{\text{ARE}}(\hat{\beta}_{_{(--)}})=\frac{\widehat{Var}(\hat{\beta}_{_{(OM)}})}{\widehat{Var}(\hat{\beta}_{_{(--)}})}$) with varying values of the three (conditional) correlations while fixing exposure effect and main study and validation study sample sizes (section 3 of SM). We find that in order to gain efficiency by adjusting $V_{8(XZ-)}$ in both outcome model and MEM, $V_{8(XZ-)}$ needs to be very weakly associated with true exposure and strongly associated with measurement error, a rare scenario in realistic settings. In the same section of SM, we also plot the $\widehat{\text{ARE}}$ under DAG 1 where we vary the value of $V-Y$ correlation condition on $X$ (i.e. $\rho_{_{V,Y|X}}$) and of $\rho_{_{X,Z|V}}$ while fixing the sample sizes. The result suggests that we gain more efficiency by adjusting for $V_{1(-{}-{}Y)}$ as $\rho_{_{X,Z|V}}$ and $\rho_{_{V,Y|X}}$ become greater.

\subsection{3.4 Approximation in Generalized Linear Models with Logistic Link} \label{s34}
For binary outcomes modeled by logistic regression, the RSW estimators will approximate the true log odds ratio with minimal bias (e.g. percent bias less than 12\%) if one of the following conditions is met: (I) $\operatorname{Var}(X|Z,V)\beta^2$ is small (e.g. less than 0.5), where $\operatorname{Var}(X|Z,V)$ can be estimated as the mean squared error from MEM and $\beta$ is the true causal parameter of interest \cite{kuha1994corrections,spiegelman2000estimation} or (II) the disease is rare (e.g. less than 5\%) and the measurement error in the MEM is homoskedastic\cite{RSW2}.

We note that the analytical efficiency results for continuous outcomes under linear models do not necessarily hold for binary outcomes modeled by logistic regression. Specifically, under DAG 1, Neuhaus et al.\cite{neuhaus1993geometric} showed that the coefficient estimate for $Z$ (i.e. $\hat{\gamma_1}^*$) when regressing a binary $Y$ on $Z$ with logistic link not conditioning on $V_{1(-{}-{}Y)}$ is always closer to null than the $V_{1(-{}-{}Y)}$-adjusted coefficient estimate for $Z$ (i.e. $\hat{\gamma_1}$), a phenomenon known as non-collapsibility of odds ratio\cite{sjolander2016note,robinson1991some}. Thus, under DAG 1, the efficiency of RSW estimators, which depends on $\hat{\gamma}_1$ and $\hat{\gamma}_1^*$, do not necessarily follow the linear regression results.

\section{SIMULATION}
We performed a series of Monte Carlo simulation experiments to empirically evaluate the finite sample validity and efficiency of the estimators for both continuous and binary outcomes, with varying model parameterization and sample sizes under each DAG.

\subsection{4.1 Simulation Study Design}
All data were generated using the following data generating process with 1,000 samples: 

\begin{enumerate}[noitemsep,topsep=0pt,parsep=0pt,partopsep=0pt]
    \item $X = \eta_{v} V + \epsilon_x$, $\epsilon_x \sim N(0,1)$, 
    \item $Z = \theta_{x} X + \theta_{v} V + \epsilon_z, \epsilon_z \sim N(0,0.5)$, 
    \item $Y = \beta_{x} X + \beta_{v}V  + \epsilon_y,\epsilon_y \sim N(0,1)$ for continuous outcome $Y$, and 
    \item $Y \sim \operatorname{Bern}(p), \text{ where } \log(p/(1-p))= -5 + \beta_{x} X + \beta_{v} V$ for binary outcome $Y$. 
\end{enumerate}

For the scenarios with binary outcome, all simulations satisfy the condition that $\widehat{\operatorname{Var}}(X|Z,V)\beta_x^2<0.5$\cite{kuha1994corrections}. 
As the base case for the 8 simulation experiments corresponding to the 8 DAGs, we generated samples of size $n=5000$ for continuous outcome $Y$ and $10,000$ for binary $Y$, where a subset of size $n_{VS}=400$ is randomly sampled as external validation data with $(V, X, Z)$ retained. The remaining $n_{MS}=4,600$ or $9,600$ was kept as main study data where only $(V, Z, Y)$ is observed. Smaller sample sizes were not used (e.g. $2,000$ for binary outcome) because modern epidemiological cohorts that use RC method often have a sample size greater than $10,000$ \cite{bao2016origin,women1998design,gu2022dietary} and the asymptotic statistical properties may not be evident when rare diseases are studied under smaller sample sizes. Covariate $V$ is distributed as $N(0,1)$ and the true causal effect of $X$ on $Y$ is set to be $\beta_x = 0.5$, additive scale for continuous outcome and logit scale for binary outcome (equivalent to an odds ratio of approximately $1.65$). For other parameters, we used all possible combinations of the following parameter values: $\eta_{v} \in (0.4,0), \theta_{x} \in 0.5, \theta_{v} \in (0.1, 0), \beta_{v} \in (0.8,0)$, which corresponds to each of the scenarios in the $2^3=8$ DAGs in Figure \ref{fig1}, with zero value representing the removal of a specific arrow on the DAG. Details for the simulation scenarios are available in section 4 of the SM. 

\subsection{4.2 Results}
The simulation results for the point estimates and empirical variances of the valid estimators for the base case are reported in Table \ref{tab:basecase} and Table \ref{tab:basecase_efficiency}, respectively. For the point estimates, we present percent biases. For efficiency comparison of the valid estimators, we reported both analytical and empirical relative efficiency ($\widehat{\text{ARE}}$ and ERE respectively) under continuous outcome, calculated as the analytical or empirical variance of $\hat{\beta}_{(OM)}$ over the analytical or empirical variance of other valid estimators, as well as analytical and empirical variance for the valid point estimates obtained over simulation replicates. For binary outcome, only empirical results are presented. Analytical and simulation results for all scenarios can be found in sections 5 and 6 of SM, which is briefly summarized in the remainder of this section.

\begin{table*}[!t]
  \centering
  \begin{threeparttable}
  \caption{Percent Bias for Point Estimates in the Simulation Study under Base Case \label{tab:basecase}}

\begin{tabular}{| l |  c c c c | c c c c |}
\hline
 &  \multicolumn{8}{ c |}{\textbf{Percent Bias (\%)}} \\
\textbf{$V_j$ as in} &  \multicolumn{4}{ c |}{\textbf{Continuous Outcome}} &  \multicolumn{4}{ c |}{\textbf{Binary outcome}\tnote{a}} \\
& \textbf{$\hat{\beta}_{(OM)}$} & \textbf{$\hat{\beta}_{(--)}$} & \textbf{$\hat{\beta}_{(-M)}$} & \textbf{$\hat{\beta}_{(O-)}$} & \textbf{$\hat{\beta}_{(OM)}$} & \textbf{$\hat{\beta}_{(--)}$} & \textbf{$\hat{\beta}_{(-M)}$} & \textbf{$\hat{\beta}_{(O-)}$} \\
\hline 
                         $V_{1,(-{}-Y)}$                                 & 0 & 0  & 0   & 0  & 0 & -1 & -1  & 0 \\
                         $V_{2,(-{}ZY)}$                                 & 0 & 32 & 30   & 2 & 0 & 31 & 28   & 2 \\
                         $V_{3,(X{}-{}Y)}$                               & 0 & 55 & 67  & -7  & 0 & 52 & 64  & -7 \\
                         $V_{4,(X{}Z{}Y)}$                               & 0 & 78 & 87  & -5 & 0 & 75 & 84  & -5 \\
                         $V_{5,(-{}-{}-)}$                               & 0 & 0  & 0   & 0   & 1 & 1 & 1  & 1\\
                         $V_{6,(-{}Z{}-)}$                               & 0 & 0  & -2 & 2  & 1 & 1 & -1 & 3\\
                         $V_{7,(X{}-{}-)}$                               & 0 & 0  & 8   & -7 & 1 & 2 & 9   & -6  \\
                         $V_{8,(X{}Z{}-)}$                               & 0 & 0  & 5 & -5  & 1 & 2 & 7 & -4 \\
\hline
\end{tabular}
         \begin{tablenotes}
        \item[a] Satisfies small measurement error condition $\widehat{var}(X|Z,V)\beta^2<0.5)$.
    \end{tablenotes}
 \end{threeparttable}
\end{table*}

For continuous outcomes, all simulation results are consistent with analytical expectations. For binary outcomes with logistic outcome model and linear MEM, the percent bias is less than 6\% across all scenarios except one (where percent bias is 12\%). However, we noticed that adjusting for $V_{1(--Y)}$ in either outcome model only or in both models resulted in similar variances as when adjusting for $V_{1(--Y)}$ as in either $\beta_{(-M)}$ or $\beta_{(--)}$ (section 6 of SM). This could be due to that the odds ratios are non-collapsible and that in finite sample, the efficiency gain from including the covariate cannot compensate the loss of efficiency due to estimation of the additional parameters.

For covariates $V_{2(-ZY)}$, $V_{3(X-Y)}$ and $V_{4(-ZY)}$, adjusting for them in the outcome model only results in much less biased estimates (e.g. percent bias $< 10\%$ in most cases) than would be obtained if one does not adjust for them in any model (e.g. percent bias $\ge 30\% $ in most cases) provided that the correlation between these covariate(s) and measurement error is small (e.g. $\rho_{v,z|x} \le 0.2$). 

\begin{table*}[!t]
  \centering
  \begin{threeparttable}
  \caption{Empirical Asymptotic Relative Efficiency (ERE) and Variance in the Base Case Simulation Study When More Than One Estimator is Valid  \label{tab:basecase_efficiency}}

\begin{tabular}{| l l | c c c c |}
\hline
\textbf{Outcome Type} & \textbf{$V_j$ as in} &  \multicolumn{4}{ c |}{\textbf{ERE \tnote{a} (Empirical Variance \tnote{b} )}} \\
& & \textbf{$\hat{\beta}_{(OM)}$}   & \textbf{$\hat{\beta}_{(--)}$} & \textbf{$\hat{\beta}_{(-M)}$} & \textbf{$\hat{\beta}_{(O-)}$} \\
\hline
Continuous & $V_{1,(-{}-Y)}$ &  1 (1.14) & 0.79 (1.44) & 0.79 (1.44) & 1.00 (1.14) \\
           & $V_{5,(-{}-{}-)}$ &  1 (1.14) & 1.00 (1.14)    & 1.00 (1.14)    & 1.00 (1.14) \\
           & $V_{6,(-{}Z{}-)}$ & 1 (1.14) & 0.97 (1.18) &             &          \\
           & $V_{7,(X{}-{}-)}$ & 1 (1.14) & 1.19 (0.96) &             &          \\
           & $V_{8,(X{}Z{}-)}$ & 1 (1.14) & 1.28 (0.89) &             &          \\
\hline
Binary \tnote{c}     & $V_{1,(-{}-Y)}$ & 1 (2.11) & 1.01 (2.09) & 1.01 (2.09) & 1 (2.11) \\
           & $V_{5,(-{}-{}-)}$ & 1 (2.90)  & 1.00 (2.91) & 1.00 (2.90)     & 1 (2.91) \\
           & $V_{6,(-{}Z{}-)}$ & 1 (2.90)  & 0.97 (2.99) &             &          \\
           & $V_{7,(X{}-{}-)}$ & 1 (2.84) & 1.25 (2.28) &             &          \\
           & $V_{8,(X{}Z{}-)}$ & 1 (2.84) & 1.31 (2.17) &             &  \\
\hline
\end{tabular}
         \begin{tablenotes}
        \item[a] Calculated as the empirical variance of $\hat{\beta}_{(OM)}$ over the empirical variance of each estimator $\hat{\beta}_{(\cdot)}$ under consideration, i.e. $\text{ERE} (\hat{\beta}_{(\cdot)}) = \frac{Var(\hat{\beta}_{_{(OM)}})}{Var(\hat{\beta}_{{(\cdot)}})}$, where $\hat{\beta}_{(\cdot)}$ could be any of the four estimators $\hat{\beta}_{(OM)}$, $\hat{\beta}_{(--)}$, $\hat{\beta}_{(-M)}$ or $\hat{\beta}_{(O-)}$.
        \item[b] Calculated as the variance of the point estimates $\hat{\beta}_{(\cdot)}$ over simulation replicates for each estimator. $\times 10^{-3}$ for continuous outcome and $\times 10^{-2}$ for binary outcome.  
        \item[c] Satisfies small measurement error condition $\widehat{var}(X|Z,V)\beta^2<0.5$.
    \end{tablenotes}
 \end{threeparttable}
\end{table*}

\section{REAL-DATA EXAMPLE}
\subsection{5.1 Methods}

We applied the proposed covariate selection framework to the study of effect of total daily fiber intake on the risk of cardiovascular event, defined as either myocardial infarction or angina, among the HPFS. HPFS is an ongoing prospective study of 51,529 U.S. male health professionals 40 to 75 years of age at enrollment in 1986\cite{gu2022dietary,kim2014longitudinal}. The surrogate exposure, fiber intake (g/day), was measured using a food frequency questionnaire (FFQ) every 4 years since 1986 \cite{gu2022dietary}. A validation study of size 651 using dietary records (DR) was conducted in 2012 among the cohort participants. The participants were followed up from 1990 to 2016  \cite{cahill2013prospective,liu2002prospective}. For this example, we created a cohort consisting of participants who completed the sleep duration question in 1987 and sunscreen use question in 1992. We excluded participants who died, reported a CVD event, or any cancer diagnosis other than melanoma prior to 1990, when follow-up started. The final analysis includes 22,379 participants, among whom 409 participants were in the validation study and the remaining (n = 21,970) were the main study participants. 

We included the following covariates in our analysis, with their corresponding covariate set indicated in parentheses: family history of CVD ($V_{1(-{}-Y)}$), energy intake, baseline hypertension, diabetes and hypercholesterolemia, smoking, marital status ($V_{3(X-Y)}$), age, physical exercise/activity\cite{liu2002prospective, mandolesi2018effects}, body mass index (BMI), use of multivitamin supplements\cite{liu2002prospective, grima2012effects}, alcohol intake \cite{brennan2020long}, depression status\cite{mcdermott2009meta,van2007depression}, sleep duration\cite{ma2020association,hoevenaar2011sleep}($V_{4(XZY)}$), and frequency of sunscreen use ($V_{7(X-{}-)}$). We encoded these assumptions into the DAG in Figure \ref{fig2}. We allow correlation between covariate sets through the unmeasured common cause $U$. For example, physical exercise and fiber intake could be correlated due to an unmeasured common cause related to healthy lifestyle. Note that sunscreen use cannot reasonably be assumed to directly cause changes in fiber intake. Therefore, we show the backdoor path through the unmeasured $U$ (e.g. healthy lifestyle), connecting $V_{7(X-{}-)}$ and $X$. In this way, sunscreen use has the essential feature that it is not a direct cause of cardiovascular incidence and that it is conditionally independent of measurement error. See section 7 of the SM for more details. 

\usetikzlibrary{shapes,decorations,arrows,calc,arrows.meta,fit,positioning}
\begin{figure*}[!t]
    \begin{center}
     \caption{\label{fig2}Proposed DAG for the Study of Effect of Fiber Intake on Cardiovascular Disease Incidence in HPFS}
             \begin{tikzpicture}[>= stealth, shorten >= 1pt, auto, node distance = 0.8 cm, semithick]
            \draw (0,0) node[above] (x) {fiber, DR (X)};
            \draw (4,0) node[above] (z) {fiber, FFQ (Z)};
            \draw (7,0) node[above] (y) {CVD (Y)};

            \draw (0, -3) node[below,text width=7cm] (v3) {{$V_{3(X-Y)}$: energy intake, hypertension, diabetes, hypercholesterolemia, smoking status}};
            \draw (-5, 3) node[below,text width=4cm] (v1) {{$V_{1(-{}-Y)}:$ marital status, family history}};
            \draw (-4,-1) node[below,text width=6cm] (v4) {{$V_{4(XZY)}:$ age, physical exercise, BMI, multivitamin use, alcohol use, sleep duration, depression}};
            \draw (-5,1) node[below] (v7) {{$V_{7(X{}-{}-)}:$ sunscreen use}};
            \draw (-8,-1) node[below] (u) {{$U$}};
                \path[->] (x) edge node {} (z);
                \path[->] (x) edge [out=45, in=135] node {} (y);
                \path[->] (v1) edge [out=0, in=135] node {} (y);
                \path[->] (v3) edge node {} (x);
                \path[->] (v3) edge [out= 0, in=270] node {} (y);
                \path[->] (v4) edge node {} (z);
                \path[->] (v4) edge node {} (x);
                \path[->] (v4) edge [out=0, in=240] node {} (y);
                \path[->] (u) edge [dashed, out=30, in=180] node {} (x);

               \path[->] (u) edge [dashed, out=270, in=180] node {} (v3);
                \path[->] (u) edge [dashed, out=270, in=180] node {} (v4);
                \path[->] (u) edge [dashed, out=270, in=180] node {} (v7);

                


            \end{tikzpicture}
    \end{center}
\end{figure*}

 In accordance with the theoretical results, we adjusted for covariates sets $V_{3(X-Y)}$ and $V_{4(XZY)}$, the minimal covariate adjustment set in the absence of $V_{2(-ZY)}$, in both the outcome model and MEM ($\hat{\beta}_{(OM)}$). We also adjusted for $V_{1(-{}-Y)}$ in the outcome model only ($\hat{\beta}_{(O-)}$) and $V_{7(X{}-{}-)}$ in both outcome and MEM ($\hat{\beta}_{(OM)}$) in addition to including $V_{3(X-Y)}, V_{4(XZY)}$ in both MEM and the outcome model. We additionally evaluated the bias induced by omitting $V_{3(X-Y)}$ or $V_{4(XZY)}$ from the MEM alone or from both MEM and outcome model. 

\subsection{5.2 Results}
We summarize the effect of 10g in daily fiber intake on incidence of CVD in Table \ref{tab:application}. Over the 16 years of follow-up, 2,090 (9\%) cardiovascular events were reported. The standard analysis regressing CVD event on total fiber adjusting for $V_{3(X-Y)}$ and $V_{4(XZY)}$ produces an attenuated odds ratio estimate of 0.93 associated with 10g increase in daily total fiber intake. Adjusting for this set of covariates in both MEM and outcome model gives a corrected odds ratio of $0.83$, so do all other valid estimators. Consistent with theory, we lose efficiency by additionally adjusting for $V_{7(X{}-{}-)}$ in both outcome model and MEM while the point estimate remained similar, increasing the standard error of the parameter of interest from $9.46 \times 10^{-2}$ when adjusting only for ${V_{3(X-Y)}, V_{4(XZY)}}$ to $9.54  \times 10^{-2}$. 
Omitting $V_{4(XZY)}$ from both MEM and the outcome model produced the most biased point estimate of 1.03. As long as $V_{3(X-Y)}$ or $V_{4(XZY)}$ are adjusted for in the outcome model, the bias resulting from omitting these variables from the MEM is very minimal. This result is consistent with those from our simulation as the empirical partial correlation between each of ($V_{3(X-Y)}$, $V_{4(XZY)}$) and $Z$ (conditional on $X$ and other covariates) are all less than 0.17.

\begin{table*}[!t]
  \begin{threeparttable}
  \caption{Effect Estimates of 10g/d Increase in Fiber Intake on CVD Risk in HPFS \label{tab:application}}
  
  \small
    \begin{tabular}{| r | c c | c c c |}
        \hline
        RSW Adjusted Sets (Models) \tnote{a} & \multicolumn{2}{c|}{Uncorrected Analysis}  & \multicolumn{3}{c|}{Corrected Analysis} \\
        {} & Uncorrected OR & 95\% CI & Corrected OR & 95\% CI & SE ($10^{-2}$) \\
        \hline
        \multicolumn{6}{|l|}{Theoretically Valid RSW Estimators} \\
        \hline
        $V_{3(X-Y)}, V_{4(XZY)}$ (OM) & 0.93 & (0.86, 1.00) & 0.83 & (0.69, 1.00) &  9.46\\
        $V_{1(-{}-{}Y)}, V_{3(X-Y)}, V_{4(XZY)}$ (OM) & 0.93 & (0.86, 1.00) & 0.83 & (0.69, 1.00) &  9.45\\
        $V_{1(-{}-{}Y)}$ (-M) , $V_{3(X-Y)}, V_{4(XZY)}$ (OM) & 0.93 & (0.86, 1.00) & 0.83 & (0.69, 1.00) &  9.44\\
        $V_{1(-{}-Y)}$ (O-), $V_{3(X-Y)}, V_{4(XZY)}$ (OM) & 0.93 & (0.86, 1.00) & 0.83 & (0.69, 1.00) & 9.47\\
        $V_{3(X-Y)}, V_{4(XZY)}, V_{7(X{}-{}-)}$ (OM) & 0.93 & (0.87, 1.00) & 0.83 & (0.69, 1.01) & 9.54\\
        \hline
        \multicolumn{6}{|l|}{Theoretically Biased RSW Estimators}\\
        \hline
        $V_{3(X-Y)}$ (OM) & 1.01 & (0.86, 1.00) & 1.03 & (0.89, 1.20) &  7.68\\
        $V_{4(XZY)}$ (OM) & 0.92 & (0.86, 1.00) & 0.81 & (0.67, 0.97) &  9.40\\
        $V_{4(XZY)}$ (O-), $V_{3(XZY)}$ (OM) & 0.93 & (0.86, 1.00) & 0.85 & (0.71, 1.00) &  8.54\\
        $V_{3(X-Y)}$ (O-), $V_{4(XZY)}$ (OM) & 0.93 & (0.86, 1.00) & 0.83 & (0.69, 1.00) &  9.38\\
        \hline
    \end{tabular}
        \begin{tablenotes}
        \item[a] (OM), (-M) and (O-) indicate whether the covariate(s) are adjusted in both outcome model and MEM, MEM only or outcome model only. For example, $V_{1(-{}-{}Y)}$ (-M) , $V_{3(X-Y)}, V_{4(XZY)}$ (OM) indicates that $V_{1(-{}-{}Y)}$ is adjusted in MEM only and $V_{3(X-Y)}, V_{4(XZY)}$ are adjusted in both MEM and outcome model. 
    \end{tablenotes}
    \end{threeparttable}
\end{table*}

\section{DISCUSSION}
\subsection{6.1 Summary of Main Results}
First and foremost, we showed that $V=(V_{2(-ZY)}, V_{3(X-Y)}, V_{4(XZY)})$, i.e. any common causes (1) of true exposure and outcome and (2) of measurement error and outcome, are the minimal covariate set and should be adjusted for in both outcome model and MEM. When designing a study that includes exposure measurement error correction, investigators should ensure the collection of $V$ in both the main \emph{and} validation studies. One may find it counter-intuitive that confounders not contributing to measurement error (i.e. $V_{3(X-Y)}$) must be included in the MEM, with the rationale that under DAG 3, $Z\perp V_{3(X-Y)}|X$ and therefore $E[Z|X] = E[Z|X, V_{3(X-Y)}]$, which could suggest that $V_{3(X-Y)}$ is not necessary in the MEM. However, note that the MEM used in RC method models the mean of $X$ rather than $Z$ and thus bias is induced when $V_{3(X-Y)}$ is excluded. For $V_{2(-ZY)}$, while they are not confounding variables for $X$'s effect on $Y$, by conditioning on $Z$, a collider for $X$ and $V_{2(-ZY)}$ in both models, we induce an association between $X$ and $V_{2(-ZY)}$, making $V_{2(-ZY)}$ de facto confounders for the $X-Y$ relationship.

Second, our results suggest that the RSW RC method can be relaxed to allow for the adjustment of risk factors that are conditionally independent of both true exposure and measurement error in the outcome model only. This provides an opportunity to gain statistical efficiency when such variables are only available in the main study. Another opportunity for increasing efficiency would be to adjust for non-risk-factors that are strong determinants of measurement error but have no association with true exposure (e.g.$V_{6(-{}Z{}-)}$). In short, to maximize efficiency with validity guaranteed, one should adjust for variables of types $V_{1(-{}-Y)}$ through $V_{6(-{}Z{}-)}$ in both MEM and outcome models. 

Last, we recommend that when covariates $V=(V_{2(-ZY)}, V_{3(X-Y)}, V_{4(XZY)})$ are not available in the validation study, in order to obtain a less biased estimate, they should still be adjusted for at least in the outcome model, provided that these covariates are not strongly correlated with the measurement error, as in most realistic settings. We also recommend that researchers leave out any non-risk factors that are even weakly associated with the true exposure, as adjustment for these covariates in both outcome model and MEM almost always result in efficiency loss. 

\subsection{6.2 Main Study/Internal Validation Study Design}

For internal validation studies, the true exposure $X$ is partially observed in the main study. Spiegelman et al.\cite{spiegelman2001efficient} proposed an inverse-variance-weighted estimator (IVWE) that combines the corrected RSW estimate using external validation study with the additional regression coefficient estimate obtained from regressing outcome on the true exposure in the main study, with the IVWE being asymptotically nearly unbiased and efficient as the optimal maximum-likelihood estimator. 

\subsection{6.3 Effect Modification by Covariate $V$} \label{s35}
The identification formula \eqref{identify_formula_V} allows for effect modification by $V$ and can be estimated non-parametrically or parametrically. However, parametric estimation of $\beta(V)$ requires special attention. For example, we demonstrate in section 8.1 of SM that under an additive-scale $X-V$ interaction, one would need to add a $Z-V$ product term as well as a $V^2$ term in the outcome model, i.e. $E[Y|Z,V] = \beta_0^* + \beta_z^* Z + \beta_v^* V + \beta_{zv}^* ZV + \beta_{v^2}^* V^2$. Together with MEM in \eqref{eq1}, this gives CATE estimator $\hat{\beta} (V) = \frac{\hat{\beta}_z^* + \hat{\beta}_{zv}^* V}{\hat{\alpha}_1}$.  

More generally, we can extend the parametric RSW estimators in section \ref{parametricRSWEstimators} to the following non-parametric estimators:

\begin{equation} \label{eq13}
\hat{\beta}_{(OM)} (V) = \frac{\hat{E} [Y|Z=z,V] - \hat{E}[Y|Z=z' , V]}{\hat{E}[X|Z=z,V] - \hat{E}[X|Z=z' , V]}
\end{equation}
\begin{equation} \label{eq14}
\hat{\beta}_{(-M)} (V) = \frac{\hat{E}[Y|Z=z] - \hat{E}[Y|Z=z']}{\hat{E}[X|Z=z,V] - \hat{E}[X|Z=z' , V]}
\end{equation}
\begin{equation} \label{eq15}
\hat{\beta}_{(O-)} (V) = \frac{\hat{E}[Y|Z=z,V] - \hat{E}[Y|Z=z' , V]}{\hat{E}[X|Z=z] - \hat{E}[X|Z=z' ]},
\end{equation}

We also show in section 8.2 of SM that the validity result in \ref{theoreticalResults} still apply in all cases except that $\hat{\beta}_{_{(-M)}}(V)$ can no longer estimate CATE when $V_{1(-{}-{}Y)}$ is an effect modifier. The evaluation of (relative) efficiency for the above estimators is beyond the scope of this study.

\subsection{6.4 Other Measurement Error Structures}
\begin{figure*}[!t]
     \caption{Other Measurement Error Structures}
   \begin{subfigure}[b]{0.5\textwidth} 
         \begin{tikzpicture}[>= stealth, shorten >= 1pt, auto, node distance = 0.8 cm, semithick]
            \node[text centered] (x) {$X$};
            \node[right= of x] (v) {$V$ (Biomarker)};
            \node[right= of v] (y) {$Y$};
            \node[below= of v] (z) {$Z$};
            \path[->] (x) edge node {} (z);
            \path[->] (x) edge [out=45, in=135] node {} (y);
            \path[->] (v) edge node {} (y);
            \path[->] (x) edge node {} (v);
        \end{tikzpicture}
        \caption{\label{fig3a}}
     \end{subfigure}
  \begin{subfigure}[b]{0.5\textwidth} 
         \begin{tikzpicture}[>= stealth, shorten >= 1pt, auto, node distance = 0.8 cm, semithick]
            \node[text centered] (x) {$X$};
            \node[right= of x] (v) {$V$ (BMI)};
            \node[right= of v] (y) {$Y$};
            \node[below= of v] (z) {$Z$};
            \path[->] (x) edge node {} (z);
            \path[->] (x) edge [out=45, in=135] node {} (y);
            \path[->] (v) edge node {} (y);
            \path[->] (x) edge node {} (v);
            \path[->] (v) edge node {} (z);
        \end{tikzpicture}
        \caption{\label{fig3b}}
     \end{subfigure}
\end{figure*}
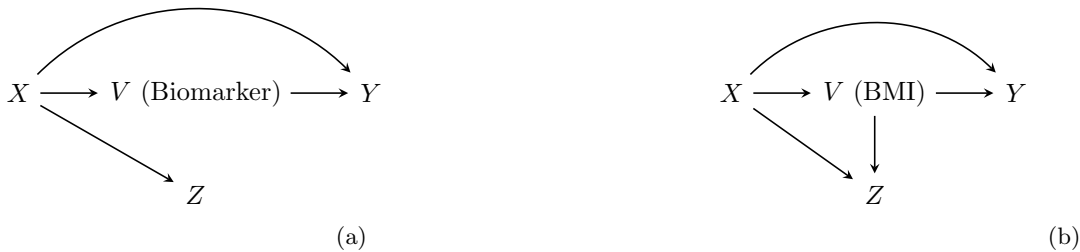

We are aware that the measurement error structures in Figure \ref{fig1} do not cover all research contexts. For example, instead of being a confounder, a variable such as a biomarker or BMI may be a mediator between dietary intake and outcome, with BMI potentially inducing exposure measurement error, as depicted in DAGs \ref{fig3a} and \ref{fig3b} respectively. Under these scenarios, it is unclear what meaningful causal estimand, if any, the RSW estimators correspond to. We will discuss these measurement error structures in new papers. 
 
Finally, we have devoted a separate manuscript to the covariate selection issue using CRS RC method, as it does not always require the same covariate adjustment rules as the RSW RC method.

\section{ACKNOWLEDGEMENT}
\textit{We thank Drs Jessica Young, James Robins and Christopher Boyer (Harvard University) for helpful comments on different versions of this article and its supplemental material.}

\small
\bibliographystyle{unsrt}
\bibliography{sample}

\begin{thebibliography}{10}

\bibitem{RSWapp1}
Eric~B. Rimm, Edward~L. Giovannucci, Meir~J. Stampfer, Graham~A. Colditz,
  Lisa~B. Litin, and Walter~C. Willett.
\newblock {Reproducibility and Validity of an Expanded Self-Administered
  Semiquantitative Food Frequency Questionnaire among Male Health
  Professionals}.
\newblock {\em American Journal of Epidemiology}, 135(10):1114--1126, 05 1992.

\bibitem{RSWapp2}
Edward Giovannucci, Yan Liu, Eric~B. Rimm, Bruce~W. Hollis, Charles~S. Fuchs,
  Meir~J. Stampfer, and Walter~C. Willett.
\newblock {Prospective Study of Predictors of Vitamin D Status and Cancer
  Incidence and Mortality in Men}.
\newblock {\em JNCI: Journal of the National Cancer Institute}, 98(7):451--459,
  04 2006.

\bibitem{RSWapp3}
Mihaela Tanasescu, Michael~F. Leitzmann, Eric~B. Rimm, Walter~C. Willett,
  Meir~J. Stampfer, and Frank~B. Hu.
\newblock {Exercise Type and Intensity in Relation to Coronary Heart Disease in
  Men}.
\newblock {\em JAMA}, 288(16):1994--2000, 10 2002.

\bibitem{CRSapp1}
Robert Kaestner and Neeraj Kaushal.
\newblock Effect of immigrant nurses on labor market outcomes of us nurses.
\newblock {\em Journal of Urban Economics}, 71(2):219 -- 229, 2012.

\bibitem{CRSapp2}
Lori~D. Bash, Thomas~P. Erlinger, Josef Coresh, Jane Marsh-Manzi, Aaron~R.
  Folsom, and Brad~C. Astor.
\newblock Inflammation, hemostasis, and the risk of kidney function decline in
  the atherosclerosis risk in communities (aric) study.
\newblock {\em American Journal of Kidney Diseases}, 53(4):596 -- 605, 2009.

\bibitem{CRSapp3}
Clairea~Philippat Célinea~Vernet and Lydianea Agier.
\newblock An empirical validation of the within-subject biospecimens pooling
  approach to minimize exposure misclassification in biomarker-based studies.
\newblock {\em Epidemiology}, 30:756 -- 767, 2019.

\bibitem{RSW1}
B.~Rosner, W.~C. Willett, and D.~Spiegelman.
\newblock Correction of logistic regression relative risk estimates and
  confidence intervals for systematic within-person measurement error.
\newblock {\em Statistics in Medicine}, 8(9):1051--1069, 1989.

\bibitem{RSW2}
B.~Rosner, D.~Spiegelman, and W.~C. Willet.
\newblock {Correction of Logistic Regression Relative Risk Estimates and
  Confidence Intervals for Measurement Error: the Case of Multiple Covaraites
  Measured with Error}.
\newblock {\em American Journal of Epidemiology}, 132(4):734--745, 10 1990.

\bibitem{liao2011survival}
Xiaomei Liao, David~M Zucker, Yi~Li, and Donna Spiegelman.
\newblock Survival analysis with error-prone time-varying covariates: A risk
  set calibration approach.
\newblock {\em Biometrics}, 67(1):50--58, 2011.

\bibitem{white2006commentary}
Ian~R White.
\newblock Commeagentary: Dealing with measurement error: multiple imputation or
  regression calibration?
\newblock {\em International Journal of Epidemiology}, 35(4):1081--1082, 2006.

\bibitem{CRS_book}
Stefanski~L. Carroll~R., Ruppert~D. and C.~Crainiceanu.
\newblock {\em Measurement Error in Nonlinear Models: A Modern Perspective}.
\newblock New York: Chapman and Hall/CRC, 2 edition, 2006.
\newblock https://doi.org/10.1201/9781420010138.

\bibitem{CRS1}
James~W. Hardin, Henrik Schmiediche, and Raymond~J. Carroll.
\newblock The regression-calibration method for fitting generalized linear
  models with additive measurement error.
\newblock {\em The Stata Journal}, 3(4):361--372, 2003.

\bibitem{ambrose2004pathophysiology}
John~A Ambrose and Rajat~S Barua.
\newblock The pathophysiology of cigarette smoking and cardiovascular disease:
  an update.
\newblock {\em Journal of the American college of cardiology},
  43(10):1731--1737, 2004.

\bibitem{mandolesi2018effects}
Laura Mandolesi, Arianna Polverino, Simone Montuori, Francesca Foti, Giampaolo
  Ferraioli, Pierpaolo Sorrentino, and Giuseppe Sorrentino.
\newblock Effects of physical exercise on cognitive functioning and wellbeing:
  biological and psychological benefits.
\newblock {\em Frontiers in psychology}, 9:509, 2018.

\bibitem{mcdermott2009meta}
Lisa~M McDermott and Klaus~P Ebmeier.
\newblock A meta-analysis of depression severity and cognitive function.
\newblock {\em Journal of affective disorders}, 119(1-3):1--8, 2009.

\bibitem{ma2020association}
Yanjun Ma, Lirong Liang, Fanfan Zheng, Le~Shi, Baoliang Zhong, and Wuxiang Xie.
\newblock Association between sleep duration and cognitive decline.
\newblock {\em JAMA network open}, 3(9):e2013573--e2013573, 2020.

\bibitem{cahill2013prospective}
Leah~E Cahill, Stephanie~E Chiuve, Rania~A Mekary, Majken~K Jensen, Alan~J
  Flint, Frank~B Hu, and Eric~B Rimm.
\newblock Prospective study of breakfast eating and incident coronary heart
  disease in a cohort of male us health professionals.
\newblock {\em Circulation}, 128(4):337--343, 2013.

\bibitem{rosner1990correction}
B~Rosner, Donna Spiegelman, and Walter~C Willett.
\newblock Correction of logistic regression relative risk estimates and
  confidence intervals for measurement error: the case of multiple covariates
  measured with error.
\newblock {\em American journal of epidemiology}, 132(4):734--745, 1990.

\bibitem{LoganSpiegelman2012blinplus}
Roger Logan and Donna Spiegelman.
\newblock Sas: The sas \%blinplus macro.
\newblock \url{https://ysph.yale.edu/cmips/research/software/blinplus/}, 2012.
\newblock Accessed: 2022-09-30.

\bibitem{vanderweele2012results}
Tyler~J VanderWeele and Miguel~A Hern{\'a}n.
\newblock Results on differential and dependent measurement error of the
  exposure and the outcome using signed directed acyclic graphs.
\newblock {\em American journal of epidemiology}, 175(12):1303--1310, 2012.

\bibitem{weinberg1994will}
Clarice~A Weinberg, David~M Umbach, and Sander Greenland.
\newblock When will nondifferential misclassification of an exposure preserve
  the direction of a trend?
\newblock {\em American Journal of Epidemiology}, 140(6):565--571, 1994.

\bibitem{HernanCole2009}
Miguel~A. Hernán and Stephen~R. Cole.
\newblock {Invited Commentary: Causal Diagrams and Measurement Bias}.
\newblock {\em American Journal of Epidemiology}, 170(8):959--962, 09 2009.

\bibitem{VanderweeleHernan2012}
Tyler~J. VanderWeele and Miguel~A. Hernán.
\newblock {Results on Differential and Dependent Measurement Error of the
  Exposure and the Outcome Using Signed Directed Acyclic Graphs}.
\newblock {\em American Journal of Epidemiology}, 175(12):1303--1310, 05 2012.

\bibitem{weisskopf2017trade}
Marc~G Weisskopf and Thomas~F Webster.
\newblock Trade-offs of personal vs. more proxy exposure measures in
  environmental epidemiology.
\newblock {\em Epidemiology (Cambridge, Mass.)}, 28(5):635, 2017.

\bibitem{zivichuse}
Paul~N Zivich, Bonnie~E Shook-Sa, Jessie~K Edwards, Daniel Westreich, and
  Stephen~R Cole.
\newblock On the use of covariate supersets for identification conditions.
\newblock {\em Epidemiology}, pages 10--1097, 2022.

\bibitem{HernanRobinsWhatIf}
Miguel Hernán and James Robins.
\newblock {\em Causal Inference: What If}.
\newblock Boca Raton: Chapman and Hall/CR, 1 edition, 2020.

\bibitem{Richardson2013SWIG}
Thomas~S. Richardson and James Robins.
\newblock Single world intervention graphs (swigs) : A unification of the
  counterfactual and graphical approaches to causality.
\newblock University of Washington, 2013.

\bibitem{kuha1994corrections}
Jouni Kuha.
\newblock Corrections for exposure measurement error in logistic regression
  models with an application to nutritional data.
\newblock {\em Statistics in medicine}, 13(11):1135--1148, 1994.

\bibitem{spiegelman2000estimation}
Donna Spiegelman, Bernard Rosner, and Roger Logan.
\newblock Estimation and inference for logistic regression with covariate
  misclassification and measurement error in main study/validation study
  designs.
\newblock {\em Journal of the American Statistical Association},
  95(449):51--61, 2000.

\bibitem{neuhaus1993geometric}
John~M Neuhaus and Nicholas~P Jewell.
\newblock A geometric approach to assess bias due to omitted covariates in
  generalized linear models.
\newblock {\em Biometrika}, 80(4):807--815, 1993.

\bibitem{sjolander2016note}
Arvid Sj{\"o}lander, Elisabeth Dahlqwist, and Johan Zetterqvist.
\newblock A note on the noncollapsibility of rate differences and rate ratios.
\newblock {\em Epidemiology}, 27(3):356--359, 2016.

\bibitem{robinson1991some}
Laurence~D Robinson and Nicholas~P Jewell.
\newblock Some surprising results about covariate adjustment in logistic
  regression models.
\newblock {\em International Statistical Review/Revue Internationale de
  Statistique}, pages 227--240, 1991.

\bibitem{bao2016origin}
Ying Bao, Monica~L Bertoia, Elizabeth~B Lenart, Meir~J Stampfer, Walter~C
  Willett, Frank~E Speizer, and Jorge~E Chavarro.
\newblock Origin, methods, and evolution of the three nurses’ health studies.
\newblock {\em American journal of public health}, 106(9):1573--1581, 2016.

\bibitem{women1998design}
Women's Health Initiative~Study Group et~al.
\newblock Design of the women's health initiative clinical trial and
  observational study.
\newblock {\em Control Clin Trials}, 19:61--109, 1998.

\bibitem{gu2022dietary}
Xiao Gu, Dong~D Wang, Teresa~T Fung, Dariush Mozaffarian, Luc Djouss{\'e},
  Bernard Rosner, Frank~M Sacks, and Walter~C Willett.
\newblock Dietary quality and risk of heart failure in men.
\newblock {\em The American Journal of Clinical Nutrition}, 116(2):378--385,
  2022.

\bibitem{kim2014longitudinal}
Hyun~Ja Kim, Edward Giovannucci, Bernard Rosner, Walter~C Willett, and Eunyoung
  Cho.
\newblock Longitudinal and secular trends in dietary supplement use: nurses'
  health study and health professionals follow-up study, 1986-2006.
\newblock {\em Journal of the Academy of Nutrition and Dietetics},
  114(3):436--443, 2014.

\bibitem{liu2002prospective}
Simin Liu, Julie~E Buring, Howard~D Sesso, Eric~B Rimm, Walter~C Willett, and
  JoAnn~E Manson.
\newblock A prospective study of dietary fiber intake and risk of
  cardiovascular disease among women.
\newblock {\em Journal of the American College of Cardiology}, 39(1):49--56,
  2002.

\bibitem{grima2012effects}
Natalie~A Grima, Matthew~P Pase, Helen Macpherson, and Andrew Pipingas.
\newblock The effects of multivitamins on cognitive performance: a systematic
  review and meta-analysis.
\newblock {\em Journal of Alzheimer's Disease}, 29(3):561--569, 2012.

\bibitem{brennan2020long}
Sue~E Brennan, Steve McDonald, Matthew~J Page, Jane Reid, Stephanie Ward,
  Andrew~B Forbes, and Joanne~E McKenzie.
\newblock Long-term effects of alcohol consumption on cognitive function: a
  systematic review and dose-response analysis of evidence published between
  2007 and 2018.
\newblock {\em Systematic reviews}, 9(1):1--39, 2020.

\bibitem{van2007depression}
Koen Van~der Kooy, Hein Van~Hout, Harm Marwijk, Haan Marten, Coen Stehouwer,
  and Aartjan Beekman.
\newblock Depression and the risk for cardiovascular diseases: systematic
  review and meta analysis.
\newblock {\em International Journal of Geriatric Psychiatry: A journal of the
  psychiatry of late life and allied sciences}, 22(7):613--626, 2007.

\bibitem{hoevenaar2011sleep}
Marieke~P Hoevenaar-Blom, Annemieke~MW Spijkerman, Daan Kromhout, Julia~F
  van~den Berg, and WM3198203 Verschuren.
\newblock Sleep duration and sleep quality in relation to 12-year
  cardiovascular disease incidence: the morgen study.
\newblock {\em Sleep}, 34(11):1487--1492, 2011.

\bibitem{spiegelman2001efficient}
Donna Spiegelman, Raymond~J Carroll, and Victor Kipnis.
\newblock Efficient regression calibration for logistic regression in main
  study/internal validation study designs with an imperfect reference
  instrument.
\newblock {\em Statistics in medicine}, 20(1):139--160, 2001.

\end{thebibliography}
\end{multicols}
\end{document}


\title{Causally Select Covariates in Regression Calibration for Mismeasured Continuous Exposure: Web Supplemental Material}


\section{Validity of RSW Estimators}

\subsection{Data Set-up}\label{A1}
As specified in the manuscript, we assume that the underlying data generating law, $P(\mathbf{O})$, as consistent with each DAG, gives data $\mathbf{O}=(X,Y,Z,V)$, where $X$ is true exposure, $Z$ is mismeasured exposure, $V$ is a set of covariates, $Y$ outcome and $(X,Y,V)$ and $(X,Z,V)$ are observed in main study and validation study respectively, following a MS/EVS design. Because the main and validation study data are both generated by $\mathbf{O}=(X,Y,Z,V)$, transportability is guaranteed for all RSW estimators considered in the paper.  This automatically gives us transportability between the main and validation studies for all RSW estimators. All statistical models in section \ref{A3} are defined with respect to law $P(\mathbf{O})$. 

For simplicity, we treat $V$ as univariate but it can be easily extended to multvariate case and we use $V$ to generically indicate the covariate set $V_j$ in each DAG where $j=1,2,\dots, 8$.

\subsection{Identification} \label{A2}
\subsubsection{Identification under DAG 1 through 8} \label{A21}
Same as the manuscript, we are interested in either the conditional average treatment effect $E[Y^x - Y^{x^{'}}|V]$ or to a lesser extent the (marginal) average treatment effect $E[Y^x - Y^{x^{'}}]$, where they are connected by the equation $E[Y^x - Y^{x^{'}}] = \sum_v E[Y^x - Y^{x^{'}}|V=v] P(V=v)$. We start with conditional average treatment effect:

\begin{align*} \tag{A1} \label{target_x}
 E[Y^x - Y^{x^{'}}|V].
\end{align*} 
If we are willing to further assume that $V$ does not modify effect of $X$ on $Y$ then we can write: 
\begin{align*} \tag{A2} \label{target_x_alternate}
 E[Y^x - Y^{x^{'}}|V] = E[Y^x - Y^{x^{'}}].
\end{align*}

But for now, we allow such effect modification. Let us fix reference $x'$ value to $X=0$ and rewrite the causal estimand in \eqref{target_x} as the following: 

\begin{align*}  \tag{A3}  \label{target_x_0}
E[Y^x - Y^0 | V]
\end{align*} 

\eqref{target_x_0} is identified as $E[Y|X, V] - E[Y|X=0, V]$ under exchangeabiblity condition $Y^x \perp X |V$ (equivalent to the assumption of no residual confounding conditional on $V$), positivity and consistency. However, it cannot be directly estimated because X is not measured in the main study together with $Y$. Thus we consider the following transformation with additional assumptions: 

\begin{align*}
E[Y^x - Y^0|V]    & = E[Y^x|X=x,V]  - E[Y^0|X=0, V] & \text{(C.1.1: exchangeability condition $Y^x \perp X | V$)} \\
        & = E[Y|X=x, V] - E[Y|X=0, V]& \text{(C.1.2: consistency)}\\
        & = E[Y|X=x, Z, V] - E[Y|X=0, V] & \text{(C.1.3: surrogacy assumption: $Y \perp Z |X,V$)}\\
        & = \beta(V) x & \text{(C.1.4: linear causal effect model conditional on $V$)}.
\end{align*} Since the equation is true for any realized value of $x$, we can replace $x$ with any random variable notation $X$. Taking the expectation over $X|Z,V$ on both sides gives: $\beta(V) E[X|Z,V] = E[Y|Z,V] - E[Y|X=0,V]$. where $E[Y|X=0,V]$ is a constant given $V$. This equation is true for any two realization of $Z$, $z \neq z^{'}$. This gives:
$$\beta (V)  E[X|Z=z,V] - E[Y|Z=z,V] = \beta (V) E[X|Z=z',V] - E[Y|Z=z',V].$$ Rearranging this equation we have the identifying formula for the conditional unit treatment effect parameter $\beta(V)$:

\begin{equation}  \tag{A5}  \label{identify_formula_V}
\beta (V) = \frac{E[Y|Z=z, V] - E[Y|Z=z^{'},V]}{E[X|Z=z, V] - E[X|Z=z^{'},V]}.
\end{equation}

In order for \eqref{identify_formula_V} to hold, we only have to assume that $\beta(V)$ does not depend on $Z$, which, for example, can be achieved by modeling $E[Y|Z=z, V]$ and $E[X|Z=z, V]$ respectively as $E[Y|Z=z, V] = Z \phi_1 (V)$ and $E[X|Z=z, V] = Z \phi_2 (V)$, where $\phi_1 (\cdot)$  and $\phi_2 (\cdot)$ are some generic functions. No linear modeling assumptions are required for either $E[Y|Z=z, V]$ or $E[X|Z=z, V]$. 

\subsubsection{Identification under DAG 1, 5, 6, 7 and 8} \label{A22}
Alternatively, we can be interested in the average treatment effect:

\begin{align*}  \tag{A6}  \label{target_ATE}
E[Y^x - Y^{x^{'}}].
\end{align*}

Similar to previous section, to identify \eqref{target_ATE}, if we fix $X$ to $0$ then:
\begin{align*}  \tag{A6}  \label{target_ATE_alternative}
E[Y^x - Y^{x^{0}}].
\end{align*} 
We consider the following transformation with additional assumption under DAG 1, 5, 6, 7 and 8: 
\begin{align*}
      E[Y^x - Y^0]  & = E[Y^x|X=x]  - E[Y^0|X=0] & \text{(C.2.1: exchangeability condition $Y^x \perp X$)} \\
        & = E[Y|X=x] - E[Y|X=0]& \text{(C.2.2: consistency)}\\
        & = E[Y|X=x, Z] - E[Y|X=0] & \text{(C.2.3: surrogacy assumption: $Y \perp Z |X$)}\\
        & = \beta x & \text{(C.2.4: linear effect model)}.
\end{align*} Since the equation is true for any realized value of $x$, we can replace $x$ with any random variable notation $X$. Taking the expectation over $X|Z$ on both sides gives: $\beta E[X|Z] = E[Y|Z] - E[Y|X=0]$, where $E[Y|X=0]$ is a constant. This equation is true for any two realization of $Z$, $z \neq z^{'}$. This gives:
$$\beta  E[X|Z=z] - E[Y|Z=z] = \beta (V) E[X|Z=z'] - E[Y|Z=z'].$$ Rearranging this equation we have the identification formula for the unit treatment effect parameter $\beta$:

\begin{equation}  \tag{A7}  \label{identify_formula_noV}
\beta = \frac{E[Y|Z=z] - E[Y|Z=z^{'}]}{E[X|Z=z] - E[X|Z=z^{'}]}.
\end{equation}

In order for \eqref{identify_formula_noV} to hold, we only have to assume that $\beta$ does not depend on $Z$, which, for example, can be achieved by modeling $E[Y|Z=z]$ and $E[X|Z=z]$ respectively as $E[Y|Z] = c_1 \phi(Z)$ and $E[X|Z] = c_2 \phi(Z)$, where $\phi (\cdot)$ is some generic function. No linear modeling assumptions are required for either $E[Y|Z=z]$ or $E[X|Z=z]$. 

To avoid confusion, we denote $\beta$ identified here under DAGs 1, 5, 6, 7 and 8 as $\beta^{'}$. Note that \eqref{identify_formula_V} and \eqref{identify_formula_noV} are both non-parametric in nature. 

\subsubsection{Statistical Models and RSW Estimators} \label{A3}
Now suppose the quantities $E[Y|Z,V]$ and $E[X|Z,V]$ shown in identification formulas \eqref{identify_formula_V} and \eqref{identify_formula_noV} can be modeled with models given in this section. For example, we can assume as in the Rosner et al's original paper \cite{RSW1} and their follow-up discussion \cite{spiegelman1997regression} that the conditional mean of $X$ given $(Z,V)$ can be modeled as: 
\begin{equation}  \tag{A8}  \label{EX_ZV}
    E[X|Z,V] = \alpha_0 + \alpha_1 Z + \alpha_2 V \text{ (same as equation (1) in main text)}.
\end{equation}

If $E[V|Z]$ is a linear function of $Z$:
\begin{equation}  \tag{A9} \label{EV_Z}
    E[V|Z] = \lambda_0 + \lambda_1 Z,
\end{equation} then the conditional mean of $X$ model given $Z$ is:
\begin{equation} \tag{A10}  \label{EX_Z}
    E[X|Z] = \alpha_0^* + \alpha_1^* Z \text{ (same as equation (3) in main text)}.
\end{equation}

The conditional mean of $Y$ given $(Z,V)$ can be modeled as: 
\begin{equation}  \tag{A11}  \label{EY_ZV}
    E[Y|Z,V] = \gamma_0 + \gamma_1 Z + \gamma_2 V \text{ (same as equation (2) in main text)}.
\end{equation} If model \eqref{EV_Z} holds, then: 
\begin{equation}  \tag{A12}  \label{EY_Z}
    E[Y|Z] = \gamma_0^* + \gamma_1^* Z \text{ (same as equation (4) in main text)}.
\end{equation}

From \eqref{EX_ZV} to \eqref{EY_Z} we have the following four RSW estimators and convergence by Slutsky's theorem: 

\begin{align}
\hat{\beta}_{_{(OM)}} = \nicefrac{\hat{\gamma}_1}{\hat{\alpha}_1} \overset{p}{\to}  \nicefrac{\gamma_1}{\alpha_1} ,\\
\hat{\beta}_{_{(--)}} = \nicefrac{\hat{\gamma}_1^*}{\hat{\alpha}_1^*} \overset{p}{\to}  \nicefrac{\gamma_1^*}{\alpha_1^*} ,\\
\hat{\beta}_{_{(-M)}} = \nicefrac{\hat{\gamma}_1^*}{\hat{\alpha}_1} \overset{p}{\to}  \nicefrac{\gamma_1^*}{\alpha_1} ,\\
\hat{\beta}_{_{(O-)}} = \nicefrac{\hat{\gamma}_1}{\hat{\alpha}_1^*} \overset{p}{\to}  \nicefrac{\gamma_1}{\alpha_1^*},  
\end{align} where $\gamma_1$, $\gamma_1^*$ are parameters from models \eqref{EY_ZV} and \eqref{EY_Z} respectively and $\alpha_1$,$\alpha_1^*$ are from models \eqref{EX_ZV} and \eqref{EX_Z} respectively and the parameters with subscript * indicate that $V$ is excluded from the model, that is, models marginal to $V$. 

\subsection{Validity of RSW estimators}

We note that consistent with all eight DAGs, we assume that there is at least some correlation between the true exposure $X$ and its surrogate $Z$ conditional on covariate $V$, i.e. $Cov(X,Z|V)\ne 0$. This is generally true if the surrogate contains any useful information about the true exposure. 

\subsubsection{Estimator 1: $\hat{\beta}_{_{(OM)}} = \frac{\hat{\gamma}_1}{\hat{\alpha}_1}$}

For DAGs 1 through 8, under linear models assumed in \eqref{EX_ZV} and \eqref{EY_ZV}, the causal parameter of interest $\beta (V)$ reduces to:

\begin{equation}  \tag{A13}  \label{identify_formula_ATE_reduce}
\beta (V) = \frac{\gamma_0 + \gamma_1 z + \gamma_2 V - (\gamma_0 + \gamma_1 z^{'} + \gamma_2 V)}{\alpha_0 + \alpha_1 z + \alpha_2 V - (\alpha_0 + \alpha_1 z^{'}+ \alpha_2 V)} = \frac{\gamma_1}{\alpha_1} =\beta,
\end{equation} where the unit conditional treatment effect $\beta(V)$ no longer depends on covariate value $V$ thus equals to the unit average treatment effect $\beta$. For \textbf{DAGs 1 through 8}, under models \eqref{EX_ZV} and \eqref{EY_ZV}, $\hat{\gamma}_1$ and $\hat{\alpha}_1$ are consistent estimators of ${\gamma}_1$ and $\alpha_1$ respectively. By results from \ref{A2}, $\hat{\beta}_{_{(OM)}} = \frac{\hat{\gamma}_1}{\hat{\alpha}_1} \overset{p}{\to} \frac{\gamma_1}{\alpha_1} = \beta$.

\subsubsection{Estimator 2: $\hat{\beta}_{_{(--)}} = \frac{\hat{\gamma}_1^*}{\hat{\alpha}_1^*}$}

For \textbf{DAGs 1, 5, 6, 7 and 8}, under the models \eqref{EX_Z} and \eqref{EY_Z}, $\hat{\gamma}_1^*$ and $\hat{\alpha}_1^*$ are consistent estimators of ${\gamma}_1^*$ and $\alpha_1^*$ respectively. By results from \ref{A2}, $\hat{\beta}_{_{(--)}} = \frac{\hat{\gamma}_1^*}{\hat{\alpha}_1^*} \overset{p}{\to} \frac{\gamma_1^*}{\alpha_1^*} = \beta^{'}$.

Now we show that $\hat{\beta}_{_{(--)}}$ does not consistently estimate the valid causal effect under DAG 2, 3 and 4. To do so, we can reparameterize $\gamma_1$ as functions of $\gamma_1, \gamma_2, \alpha_1, \alpha_2$ and $\lambda_1$:

\begin{align*}
    \hat{\beta}_{_{(--)}} \overset{p}{\to} \beta^{'} & = \frac{E[Y|Z=z] - E[Y|Z=z^{'}]}{E[X|Z=z] - E[X|Z=z^{'}]}\\
    & = \frac{E[E[Y|Z=z,V]|Z=Z] - E[E[Y|Z=z^{'},V]|Z=z']}{E[E[X|Z=z,V]|Z=z] - E[E[X|Z=z^{'},V]|Z=z']} \\
    & = \frac{\gamma_1(z-z') + \gamma_2 (E[V|Z=z] - E[V|Z=z'])}{\alpha_1(z-z') + \alpha_2 (E[V|Z=z] - E[V|Z=z'])} \\
    & = \frac{\gamma_1 + \gamma_2 \lambda_1}{\alpha_1 + \alpha_2 \lambda_1}.
\end{align*}

We note that  $\hat{\beta}_{_{(--)}} \overset{p}{\to} \beta$ if and only if $\beta' = \beta =\nicefrac{\gamma_1}{\alpha_1}$ and assessing $\beta' = \beta$ is equivalent to assessing $\alpha_1 \gamma_2 - \alpha_2 \gamma_1 = 0$. Under DAG 2, 3 and 4, we have $\alpha_1 \ne 0$ (i.e. $Cov(X,Z|V)\ne0)$), $\alpha_2 \ne 0$ (i.e. $Cov(X,V|Z)\ne0)$), $\gamma_1 \ne 0$ (i.e. $Cov(Z,Y|V)\ne 0$) and $\gamma_2 \ne 0$ (i.e. $Cov(Y,V|Z)\ne 0$). These nonzero covariance is a direct consequence of the conditional dependencies implied by DAG 2, 3 and 4 under faithfulness. 

\begin{itemize}
    \item For \textbf{DAG 2}, $Cov(X,V)=0$. From model \eqref{EX_ZV}, we have $Cov(X,V) = \alpha_1 Cov(V,Z) + \alpha_2 Var(V) = 0$. Replacing $\alpha_2 = \nicefrac{-\alpha_1 Cov(V,Z)}{Var(V)}$ in $\alpha_1 \gamma_2 - \alpha_2 \gamma_1 = 0$ gives $\gamma_2 Var(V) + \gamma_1 Cov(V,Z) = 0$. But \eqref{EY_ZV} also gives $\gamma_2 Var(V) + \gamma_1 Cov(V,Z) = Cov(Y,V) \ne 0$. Therefore $\alpha_1 \gamma_2 - \alpha_2 \gamma_1 \ne 0$ and $\beta' \ne \beta$. 
    
    \item For \textbf{DAG 3 and 4}, we use counter examples to prove that $\alpha_1 \gamma_2 - \alpha_2 \gamma_1 = 0$ is not always true. Consistent with the both DAG 3 and 4, we assume the following data generating process:
    
    \begin{align*}
        V & \sim N(0,1) \\
        X & = V + e_x, e_x \sim N(0,1), e_x \perp V \\
        Z & = X + a V + e_z, e_z \sim N(0,1), e_z \perp (X,V) \\
        Y & = bX + V + e_y, e_y \sim N(0,1), e_y \perp (X,V,Z)
    \end{align*}
\end{itemize}
    We can immediately find the joint multivariate distribution of $(X,Z,V)$ given the above data generating process as:
    
    \[\begin{pmatrix}
    X \\
    Z \\
    V
    \end{pmatrix}\sim N\left(\begin{pmatrix}
    0 \\
    0 \\
    0 
    \end{pmatrix},\begin{pmatrix}
    2 & a + 2 & 1 \\
    a + 2 & a^2 + 2a + 3 & a + 1 \\
    1 & a + 1 & 1
    \end{pmatrix}\right).
    \]
    And we can obtain: 
    
    \begin{align*}
        E[X|Z,V] & = \begin{pmatrix} a+2 & 1 \end{pmatrix} \begin{pmatrix} a^2 + 2a + 3 & a+1 \\ 
        a+1 & 1 \end{pmatrix}^{-1} \begin{pmatrix} Z \\ 
        V \end{pmatrix} \\
        & = \frac{1}{2} Z + \frac{1}{2} (1-a) V.
    \end{align*}

  The above gives:
  \begin{align*}
      \frac{\alpha_1}{\alpha_2} = \frac{1}{1-a}.
  \end{align*}
    
    Similarly, we can obtain the joint distribution of $(Y,Z,V)$ as: 
    
      \[\begin{pmatrix}
    Y \\
    Z \\
    V
    \end{pmatrix}\sim N\left(\begin{pmatrix}
    0 \\
    0 \\
    0 
    \end{pmatrix},\begin{pmatrix}
    Var(Y) &  ab + a + 2b + 1 & b + 1 \\
    ab + a + 2b + 1 & a^2 + 2a + 3 & a + 1 \\
    b + 1 & a + 1 & 1
    \end{pmatrix}\right).
    \]
    And we can obtain: 
    
    \begin{align*}
        E[Y|Z,V] & = \begin{pmatrix} ab + a + 2b + 1 & b + 1 \end{pmatrix} \begin{pmatrix} a^2 + 2a + 3 & a+1 \\ 
        a+1 & 1 \end{pmatrix}^{-1} \begin{pmatrix} Z \\ 
        V \end{pmatrix} \\
        & = \frac{1}{2} bZ + \frac{1}{2} (-ab + b + 2) V.
    \end{align*}
  
  The above gives:
  \begin{align*}
      \frac{\gamma_1}{\gamma_2} = \frac{b}{-ab + b + 2}.
  \end{align*}
    
   Clearly, under DAG 3 and 4, $\frac{\gamma_1}{\gamma_2} = \frac{\alpha_1}{\alpha_2}$ only in special cases.

\subsubsection{Estimator 3: $\hat{\beta}_{_{(-M)}} = \frac{\hat{\gamma}_1^*}{\hat{\alpha}_1}$}

Under the models specified in section \ref{A3}, we can reparameterize $\frac{\gamma_1^*}{\alpha_1}$ as follows: 

\begin{align*}
    \hat{\beta}_{_{(-M)}} \overset{p}{\to} \frac{\gamma_1^*}{\alpha_1} & = \frac{E[Y|Z=z] - E[Y|Z=z^{'}]}{E[X|Z=z,V] - E[X|Z=z^{'},V]}\\
    & = \frac{E[E[Y|Z=z,V]|Z=Z] - E[E[Y|Z=z^{'},V]|Z=z']}{E[X|Z=z,V] - E[X|Z=z^{'},V]} \\
    & = \frac{\gamma_1(z-z') + \gamma_2 (E[V|Z=z] - E[V|Z=z'])}{\alpha_1(z-z')} \\
    & = \frac{\gamma_1 + \gamma_2 \lambda_1}{\alpha_1}.
\end{align*}

Thus $ \hat{\beta}_{_{(-M)}} \overset{p}{\to} \beta$ if and only if (1) $\gamma_2 =0$ or (2) $\lambda_1=0$, that is, iff $Cov(Y,V|Z)=0$ or $Cov(V,Z)=0$, respectively. Among the DAGs under consideration,

\begin{itemize}
    \item DAG 1 satisfies condition (2), thus $\hat{\beta}_{_{(-M)}}$ is a valid estimator \footnote{In fact, we note that under DAG 1 and 5 $X\perp V|Z$ thus:
\begin{align*}
    \beta & = \frac{E[Y|Z=z] - E[Y|Z=z^{'}]}{E[X|Z=z] - E[X|Z=z^{'}]}\\
              & = \frac{E[Y|Z=z] - E[Y|Z=z^{'}]}{E[X|Z=z,V] - E[X|Z=z^{'},V]}
\end{align*}. Therefore, linear model specification as in \eqref{EX_Z} is not necessary and one can still estimate $\beta$ (where we assume no effect modification of $V$) consistently by fitting models for $E[X|Z=z]$ with non-linear terms for $Z$.}. 
    \item DAG 5 satisfies condition (1) and (2), thus $\hat{\beta}_{_{(-M)}}$ is a valid estimator. 
    \item DAGs 2, 3, 4, 6, 7 and 8 violate conditions (1) and (2), thus $\hat{\beta}_{_{(-M)}}$ is a not valid under these DAGs. 
\end{itemize}

\subsubsection{Estimator 4: $\hat{\beta}_{_{(O-)}} = \frac{\hat{\gamma}_1}{\hat{\alpha}_1^*}$}

Under the models specified in section \ref{A3}, we can reparameterize $\frac{\gamma_1}{\alpha_1^*}$ as follows: 

\begin{align*}
    \hat{\beta}_{_{(O-)}} \overset{p}{\to} \frac{\gamma_1}{\alpha_1^*}& = \frac{E[Y|Z=z,V] - E[Y|Z=z^{'},V]}{E[X|Z=z] - E[X|Z=z^{'}]}\\
    & = \frac{[E[Y|Z=z,V] - [E[Y|Z=z^{'},V]}{E[E[X|Z=z,V]|Z=z] - E[E[X|Z=z^{'},V]|Z=z']} \\
    & = \frac{\gamma_1(z-z')}{\alpha_1(z-z') + \alpha_2 (E[V|Z=z] - E[V|Z=z'])} \\
    & = \frac{\gamma_1}{\alpha_1 + \alpha_2 \lambda_1}.
\end{align*}

Thus $ \hat{\beta}_{_{(O-)}} \overset{p}{\to} \beta$ if and only if (1) $\alpha_2 =0$ or (2) $\lambda_1=0$, that is, iff $Cov(X,V|Z)=0$ or $Cov(V,Z)=0$, respectively. Among the DAGs under consideration:

\begin{itemize}
    \item DAG 1 and 5 satisfies both conditions (1) and (2), thus $\hat{\beta}_{_{(O-)}}$ is a valid estimator\footnote{In fact, we note that under DAG 1 and 5 $X\perp V|Z$ thus:
\begin{align*}
    \beta (V) & = \frac{E[Y|Z=z, V] - E[Y|Z=z^{'},V]}{E[X|Z=z, V] - E[X|Z=z^{'},V]}\\
              & = \frac{E[Y|Z=z, V] - E[Y|Z=z^{'},V]}{E[X|Z=z] - E[X|Z=z^{'}]}
\end{align*}

Therefore, linear model specification as in \eqref{EY_ZV} is not necessary and one can still estimate $\beta (V)$ consistently by fitting models for $E[Y|Z=z, V]$ with non-linear terms for either $Z$ or $V$.}. 
    \item DAGs 2, 3, 4, 6, 7 and 8 violate conditions (1) and (2), thus $\hat{\beta}_{_{(O-)}}$ is a not valid under these DAGs. 
\end{itemize}

\subsection{Proof When More Than One Set of Covariates Are Present}
Previous results in this section investigates each covariate set assuming the absence of other covariate sets. Here, we demonstrate, using one particular example, how to generalize the results to investigate whether a covariate set is needed in MEM or outcome model when other covariate sets are present. This proof closely follws the logic of the simpler case, but is much more complex, making it more difficult to discern the key ideas.

We assume that the DAG includes $W = (V_{3(X-Y)}, V_{4(XZY)})$, with nodes and arrows positioned in the same way as in figure 1 in the manuscript. Our goal is to investigate when $V_{3(X-Y)}, V_{4(XZY)}$ is needed in the MEM and/or the outcome model. 

\subsubsection{Identification}

First we show identification. We notice that under this new DAG, the exchangeability and surrogacy assumptions can be restated as $Y^x \perp X| W$ and $Y \perp Z|X, W$ respectively. Then, we have the identification: 
\begin{align*}
E[Y^x - Y^0|W]    & = E[Y^x|X=x,W]  - E[Y^0|X=0, W] & \text{(C.3.1: exchangeability condition $Y^x \perp X | W$)} \\
        & = E[Y|X=x, W] - E[Y|X=0, W]& \text{(C.3.2: consistency)}\\
        & = E[Y|X=x, Z, W] - E[Y|X=0, W] & \text{(C.3.3: surrogacy assumption: $Y \perp Z |X,W$)}\\
        & = \beta(W) x & \text{(C.3.4: linear causal effect model conditional on $W$)}.
\end{align*} Since the equation is true for any realized value of $x$, we can replace $x$ with any random variable notation, $X$. Taking the expectation over $X|Z,W$ on both sides gives: $\beta(W) E[X|Z,W] = E[Y|Z,W] - E[Y|X=0,W]$. where $E[Y|X=0,W]$ is a constant given $W$. This equation is true for any two realization of $Z$, $z \neq z^{'}$. This gives:
$$\beta (W)  E[X|Z=z,W] - E[Y|Z=z,W] = \beta (V) E[X|Z=z',W] - E[Y|Z=z',W].$$ Rearranging this equation, we obtain the identifying formula for the conditional unit treatment effect parameter $\beta(W)$:

\begin{equation}  \tag{A14}  \label{identify_formula_W}
\beta (W) = \frac{E[Y|Z=z, W] - E[Y|Z=z^{'},W]}{E[X|Z=z, W] - E[X|Z=z^{'},W]} \\ 
 = \frac{E[Y|Z=z, V_{3(X-Y)}, V_{4(XZY)}] - E[Y|Z=z^{'}, V_{3(X-Y)}, V_{4(XZY)}]}{E[X|Z=z, V_{3(X-Y)}, V_{4(XZY)}] - E[X|Z=z^{'}, V_{3(X-Y)}, V_{4(XZY)}]}.
\end{equation}

 \subsubsection{Modeling Assumptions}
Second, we lay out the modeling assumptions. Suppose the quantities $E[X|Z,W]$ and $E[Y|Z,W]$ shown in identification formulas \eqref{identify_formula_W} can be modeled by: 
\begin{equation}  \tag{A15}  \label{EX_ZW}
    E[X|Z,W] = \alpha_0 + \alpha_1 Z + \alpha_2 V_{3(X-Y)} + \alpha_3 V_{4(XZY)}, 
\end{equation}
\begin{equation}  \tag{A16}  \label{EY_ZW}
    E[Y|Z,W] = \gamma_0 + \gamma_1 Z + \gamma_2 V_{3(X-Y)} + \gamma_3 V_{4(XZY)}.
\end{equation} 

If $E[V_{4(XZY)}|Z]$ is a linear function of $Z$:
\begin{equation}  \tag{A17} \label{EV4_Z}
    E[V_{4(XZY)}|Z] = \lambda_{1,0} + \lambda_{1,1} Z,
\end{equation}
and $E[V_{3(X-Y)}|Z]$ is a linear function of $Z$:
\begin{equation}  \tag{A18} \label{EV3_Z}
    E[V_{3(X-Y)}|Z] = \lambda_{2,0} + \lambda_{2,1} Z,
\end{equation}
then the conditional mean of $X$ given $Z$ is:
\begin{equation} \tag{A19}  \label{EX_Z2}
    E[X|Z] = \alpha_{1,0}^* + \alpha_{1,1}^* Z ,
\end{equation} and the conditional mean of $Y$ given $Z$ is: 
\begin{equation} \tag{A20}  \label{EY_Z2}
    E[Y|Z] = \gamma_{1,0}^* + \gamma_{1,1}^* Z . 
\end{equation}

If $E[V_{3(X-Y)}|Z, V_{4(XZY)}]$ is a linear function of $(Z,V_{4(XZY)})$:
\begin{equation}  \tag{A21} \label{EV3_ZV4}
    E[V_{3(X-Y)}|Z, V_{4(XZY)}] = \lambda_{3,0} + \lambda_{3,1} Z +  \lambda_{3,2} V_{4(XZY)},
\end{equation} then the conditional mean of $X$ model given $(Z, V_{4(XZY)})$ is:
\begin{equation} \tag{A22}  \label{EX_ZV4}
    E[X|Z,V_{4(XZY)}] = \alpha_{2,0}^* + \alpha_{2,1}^* Z +  \alpha_{2,2}^* V_{4(XZY)}, 
\end{equation} and the conditional mean of $Y$ model given $(Z, V_{4(XZY)})$ is: 
\begin{equation}  \tag{A23}  \label{EY_ZV4}
    E[Y|Z,V_{4(XZY)}] = \gamma_{2,0}^* + \gamma_{2,1}^* Z + \gamma_{2,2}^* V_{4(XZY)}.
\end{equation}

If $E[V_{4(XZY)}|Z, V_{3(X-Y)}]$ is a linear function of $(Z,V_{3(X-Y)})$:
\begin{equation}  \tag{A24} \label{EV4_ZV3}
    E[V_{4(XZY)}|Z, V_{3(X-Y)}] = \lambda_{4,0} + \lambda_{4,1} Z +  \lambda_{4,2} V_{3(X-Y)},
\end{equation} then the conditional mean of $X$ model given $(Z, V_{3(X-Y)})$ is:
\begin{equation} \tag{A25}  \label{EX_ZV3}
    E[X|Z,V_{3(X-Y)}] = \alpha_{3,0}^* + \alpha_{3,1}^* Z +  \alpha_{3,2}^* V_{3(X-Y)}, 
\end{equation} and the conditional mean of $Y$ model given $(Z, V_{3(X-Y)})$ is: 
\begin{equation}  \tag{A26}  \label{EY_ZV3}
    E[Y|Z,V_{3(X-Y)}] = \gamma_{3,0}^* + \gamma_{3,1}^* Z + \gamma_{3,2}^* V_{3(X-Y)}.
\end{equation}

We can thus choose, separately for $V_{3(X-Y)}$ and $V_{4(XZY)}$, whether to include each in only the MEM, only the outcome model, both or neither of the models. This gives us $4 \times 4 = 16$ estimators. 

\subsubsection{An Example of the Measurement Error Process under Consideration}
Here, we provide one data generating process for the DAG under consideration, the result of which will be useful later for the proof by counter-example. 

   \begin{align*}
        V_{3(X-Y)}, V_{4(XZY)} & \sim N(0,1), Cov(V_{3(X-Y)}, V_{4(XZY)})=0\\
        X & = V_{3(X-Y)} + V_{4(XZY)} + e_x, e_x \sim N(0,1), e_x \perp (V_{3(X-Y)}, V_{4(XZY)} ) \\
        Z & = X + V_{4(XZY)} + e_z, e_z \sim N(0,1), e_z \perp (X,V_{3(X-Y)},V_{4(XZY)}) \\
        Y & = bX + V_{3(X-Y)} + V_{4(XZY)} + e_y, e_y \sim N(0,1), e_y \perp (X,V_{3(X-Y)},V_{4(XZY)},Z)
    \end{align*}

    Using the well known properties of the multivariate normal distribution, we find the joint multivariate distribution of $(X,Y,Z,V_{3(X-Y)}, V_{4(XZY)})$ given the above data generating process(details omitted here) and then write out the following conditional mean quantities using manipulation similar to those in section 1.3.2.:
    
    \begin{align*}
      E[X|Z,V_{3(X-Y)}, V_{4(XZY)}] & = 
      2 Z  - V_{3(X-Y)} + 2 V_{4(XZY)},\\
      E[Y|Z,V_{3(X-Y)}, V_{4(XZY)}] & = 
      5b Z + (-5b+1)V_{3(X-Y)} + (-10b+1) V_{4(XZY)}, \\
      E[V_{4(XZY)}|Z] & = \frac{1}{3} Z,\\ 
      E[V_{3(X-Y)}|Z] & = \frac{1}{6} Z,\\
      E[V_{3(X-Y)}|Z,V_{4(XZY)}] & = \frac{1}{2} Z - V_{4(XZY)}
    \end{align*} from which we know that for this data generating process, $\alpha_1 = 2, \alpha_2 = -1, \alpha_3 = 2, \gamma_1 = 5b, \gamma_2 = -5b + 1, \gamma_3 = -10b + 1, \lambda_{1,1} = \frac{1}{3}, \lambda_{2,1} = \frac{1}{6}, \lambda_{3,1} = \frac{1}{2}, \lambda_{3,2} = -1, \lambda_{4,1} = -\frac{2}{5}, \lambda_{4,2} = -\frac{2}{5}$. 

  Last, we demonstrate how to prove the validity of some but not all of the 16 candidate estiamtors. For the rest of the 16 estimators, we omitted the steps and only give sketch of proof as well as the results. 
  
\subsubsection{Estimator 1: $\hat{\beta}_{_{V_{3(X-Y)} (OM), V_{4(XZY)} (OM)}} = \frac{\hat{\gamma}_1}{\hat{\alpha}_1}$}
Under linear models assumed in \eqref{EX_ZW} and \eqref{EY_ZW}, the causal parameter of interest $\beta (W)$ identified from result \eqref{identify_formula_W} reduces to:

\begin{equation}  \tag{A24}  \label{identify_formula_ATE_reduce_W}
\beta (W) = \frac{\gamma_1}{\alpha_1} =\beta,
\end{equation} where the unit conditional treatment effect $\beta(W)$ no longer depends on covariate set $W$. This proves that the estimator $\hat{\beta}_{_{V_{3(X-Y)} (OM), V_{4(XZY)} (OM)}} = \nicefrac{\hat{\gamma}_1}{\hat{\alpha}_1} \overset{p}{\to}  \nicefrac{\gamma_1}{\alpha_1}$ is valid for the DAG under consideration. 

\subsubsection{Estimator 2: $\hat{\beta}_{_{V_{3(X-Y)} (--), V_{4(XZY)} (--)}} = \frac{\hat{\gamma}_{1,1}^*}{\hat{\alpha}_{1,1}^*}$}
Similar as before, under modeling assumptions \eqref{EX_ZW} through \eqref{EY_Z2}, we can reparameterize $\hat{\beta}_{_{V_{3(X-Y)} (--), V_{4(XZY)} (--)}}$ as: 

\begin{align*}
    \hat{\beta}_{_{V_{3(X-Y)} (--), V_{4(XZY)} (--)}} \overset{p}{\to} {\beta}_{_{V_{3(X-Y)} (--), V_{4(XZY)} (--)}}  & = \frac{E[Y|Z=z] - E[Y|Z=z^{'}]}{E[X|Z=z] - E[X|Z=z^{'}]}\\
    & = \frac{E [E[Y|Z=z,W]|Z=z] - E[E[Y|Z=z^{'},W]|Z=z']}{E[E[X|Z=z,W]|Z=z] - E[E[X|Z=z^{'},W]|Z=z']} \\
    & = \frac{\gamma_1 + \gamma_2 \lambda_{1,1} + \gamma_3 \lambda_{2,1}}{\alpha_1 + \alpha_2 \lambda_{1,1} + \alpha_3 \lambda_{2,1}}.
\end{align*}

In order for ${\beta}_{_{V_{3(X-Y)} (--), V_{4(XZY)} (--)}}$ to be valid, it has to equal ${\beta}_{_{V_{3(X-Y)} (OM), V_{4(XZY)} (OM)}}$. Assuming ${\beta}_{_{V_{3(X-Y)} (--), V_{4(XZY)} (--)}} = {\beta}_{_{V_{3(X-Y)} (OM), V_{4(XZY)} (OM)}}$, assessing ${\beta}_{_{V_{3(X-Y)} (--), V_{4(XZY)} (--)}} = {\beta}_{_{V_{3(X-Y)} (OM), V_{4(XZY)} (OM)}}$ is equivalent to assessing $\lambda_{1,1} (\gamma_1 \alpha_2 - \alpha_1 \gamma_2) - \lambda_{2,1} (\alpha_1 \gamma_3 - \gamma_1 \alpha_3) = 0$ as we already showed that ${\beta}_{_{V_{3(X-Y)} (OM), V_{4(XZY)} (OM)}}$ is a valid estimator. Using the results obtained from the example in section 1.4.3, this equation reduces to $20b - 3 =0$. This equation is generally not true and therefore in general  ${\beta}_{_{V_{3(X-Y)} (--), V_{4(XZY)} (--)}}$ is invalid. 

\subsubsection{Estimator 3: $\hat{\beta}_{_{V_{3(X-Y)} (--), V_{4(XZY)} (OM)}} = \frac{\hat{\gamma}_{2,1}^*}{\hat{\alpha}_{2,1}^*}$}
Similar as before, under modeling assumptions \eqref{EX_ZW} through \eqref{EY_ZV4}, we can reparameterize $\hat{\beta}_{_{V_{3(X-Y)} (--), V_{4(XZY)} (OM)}}$ as: 
\begin{align*}
    \hat{\beta}_{_{V_{3(X-Y)} (--), V_{4(XZY)} (OM)}} \overset{p}{\to} {\beta}_{_{V_{3(X-Y)} (--), V_{4(XZY)} (OM)}} & = \frac{E[Y|Z=z, V_{4(XZY)}] - E[Y|Z=z^{'}, V_{4(XZY)}]}{E[X|Z=z, V_{4(XZY)}] - E[X|Z=z^{'}, V_{4(XZY)}]}\\
    & = \frac{E [E[Y|Z=z,W]|Z=z,  V_{4(XZY)}] - E[E[Y|Z=z^{'},W]|Z=z', V_{4(XZY)}]}{E[E[X|Z=z,W]|Z=z, V_{4(XZY)}] - E[E[X|Z=z^{'},W]|Z=z', V_{4(XZY)}]} \\
    & = \frac{\gamma_1 + \gamma_2 \lambda_{3,1}}{\alpha_1 + \alpha_2 \lambda_{3,1}}.
\end{align*}

In order for ${\beta}_{_{V_{3(X-Y)} (--), V_{4(XZY)} (OM)}}$ to be valid, it has to equal ${\beta}_{_{V_{3(X-Y)} (OM), V_{4(XZY)} (OM)}}$. Assuming ${\beta}_{_{V_{3(X-Y)} (--), V_{4(XZY)} (OM)}}={\beta}_{_{V_{3(X-Y)} (OM), V_{4(XZY)} (OM)}}$ is equivalent to assessing $\alpha_1 \gamma_2 - \gamma_1 \alpha_2 = 0$ as we already showed that ${\beta}_{_{V_{3(X-Y)} (OM), V_{4(XZY)} (OM)}}$ is a valid estimator.. Using the results obtained from the example in section 1.4.3, this equation reduces to $5b - 2 =0$. This equation is generally not true and therefore in general  ${\beta}_{_{V_{3(X-Y)} (--), V_{4(XZY)} (OM)}}$ is invalid. 

\subsubsection{Other Candidate Estimators}

We can follow the exact same approach as in sections 1.4.5 and 1.4.6 to show that all the other 13 estimators are also invalid. The sketch of the proof is as follows: 

\begin{enumerate}
    \item reparameterize the candidate estimators using the parameters $\alpha_1, \alpha_2, \alpha_3, \gamma_1, \gamma_2, \gamma_3, \lambda_{1,1}, \lambda_{2,1}, \lambda_{3,1}, \lambda_{3,2},  \lambda_{4,1}, \lambda_{4,2}$; 
    \item force the reparameterized result to be equal to $\frac{\gamma_1}{\alpha_1}$ and obtain a new equality involving the parameters from previous step; 
    \item evaluate whether the (conditional) independence and zero covariance implied by the DAG means that any of the coefficients ($\alpha_1, \alpha_2, \alpha_3, \gamma_1, \gamma_2, \gamma_3, \lambda_{1,1}, \lambda_{2,1}, \lambda_{3,1}, \lambda_{3,2},  \lambda_{4,1}, \lambda_{4,2}$) is zero and whether such zero coefficients make the equality from previous step hold; 
    \item either prove that the equality holds, implying the candidate estimator is valid, or use the counter example to prove that the equation does not hold in general. 
\end{enumerate}

\subsubsection{Summary of results of valid estimators for DAG with $(V_{3(X-Y)},V_{4(XZY)})$}
Therefore, the only valid estimator when $(V_{3(X-Y)},V_{4(XZY)})$ are simultaneously present in a DAG is ${\beta}_{_{V_{3(X-Y)} (OM), V_{4(XZY)} (OM)}}$, i.e. both $V_{3(X-Y)}$ and $V_{4(XZY)}$ should be included in both MEM and outcome model. 

\newpage

\section{Comparison of Efficiency for RSW Estimators Under Linear Model}

\subsection{Set-up}

We use the same set up as in section \ref{A1}, including the modeling assumptions specified in section \ref{A3} and a main study/external validation study design. In this section, we compare the efficiency of the valid RSW estimators under each DAG analytically, with different conditional relationships implied by each DAG. We assume covariate V is a scalar but the result extends to p-dimensional $V$ where $p>1$. Variance of RSW estimators are obtained via delta method as follows:

\begin{align}
\hat{\beta}_{_{(OM)}} & = \nicefrac{\hat{\gamma}_1}{\hat{\alpha}_1}, Var(\hat{\beta}_{_{(OM)}}) \approx \nicefrac{1}{{\alpha}_1^2} Var({\hat{\gamma_1}}) + \nicefrac{{\gamma_1^2}}{{\alpha}_1^4} Var(\hat{\alpha_1}), \\
\hat{\beta}_{_{(--)}} & = \nicefrac{\hat{\gamma}_1^*}{\hat{\alpha}_1^*}, Var(\hat{\beta}_{_{(--)}}) \approx \nicefrac{1}{{{\alpha}_1^*}^2} Var(\hat{\gamma_1^*}) + \nicefrac{{\gamma_1^*}^2}{{{\alpha}^*}^4} Var(\hat{\alpha_1^*}), \\
\hat{\beta}_{_{(-M)}} & = \nicefrac{\hat{\gamma}_1^*}{\hat{\alpha}_1}, Var(\hat{\beta}_{_{(-M)}}) \approx \nicefrac{1}{{\alpha}_1^2} Var(\hat{\gamma_1^*}) + \nicefrac{{\gamma_1^*}^2}{{\alpha}_1^4} Var(\hat{\alpha_1}), \\
\hat{\beta}_{_{(O-)}} & = \nicefrac{\hat{\gamma}_1}{\hat{\alpha}_1^*}, Var(\hat{\beta}_{_{(O-)}}) \approx \nicefrac{1}{{{\alpha}_1^*}^2} Var(\hat{\gamma_1}) + \nicefrac{{\gamma_1}^2}{{{\alpha}_1^*}^4} Var(\hat{\alpha_1^*}).
\end{align} 

Note that under the external validation study design, the covariance between ($\hat{\gamma}_1$ or $\hat{\gamma}_1^*$), which are estimated in the main study, and ($\hat{\alpha}_1$ or $\hat{\alpha}_1^*$), which are estimated in the validation study, is zero by design. It has been shown that the covariance is also asymptotically zero under main study/internal validation study design \cite{spiegelman2001efficient}. 

The following results from multivariate normal distribution will be used frequently \cite{kleinbaum2013}. If $(A,B,C)$ are jointly normal then: 

\begin{align}
    \sigma_{A|B}^2 & = (1 - \rho_{AB}^2)\sigma_{A}^2, \text{ and}\\
       \rho_{A,B|C} & = \frac{\rho_{AB} - \rho_{AC}\rho_{BC}}{\sqrt{1-\rho_{AC}^2}\sqrt{1-\rho_{BC}^2}},
\end{align}

where $\sigma_{\cdot}$ is the standard error and $\rho_{\cdot}$ is the correlation coefficient. Let $n_{MS}$ and $n_{VS}$ be the sample size respectively for main study and validation study, respectively. We can then express each component of the RSW variance estimators as follows:

\begin{equation} \tag{B1} \label{B1}
\gamma_1^* = \frac{Cov(Y,Z)}{Var(Z)} = \frac{\rho_{YZ}\sigma_Y}{\sigma_Z} 
\end{equation}
\begin{equation} \tag{B2} \label{B2}
\alpha_1^* = \frac{Cov(X,Z)}{Var(Z)} = \frac{\rho_{XZ}\sigma_X}{\sigma_Z} 
\end{equation}
\begin{equation} \tag{B3} \label{B3}
Var(\hat{\gamma_1}^*) = \frac{Var(Y|Z)}{Var(Z) n_{MS}} = \frac{(1-\rho_{YZ}^2)\sigma_Y^2}{\sigma_Z^2 n_{MS}} 
\end{equation}
\begin{equation} \tag{B4} \label{B4}
Var(\hat{\alpha_1}^*) = \frac{Var(X|Z)}{Var(Z) n_{VS}} = \frac{(1-\rho_{XZ}^2)\sigma_X^2}{\sigma_Z^2 n_{VS}} 
\end{equation}
\begin{equation} \tag{B5} \label{B5}
\gamma_1 = \frac{Cov(Y,Z|V)}{Var(Z|V)} = \frac{\rho_{YZ|V}\sigma_{Y|V}}{\sigma_{Z|V}} = \frac{(\rho_{YZ} - \rho_{VY} \rho_{VZ})\sigma_{Y}}{(1-\rho_{VZ}^2)\sigma_{Z}}  
\end{equation}
\begin{equation} \tag{B6} \label{B6}
\alpha_1 = \frac{Cov(X,Z|V)}{Var(Z|V)} = \frac{\rho_{XZ|V}\sigma_{X|V}}{\sigma_{Z|V}} = \frac{(\rho_{XZ} - \rho_{VX} \rho_{VZ})\sigma_{X}}{(1-\rho_{VZ}^2)\sigma_{Z}}  
\end{equation}
\begin{equation}\tag{B7} \label{B7}
\begin{aligned}
Var(\hat{\gamma}_1) & = \frac{Var(Y|V,Z)}{Var(Z|V)n_{MS}} \\
& = \frac{(1-\rho_{YZ|V}^2)\sigma_{Y|V}^2}{\sigma_{Z|V}^2 n_{MS}} \\ 
& = \sigma_Y^2 (1-\rho_{VY}^2)\frac{(1-\rho_{VY}^2)(1-\rho_{VZ}^2) - (\rho_{YZ} - \rho_{VY}\rho_{VZ})^2}{n_{MS} \sigma_Z^2 (1-\rho_{VZ}^2)(1-\rho_{VY}^2)(1-\rho_{VZ}^2)} \\
& = \sigma_Y^2 \frac{(1-\rho_{VY}^2 -\rho_{VZ}^2 - \rho_{YZ}^2 + 2\rho_{VY}\rho_{VZ}\rho_{YZ})}{n_{MS} \sigma_Z^2 (1-\rho_{VZ}^2)^2}     
\end{aligned}
\end{equation}
\begin{equation}\tag{B8} \label{B8}
\begin{aligned}
Var(\hat{\alpha}_1) & = \frac{Var(X|V,Z)}{Var(Z|V)n_{VS}} \\
& = \frac{(1-\rho_{VX|Z}^2)\sigma_{X|Z}^2}{\sigma_{Z|V}^2 n_{VS}} \\ 
& = \sigma_X^2 (1-\rho_{XZ}^2)\frac{(1-\rho_{XZ}^2)(1-\rho_{VZ}^2) - (\rho_{VX} - \rho_{XZ}\rho_{VZ})^2}{n_{VS} \sigma_Z^2 (1-\rho_{XZ}^2)(1-\rho_{VZ}^2)(1-\rho_{VZ}^2)} \\
& = \sigma_X^2 \frac{(1-\rho_{XZ}^2 -\rho_{VZ}^2 - \rho_{VX}^2 + 2\rho_{XZ}\rho_{VZ}\rho_{VX})}{n_{VS} \sigma_Z^2 (1-\rho_{VZ}^2)^2}     
\end{aligned}
\end{equation}

\subsection{Efficiency of RSW Estimators}

\subsubsection{Under DAG 1 where $V$ is a risk factor that is not associated with exposure or measurement error, all four estimators are valid and $\rho_{VX}=0, \rho_{VZ}=0$:}
\begin{equation*}
\begin{aligned}
Var(\widehat{\beta}_{_{(OM)}}) & \approx & \nicefrac{1}{{a}_1^2} Var({\hat{\gamma_1}}) + \nicefrac{{\gamma_1^2}}{{a}_1^4} Var(\hat{\alpha_1}) \\ & = & \sigma_Y^2 \frac{\rho_{XZ}^2(1-\rho_{YZ}^2 - \rho_{VY}^2) n_{VS} + \rho_{YZ}^2(1-\rho_{XZ}^2) n_{MS}}{\sigma_X^2 \rho_{XZ}^4 n_{MS} n_{VS}} \\
Var(\widehat{\beta}_{_{(--)}}) & \approx & \nicefrac{1}{{\alpha^*_1}^2} Var({\hat{\gamma^*_1}}) + \nicefrac{{{\gamma^*_1}^2}}{{\alpha^*_1}^4} Var(\hat{\alpha^*_1})\\ & = & \sigma_Y^2 \frac{\rho_{XZ}^2(1-\rho_{YZ}^2) n_{VS} + \rho_{YZ}^2(1-\rho_{XZ}^2) n_{MS}}{\sigma_X^2 \rho_{XZ}^4 n_{MS} n_{VS}} \\
Var(\widehat{\beta}_{_{(-M)}}) & \approx & \nicefrac{1}{{\alpha_1}^2} Var({\hat{\gamma^*_1}}) + \nicefrac{{{\gamma^*_1}^2}}{{\alpha_1}^4} Var(\hat{\alpha_1})\\ & = & \sigma_Y^2 \frac{\rho_{XZ}^2(1-\rho_{YZ}^2) n_{VS} + \rho_{YZ}^2(1-\rho_{XZ}^2) n_{MS}}{\sigma_X^2 \rho_{XZ}^4 n_{MS} n_{VS}}\\ Var(\widehat{\beta}_{_{(O-)}}) & \approx & \nicefrac{1}{{\alpha_1}^2} Var({\hat{\gamma^*_1}}) + \nicefrac{{{\gamma^*_1}^2}}{{\alpha_1}^4} Var(\hat{\alpha_1})\\ & = & \sigma_Y^2 \frac{\rho_{XZ}^2(1-\rho_{YZ}^2 - \rho_{VY}^2) n_{VS} + \rho_{YZ}^2(1-\rho_{XZ}^2) n_{MS}}{\sigma_X^2 \rho_{XZ}^4 n_{MS} n_{VS}} 
\end{aligned}
\end{equation*}
Because $0<\rho_{VY}^2 \leq 1$ thus $(1-\rho_{YZ}^2 - \rho_{VY}^2) < (1-\rho_{YZ}^2)$ and: 
\begin{align*}
Var(\widehat{\beta}_{_{(OM)}})=Var(\widehat{\beta}_{_{(O-)}})<Var(\widehat{\beta}_{_{(--)}})=Var(\widehat{\beta}_{_{(-M)}}). 
\end{align*}

\subsubsection{Under DAGs 2 through 4, $\widehat{\beta}_{_{(OM)}}$ is the only valid estimator.}

\subsubsection{Under DAG 5, $\rho_{VX}=\rho_{VZ}=\rho_{VY}=0$ thus the trivial result:}
\begin{align*}
Var(\widehat{\beta}_{_{(OM)}})=Var(\widehat{\beta}_{_{(--)}})=Var(\widehat{\beta}_{_{(-M)}})=Var(\widehat{\beta}_{_{(O-)}})). 
\end{align*}

\subsubsection{Under DAG 6 where $V$ is a non-risk factor of the outcome and is associated with measurement error only, estimators $\beta_{_{(OM)}}$ and $\beta_{_{(--)}}$ are valid and $\rho_{VX}=0,\rho_{VY}=0$ and we have:}
\begin{equation*}
\begin{aligned}
Var(\widehat{\beta}_{_{(OM)}}) & \approx \nicefrac{1}{{a}_1^2} Var({\hat{\gamma_1}}) + \nicefrac{{\gamma_1^2}}{{a}_1^4} Var(\hat{\alpha_1}) \\
& = \sigma_Y^2 \frac{\rho_{XZ}^2[(1-\rho_{YZ}^2-\rho_{VZ}^2 ) n_{VS} + \rho_{YZ}^2(1-\rho_{XZ}^2)(1-\rho_{VZ}^2) n_{MS}]}{\sigma_X^2 \rho_{XZ}^4 n_{MS} n_{VS}} \\
Var(\widehat{\beta}_{_{(- -)}}) & \approx \nicefrac{1}{{\alpha^*_1}^2} Var({\hat{\gamma^*_1}}) + \nicefrac{{{\gamma^*_1}^2}}{{\alpha^*_1}^4} Var(\hat{\alpha^*_1}) \\
& = \sigma_Y^2 \frac{\rho_{XZ}^2(1-\rho_{YZ}^2) n_{VS} + \rho_{YZ}^2(1-\rho_{XZ}^2) n_{MS}}{\sigma_X^2 \rho_{XZ}^4 n_{MS} n_{VS}}
\end{aligned}    
\end{equation*}where $1-\rho_{YZ}^2-\rho_{VZ}^2 < (1-\rho_{YZ}^2)$ and $(1-\rho_{XZ}^2)(1-\rho_{VZ}^2)<(1-\rho_{XZ}^2)$. Thus: 

\begin{align*}
Var(\widehat{\beta}_{_{(OM)}})<Var(\widehat{\beta}_{_{(--)}}) 
\end{align*}

\subsubsection{For DAG 7 where $V$ is a non-risk factor of the outcome and is associated with true exposure only, estimators $\beta_{_{(OM)}}$ and $\beta_{_{(--)}}$ are valid and $\rho_{VY|X}=\rho_{VZ|X}=\rho_{YZ|X}=0$ which gives $\rho_{VY}= \rho_{VX}\rho_{XY}, \rho_{VZ}=\rho_{VX}\rho_{XZ}$ and $\rho_{YZ}=\rho_{XY}\rho_{XZ}$ following equation (10) in section 2.1:}
\begin{equation*}
\begin{aligned}
Var(\widehat{\beta}_{_{(OM)}}) & \approx \nicefrac{1}{{a}_1^2} Var({\hat{\gamma_1}}) + \nicefrac{{\gamma_1^2}}{{a}_1^4} Var(\hat{\alpha_1}) \\
& = \sigma_Y^2 \frac{(1-\rho_{YZ}^2-\rho_{VY}^2-\rho_{VZ}^2 + 2\rho_{YZ}\rho_{VY}\rho_{VZ}) n_{VS}}{\sigma_X^2 (\rho_{XZ}-\rho_{VX}\rho_{VZ})^2 n_{MS} n_{VS}} \\ & + \sigma_Y^2\frac{(\rho_{YZ} - \rho_{VY}\rho_{VZ})^2(1-\rho_{XZ}^2-\rho_{VZ}^2-\rho_{XV}^2 + 2\rho_{XZ} \rho_{VZ} \rho_{VX}) n_{MS}}{\sigma_X^2 (\rho_{XZ}-\rho_{VX}\rho_{VZ})^4 n_{MS} n_{VS}} \\
Var(\widehat{\beta}_{_{(--)}}) & \approx \nicefrac{1}{{\alpha^*_1}^2} Var({\hat{\gamma^*_1}}) + \nicefrac{{{\gamma^*_1}^2}}{{\alpha^*_1}^4} Var(\hat{\alpha^*_1}) \\
& = \sigma_Y^2 \frac{(1-\rho_{YZ}^2) n_{VS}}{\sigma_X^2 \rho_{XZ}^2 n_{MS} n_{VS}} + \sigma_Y^2\frac{\rho_{YZ}^2(1-\rho_{XZ}^2) n_{MS}}{\sigma_X^2 \rho_{XZ}^4 n_{MS} n_{VS}} 
\end{aligned}
\end{equation*}
We first focus on the quantities prior to the plus sign in the two variance formulas and show that:  
\begin{equation} \label{DAG7_efficiency_part1}
\begin{aligned}
        \frac{1-\rho_{YZ}^2}{\rho_{XZ}^2} & = \frac{1-\rho_{XY}^2\rho_{XZ}^2}{\rho_{XZ}^2} \text{ in } Var(\widehat{\beta}_{_{(--)}})\\
        & \le \frac{(1 - \rho_{XY}^2\rho_{XZ}^2 - \rho_{XY}^2\rho_{VX}^2 - \rho_{VX}^2\rho_{XZ}^2 + 2\rho_{XY}^2\rho_{XZ}^2\rho_{VX}^2)}{\rho_{XZ}^2 (1-\rho_{VX}^2)^2} \\
        & = \frac{(1-\rho_{YZ}^2-\rho_{VY}^2-\rho_{VZ}^2 + 2\rho_{YZ}\rho_{VY}\rho_{VZ})}{(\rho_{XZ}-\rho_{VX}\rho_{VZ})^2}  \text{ in } Var(\widehat{\beta}_{_{(OM)}}). \\
\end{aligned}
\end{equation}
To prove this, we note that $0<\rho_{VX}^2,\rho_{XY}^2, \rho_{XZ}^2 \le 1$ and: 
\begin{equation*}
    \begin{aligned}
    & \rho_{VX}^2[\rho_{VX}^2 -2 + \rho_{XZ}^2 + \rho_{XY}^2 (1- \rho_{VX}^2 \rho_{XZ}^2)] \le \rho_{VX}^2[\rho_{VX}^2 -2 + \rho_{XZ}^2 + (1- \rho_{VX}^2 \rho_{XZ}^2)] \text{ and}\\
    & \rho_{VX}^2[\rho_{VX}^2 -2 + \rho_{XZ}^2 + (1- \rho_{VX}^2 \rho_{XZ}^2)] = \rho_{VX}^2(1 - \rho_{VX}^2)(\rho_{XZ}^2 -1)  \le 0, \text{ thus:} \\
    & \rho_{VX}^2[\rho_{VX}^2 -2 + \rho_{XZ}^2 + \rho_{XY}^2 (1- \rho_{VX}^2 \rho_{XZ}^2)] = \rho_{VX}^4 - 2 \rho_{VX}^2 - \rho_{VX}^4 \rho_{XY}^2 \rho_{XZ}^2 + \rho_{XY}^2 \rho_{VX}^2 + \rho_{VX}^2 \rho_{XZ}^2 \le 0 \\
    \Longleftrightarrow & \rho_{VX}^4 - 2 \rho_{VX}^2 - \rho_{VX}^4 \rho_{XY}^2 \rho_{XZ}^2 \le -\rho_{XY}^2 \rho_{VX}^2 - \rho_{VX}^2 \rho_{XZ}^2 \\
    \Longleftrightarrow & 1 - \rho_{XY}^2 \rho_{XZ}^2 +  \rho_{VX}^4 - 2 \rho_{VX}^2 - \rho_{VX}^4 \rho_{XY}^2 \rho_{XZ}^2 + 2 \rho_{VX}^2 \rho_{XY}^2 \rho_{XZ}^2 \le 1 - \rho_{XY}^2 \rho_{XZ}^2  -\rho_{XY}^2 \rho_{VX}^2 - \rho_{VX}^2 \rho_{XZ}^2 + 2 \rho_{VX}^2 \rho_{XY}^2 \rho_{XZ}^2, \text{ where}\\
    & 1 - \rho_{XY}^2 \rho_{XZ}^2 +  \rho_{VX}^4 - 2 \rho_{VX}^2 - \rho_{VX}^4 \rho_{XY}^2 \rho_{XZ}^2 + 2 \rho_{VX}^2 \rho_{XY}^2 \rho_{XZ}^2 = (1-\rho_{XY}^2\rho_{XZ}^2)(1-\rho_{VX}^2)^2 \\
    \Longleftrightarrow & (1-\rho_{XY}^2\rho_{XZ}^2)(1-\rho_{VX}^2)^2 \le 1 - \rho_{XY}^2 \rho_{XZ}^2  -\rho_{XY}^2 \rho_{VX}^2 - \rho_{VX}^2 \rho_{XZ}^2 + 2 \rho_{VX}^2 \rho_{XY}^2 \rho_{XZ}^2\\
    \Longleftrightarrow & (1-\rho_{XY}^2\rho_{XZ}^2) \le \frac{1 - \rho_{XY}^2 \rho_{XZ}^2  -\rho_{XY}^2 \rho_{VX}^2 - \rho_{VX}^2 \rho_{XZ}^2 + 2 \rho_{VX}^2 \rho_{XY}^2 \rho_{XZ}^2}{(1-\rho_{VX}^2)^2} \text{, thus inequality \eqref{DAG7_efficiency_part1} is proved.}
    \end{aligned}
\end{equation*}  
We now focus on the quantities after the plus sign in the two variance formulas and show that:  
\begin{equation} \label{DAG7_efficiency_part2}
   \begin{aligned}
        \frac{\rho_{YZ}^2(1-\rho_{XZ}^2)}{\rho_{XZ}^4} &  = \frac{\rho_{XY}^2\rho_{XZ}^2(1-\rho_{XZ}^2)}{\rho_{XZ}^4}  \text{ in } Var(\widehat{\beta}_{_{(--)}})\\
        \le & \frac{\rho_{XY}^2\rho_{XZ}^2(1 - \rho_{VX}^2)^2(1-\rho_{XZ}^2-\rho_{VX}^2\rho_{XZ}^2-\rho_{VX}^2 + 2\rho_{XZ}^2 \rho_{VX}^2) }{\rho_{XZ}^4 (1 -\rho_{VX}^2)^4}\\
        = & \frac{(\rho_{YZ} - \rho_{VY}\rho_{VZ})^2(1-\rho_{XZ}^2-\rho_{VZ}^2-\rho_{VX}^2 + 2\rho_{XZ} \rho_{VZ} \rho_{VX})}{(\rho_{XZ}-\rho_{VX}\rho_{VZ})^4 }  \text{ in } Var(\widehat{\beta}_{_{(OM)}}).\\
    \end{aligned} 
\end{equation}
To prove this, we have $0<\rho_{VX}^2,\rho_{XZ}^2 \le 1$ and:
\begin{equation*}
    \begin{aligned}
         (1-\rho_{XZ}^2)(1 -\rho_{VX}^2)^4 & = (1 -\rho_{VX}^2)^2 (1-\rho_{XZ}^2)(1 -\rho_{VX}^2)^2  \textbf{} \\
         = & (1 -\rho_{VX}^2)^2 (1 - \rho_{XZ}^2  + \rho_{VX}^4 - 2\rho_{VX}^2  - \rho_{XZ}^2 \rho_{VX}^4 + 2 \rho_{XZ}^2 \rho_{VX}^2 )\\
\text{Now we compare } & \rho_{VX}^4 - 2\rho_{VX}^2  - \rho_{XZ}^2 \rho_{VX}^4 \text{ and } -\rho_{VX}^2(\rho_{XZ}^2 + 1) \text{ by noticing that: } \\
\rho_{VX}^4 - 2\rho_{VX}^2  - \rho_{XZ}^2 \rho_{VX}^4 + \rho_{VX}^2(\rho_{XZ}^2 + 1) & = \rho_{VX}^4 - \rho_{VX}^2 - \rho_{VX}^4 \rho_{XZ}^2 + \rho_{VX}^2 \rho_{XZ}^2 = \rho_{VX}^2(\rho_{VX}^2 - 1)(1- \rho_{XZ}^2) \le 0 \text{ and as a result:}\\
\rho_{VX}^4 - 2\rho_{VX}^2  - \rho_{XZ}^2 \rho_{VX}^4 &  \le - \rho_{VX}^2(\rho_{XZ}^2 + 1) \text{ and}\\
(1 -\rho_{VX}^2)^2  (1 - \rho_{XZ}^2  + \rho_{VX}^4 & - 2\rho_{VX}^2   - \rho_{XZ}^2 \rho_{VX}^4 + 2 \rho_{XZ}^2 \rho_{VX}^2 ) \le (1 - \rho_{VX}^2)^2(1-\rho_{XZ}^2-\rho_{VX}^2\rho_{XZ}^2-\rho_{VX}^2 + 2\rho_{XZ}^2 \rho_{VX}^2).\\
\Longleftrightarrow (1-\rho_{XZ}^2)(1 -\rho_{VX}^2)^4 &  \le (1 - \rho_{VX}^2)^2(1-\rho_{XZ}^2-\rho_{VX}^2\rho_{XZ}^2-\rho_{VX}^2 + 2\rho_{XZ}^2 \rho_{VX}^2)\\
\Longleftrightarrow (1-\rho_{XZ}^2) & \le \frac{(1 - \rho_{VX}^2)^2(1-\rho_{XZ}^2-\rho_{VX}^2\rho_{XZ}^2-\rho_{VX}^2 + 2\rho_{XZ}^2 \rho_{VX}^2)}{(1 -\rho_{VX}^2)^4}\text{, thus inequality \eqref{DAG7_efficiency_part2} is proved.}
\end{aligned}
\end{equation*}  

Combining inequalities \eqref{DAG7_efficiency_part1} and \eqref{DAG7_efficiency_part2} we have: 
\begin{align*}
Var(\widehat{\beta}_{_{(--)}})  \le Var(\widehat{\beta}_{_{(OM)}})
\end{align*}

\subsubsection{For DAG 8 where $V$ is a non-risk factor of the outcome and is associated with true exposure and measurement error, estimators $\beta_{_{(OM)}}$ and $\beta_{_{(--)}}$ are valid and $\rho_{VY|X}=\rho_{YZ|X}=0$ which gives $\rho_{VY}= \rho_{VX}\rho_{XY}$ and $\rho_{YZ}=\rho_{XY}\rho_{XZ}$ following equation (10) in section 2.1:}
\begin{equation*}
\begin{aligned}
Var(\widehat{\beta}_{_{(OM)}}) & \approx \nicefrac{1}{{a}_1^2} Var({\hat{\gamma_1}}) + \nicefrac{{\gamma_1^2}}{{a}_1^4} Var(\hat{\alpha_1}) \\
& = \sigma_Y^2 \frac{(1-\rho_{YZ}^2-\rho_{VY}^2-\rho_{VZ}^2 + 2\rho_{YZ}\rho_{VY}\rho_{VZ}) n_{VS}}{\sigma_X^2 (\rho_{XZ}-\rho_{VX}\rho_{VZ})^2 n_{MS} n_{VS}} \\ & + \sigma_Y^2\frac{(\rho_{YZ} - \rho_{VY}\rho_{VZ})^2(1-\rho_{XZ}^2-\rho_{VZ}^2-\rho_{VX}^2 + 2\rho_{XZ} \rho_{VZ} \rho_{VX}) n_{MS}}{\sigma_X^2 (\rho_{XZ}-\rho_{VX}\rho_{VZ})^4 n_{MS} n_{VS}} \\
Var(\widehat{\beta}_{_{(--)}}) & \approx \nicefrac{1}{{\alpha^*_1}^2} Var({\hat{\gamma^*_1}}) + \nicefrac{{{\gamma^*_1}^2}}{{\alpha^*_1}^4} Var(\hat{\alpha^*_1}) \\
& = \sigma_Y^2 \frac{(1-\rho_{YZ}^2) n_{VS}}{\sigma_X^2 \rho_{XZ}^2 n_{MS} n_{VS}} + \sigma_Y^2\frac{\rho_{YZ}^2(1-\rho_{XZ}^2) n_{MS}}{\sigma_X^2 \rho_{XZ}^4 n_{MS} n_{VS}} 
\end{aligned}
\end{equation*}

For this DAG, the conditions we have do not allow us to definitively compare $Var(\widehat{\beta}_{_{(OM)}})$ and $Var(\widehat{\beta}_{_{(--)}})$ as the inequality depend on the magnitude of $\rho_{VX}, \rho_{VZ}, \rho_{VY}$, $\rho_{XZ}$, $\rho_{YZ}$ (or equivalently $\rho_{VX}, \rho_{VZ|X}, \rho_{VY|X}, \rho_{XZ|V},\rho_{XY|V}$) and $n_{MS}, n_{VS}$.

\newpage
\section{Figures: Analytical Relative Efficiency of $\hat{\beta}_{_{(OM)}}$ and $\hat{\beta}_{_{(-{}-)}}$ under DAG 1 and 8}
\begin{figure}[!htb]
    \centering    \includegraphics[width=\textwidth,height=\textheight,keepaspectratio]{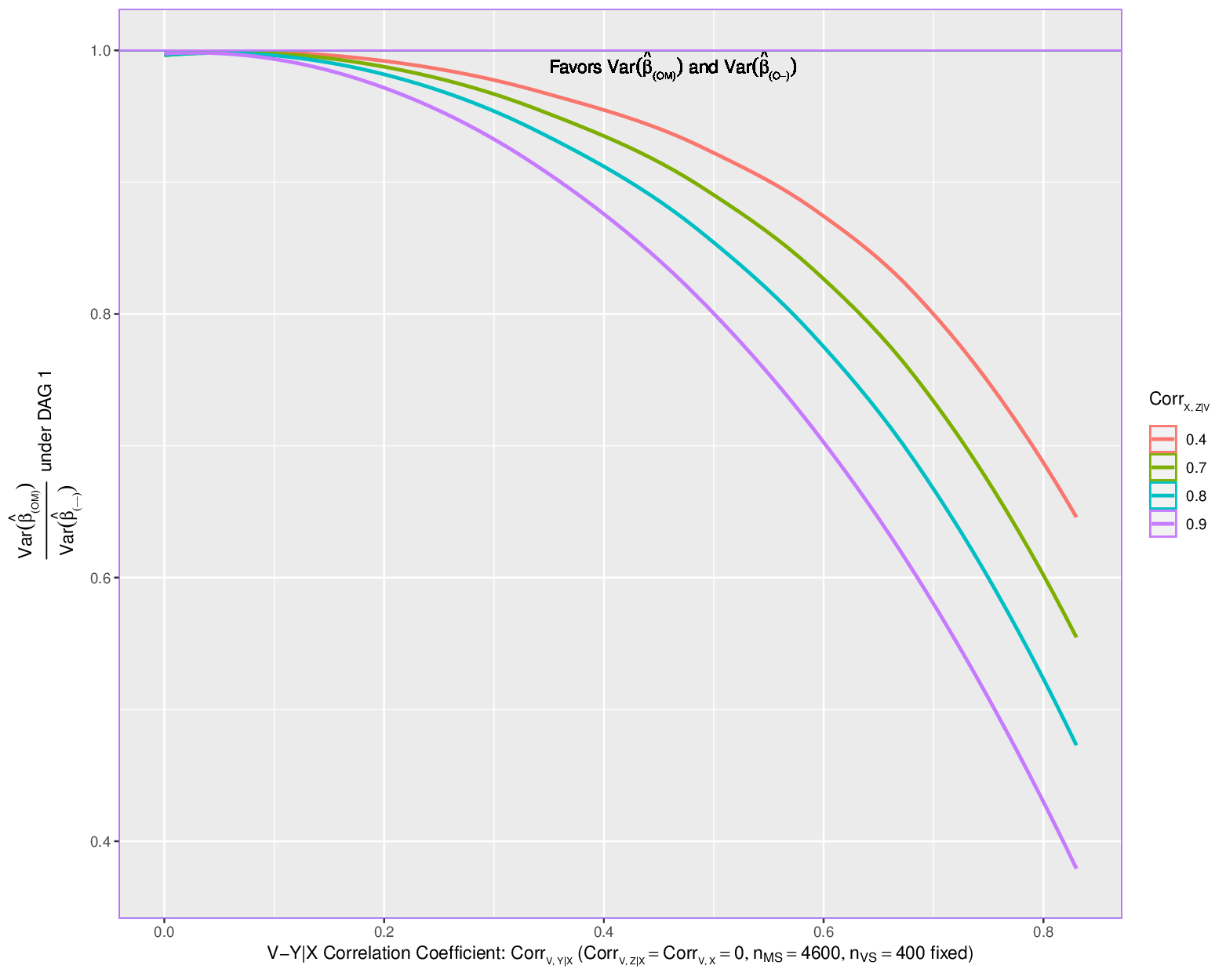}
    \caption{Analytical ARE under DAG 1 with Varying $\rho_{V,Y|X}$ and $\rho_{X,Z|V}$ Values}
    \label{fig:ARE_DAG1}
\end{figure}
\begin{figure}[!htb]
    \centering
\includegraphics[width=\textwidth,height=\textheight,keepaspectratio]{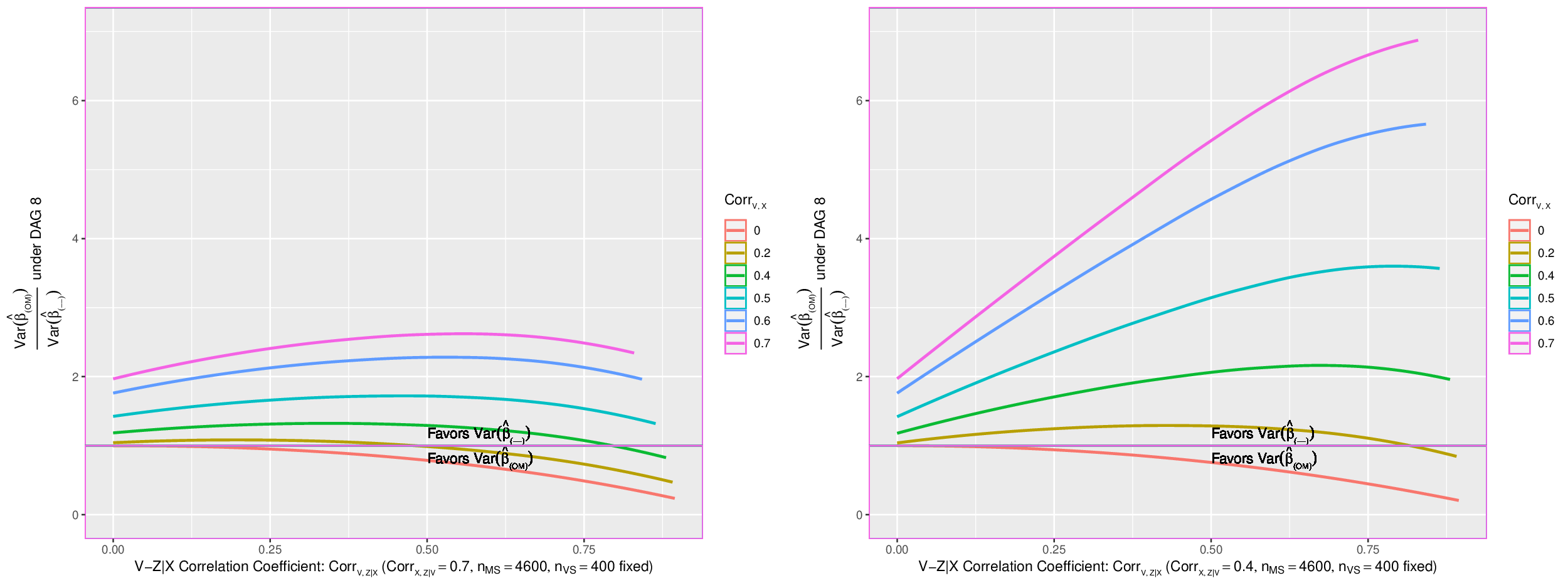}
    \caption{Analytical ARE under DAG 8 with Varying $\rho_{V,X}$, $\rho_{V,Z|X}$ and $\rho_{X,Z|V}$ Values}
    \label{fig:ARE_DAG1}
\end{figure}


\newpage
\section{Design of Simulation Studies} 
As extension for the base case, we allowed $V$ to distributed as $Bern(0.4)$ instead of $N(0,1)$ and varied values of $\eta_{v},\theta_{x},\theta_{v},\beta_{x},\beta_{v}$ to represent varying strength and direction of $\rho_{_{V,X}}, \rho_{_{X,Y|V}}, \rho_{_{X,Z|V}}, \rho_{_{V,Z|X}}$ and $\rho_{_{V,Y|X}}$. Note that when one or more of these coefficients are set to zero, the data generating process becomes compatible with a single DAG. For example, $\eta_{v}=\theta_{v}=0$ corresponds to DAG 1 ($V_{1(-{}-{}Y)}$), and $\beta_{v}=0$ in addition to $\eta_{v}=\theta_{v}=0$  corresponds to DAG 5. We summarize the possible coefficient parameterization and thus combinations of different data generating process under each DAG in the table below, where we also give their corresponding approximated (conditional) correlation coefficient $\rho_{(\cdot)}$. For the base case, the parameterization gives us $\rho_{_{V,X}} \in (0.37, 0) , \rho_{_{X,Y|V}} \approx 0.45, \rho_{_{X,Z|V}} \in 0.71, \rho_{_{V,Z|X}} \in (0.2, 0.18, 0)$ and $\rho_{_{V,Y|X}} \in (0.62, 0.6, 0)$. We also varied $n_{MS}$ and $n_{VS}$ as the variance calculation depends on them. 

We also note that for the binary outcomes generated by the logistic link function, due to non-collapsibility, the marginal effect of $X$ on $Y$ differs from the conditional effect of $X$ on $Y$ given $V$, the true data generating process under DAG 1 through 4. With a logit link and a continuous exposure, one cannot analytically obtain the marginal effect by integrating the conditional effect over covariate $V$, as the marginal effect would vary by values of $X$ and thus is not comparable to the constant conditional effect of $\beta_{x}$. However, since the simulation guarantees a rare disease with an outcome prevalence less than 5 \%, the marginal effect will be approximately $\beta_{x}$. Thus, even if for DAG 1 where $\beta_{(O-)}$ and $\beta_{(OM)}$ converge to slightly different parameter values than values that $\beta_{(--)}$ and $\beta_{(-M)}$ would converge to, we use $\beta_{x}$ as the truth for bias assessment.

\newpage 
\begin{threeparttable}
  \caption{Parameters Used in Monte Carlo Simulation}
  \label{tab:scenario}
    \scriptsize	
 \begin{tabular}{||c|c|ccccc|ccccc||}
 \hline
$V_j$ as in & Scenario \tnote{\textdagger} & $\eta_v$ & $\theta_x$ & $\theta_v$ & $\beta_x$ & $\beta_v$ & $\rho_{x,v}$ & $\rho_{x,z|v}$ & $\rho_{v,z|x}$ & $\rho_{v,y|x}$ \tnote{\textdaggerdbl} & $\rho_{x,y|v}$ \tnote{\textdaggerdbl}\\
\hline
DAG 1  & base case  & 0    & 0.5     & 0     & 0.5    & 0.8    & 0  & 0.71 & 0  & 0.62 & 0.45 \\
  &  small $\rho_{x,z|v}$     & 0    & 0.2     & 0     & 0.5    & 0.8    & 0  & 0.37 & 0  & 0.62 & 0.45 \\
  &  small effect $\beta_x$     & 0    & 0.5     & 0     & 0.1    & 0.8    & 0  & 0.71 & 0  & 0.62 & 0.1  \\
  & V is weak risk factor of Y& 0    & 0.5     & 0     & 0.5    & 0.2    & 0  & 0.71 & 0  & 0.2  & 0.45 \\
  & binary covariate V     & 0    & 0.5     & 0     & 0.5    & 0.8    & 0  & 0.71 & 0  & 0.36 & 0.45 \\
\hline
DAG 2  & base case  & 0    & 0.5     & 0.1     & 0.5    & 0.8    & 0  & 0.71 & 0.2  & 0.62 & 0.45 \\
 &  small $\rho_{x,z|v}$     & 0    & 0.2     & 0.1     & 0.5    & 0.8    & 0  & 0.37 & 0.2  & 0.62 & 0.45 \\
  &  small effect $\beta_x$     & 0    & 0.5     & 0.1     & 0.1    & 0.8    & 0  & 0.71 & 0.2  & 0.62 & 0.1  \\
  & large measurement error   $\rho_{v,z|x}$     & 0    & 0.5     & 2     & 0.5    & 0.8    & 0  & 0.71 & 0.97 & 0.62 & 0.45 \\
  & V is weak risk factor of Y    & 0    & 0.5     & 0.1     & 0.5    & 0.2    & 0  & 0.71 & 0.2  & 0.2  & 0.45 \\
  & binary covariate V     & 0    & 0.5     & 0.1     & 0.5    & 0.8    & 0  & 0.71 & 0.1  & 0.36 & 0.45 \\
\hline
DAG 3  & base case  & 0.4    & 0.5     & 0     & 0.5    & 0.8    & 0.37    & 0.71 & 0  & 0.6  & 0.45 \\
   &  small $\rho_{x,z|v}$     & 0.4    & 0.2     & 0     & 0.5    & 0.8    & 0.37    & 0.37 & 0  & 0.6  & 0.45 \\
   &  small effect $\beta_x$     & 0.4    & 0.5     & 0     & 0.1    & 0.8    & 0.37    & 0.71 & 0  & 0.6  & 0.1  \\
   & negative $\rho_{v,x}$     & -0.4   & 0.5     & 0     & 0.5    & 0.8    & -0.37     & 0.71 & 0  & 0.6  & 0.45 \\
   & small $\rho_{v,x}$     & 0.2    & 0.5     & 0     & 0.5    & 0.8    & 0.20    & 0.71 & 0  & 0.62 & 0.45 \\
   & V is weak risk factor of Y    & 0.4    & 0.5     & 0     & 0.5    & 0.2    & 0.37    & 0.71 & 0  & 0.18 & 0.45 \\
   & binary covariate V     & 0.4    & 0.5     & 0     & 0.5    & 0.8    & 0.19    & 0.71 & 0  & 0.36 & 0.45 \\
\hline
DAG 4  & base case  & 0.4    & 0.5     & 0.1     & 0.5    & 0.8    & 0.37    & 0.71 & 0.18 & 0.6  & 0.45 \\
   &  small $\rho_{x,z|v}$     & 0.4    & 0.2     & 0.1     & 0.5    & 0.8    & 0.37    & 0.37 & 0.18 & 0.6  & 0.45 \\
   &  small effect $\beta_x$     & 0.4    & 0.5     & 0.1     & 0.1    & 0.8    & 0.37    & 0.71 & 0.18 & 0.6  & 0.1  \\
   & negative $\rho_{v,x}$     & -0.4   & 0.5     & 0.1     & 0.5    & 0.8    & -0.37     & 0.71 & 0.18 & 0.6  & 0.45 \\
   & small $\rho_{v,x}$     & 0.2    & 0.5     & 0.1     & 0.5    & 0.8    & 0.20    & 0.71 & 0.19 & 0.62 & 0.45 \\
   & large measurement error   $\rho_{v,z|x}$     & 0.4    & 0.5     & 2     & 0.5    & 0.8    & 0.37    & 0.71 & 0.97 & 0.6  & 0.45 \\
   & V is weak risk factor of Y    & 0.4    & 0.5     & 0.1     & 0.5    & 0.2    & 0.37    & 0.71 & 0.18 & 0.18 & 0.45 \\
   & binary covariate V     & 0.4    & 0.5     & 0.1     & 0.5    & 0.8    & 0.19    & 0.71 & 0.1  & 0.36 & 0.45 \\
\hline
DAG 5  & base case  & 0    & 0.5     & 0     & 0.5    & 0    & 0  & 0.71 & 0  & 0  & 0.45 \\
   &  small $\rho_{x,z|v}$     & 0    & 0.2     & 0     & 0.5    & 0    & 0  & 0.37 & 0  & 0  & 0.45 \\
   &  small effect $\beta_x$     & 0    & 0.5     & 0     & 0.1    & 0    & 0  & 0.71 & 0  & 0  & 0.1  \\
   & binary covariate V     & 0    & 0.5     & 0     & 0.5    & 0    & 0  & 0.71 & 0  & 0  & 0.45 \\
\hline
DAG 6  & base case  & 0    & 0.5     & 0.1     & 0.5    & 0    & 0  & 0.71 & 0.2  & 0  & 0.45 \\
   &  small $\rho_{x,z|v}$     & 0    & 0.2     & 0.1     & 0.5    & 0    & 0  & 0.37 & 0.2  & 0  & 0.45 \\
   &  small effect $\beta_x$     & 0    & 0.5     & 0.1     & 0.1    & 0    & 0  & 0.71 & 0.2  & 0  & 0.1  \\
   & large measurement error   $\rho_{v,z|x}$     & 0    & 0.5     & 2     & 0.5    & 0    & 0  & 0.71 & 0.97 & 0  & 0.45 \\
   & binary covariate V     & 0    & 0.5     & 0.1     & 0.5    & 0    & 0  & 0.71 & 0.1  & 0  & 0.45 \\
\hline
DAG 7  & base case  & 0.4    & 0.5     & 0     & 0.5    & 0    & 0.37    & 0.71 & 0  & 0  & 0.45 \\
   &  small $\rho_{x,z|v}$     & 0.4    & 0.2     & 0     & 0.5    & 0    & 0.37    & 0.37 & 0  & 0  & 0.45 \\
   &  small effect $\beta_x$     & 0.4    & 0.5     & 0     & 0.1    & 0    & 0.37    & 0.71 & 0  & 0  & 0.1  \\
   & negative $\rho_{v,x}$     & -0.4   & 0.5     & 0     & 0.5    & 0    & -0.37     & 0.71 & 0  & 0  & 0.45 \\
   & small $\rho_{v,x}$     & 0.2    & 0.5     & 0     & 0.5    & 0    & 0.20    & 0.71 & 0  & 0  & 0.45 \\
   & binary covariate V     & 0.4    & 0.5     & 0     & 0.5    & 0    & 0.19    & 0.71 & 0  & 0  & 0.45 \\
\hline
DAG 8  & base case  & 0.4    & 0.5     & 0.1     & 0.5    & 0    & 0.37    & 0.71 & 0.18 & 0  & 0.45 \\
   &  small $\rho_{x,z|v}$     & 0.4    & 0.2     & 0.1     & 0.5    & 0    & 0.37    & 0.37 & 0.18 & 0  & 0.45 \\
   &  small effect $\beta_x$     & 0.4    & 0.5     & 0.1     & 0.1    & 0    & 0.37    & 0.71 & 0.18 & 0  & 0.1  \\
   & negative $\rho_{v,x}$     & -0.4   & 0.5     & 0.1     & 0.5    & 0    & -0.37     & 0.71 & 0.18 & 0  & 0.45 \\
   & small $\rho_{v,x}$     & 0.2    & 0.5     & 0.1     & 0.5    & 0    & 0.20    & 0.71 & 0.19 & 0  & 0.45 \\
   & large measurement error   $\rho_{v,z|x}$     & 0.4    & 0.5     & 2     & 0.5    & 0    & 0.37    & 0.71 & 0.97 & 0  & 0.45 \\
   & binary covariate V     & 0.4    & 0.5     & 0.1     & 0.5    & 0    & 0.19    & 0.71 & 0.1  & 0  & 0.45 \\
\hline
    \end{tabular}
        \begin{tablenotes}
        \item[\textdagger] We also varied the sample size: For main study, we reduced the sample size from $n_{MS} = 5,000$ to $n_{MS} = 2,000$ for continuous outcome and from $n_{MS} = 10,000$ to $n_{MS} = 5,000$ for binary outcome. For validation study, we reduced the sample size from $n_{VS}=400$ to $n_{VS} = 150$ for both continuous and binary outcome. 
        \item[\textdaggerdbl] These correlations only apply to continuous outcome generated under linear model. 
    \end{tablenotes}
 \end{threeparttable}

\newpage
\section{All Simulations Results: Percent Bias for Point Estimates in the Simulation Study}
\scriptsize	
\begin{longtable}{|| ll| lrrrrr | c c c c ||}
\caption{Percent Bias for Point Estimates}\\
\hline
\textbf{Y Type} & \textbf{DAG} & \textbf{Scenario} & $\rho_{v,x}$ & $\rho_{x,z|v}$ & $\rho_{v,z|x}$ & $\beta_x$ & $\beta_v$ & \textbf{$\beta_{(OM)}$} & \textbf{$\beta_{(--)}$} & \textbf{$\beta_{(-M)}$} & \textbf{$\beta_{(O-)}$} \\
\hline
\endhead
Cont'  & DAG 1   & base case  & 0 & 0.71& 0   & 0.5  & 0.8  & 0   & 0   & 0   & 0   \\
   & & small $\rho_{x,z|v}$ & 0 & 0.37& 0   & 0.5  & 0.8  & 2   & 1   & 2   & 2   \\
   & & small effect $\beta_x$  & 0 & 0.71& 0   & 0.1  & 0.8  & 1   & 0   & 0   & 1   \\
   & & weak risk factor V, small $\beta_v$ & 0 & 0.71& 0   & 0.5  & 0.2  & 0   & 0   & 0   & 0   \\
   & & binary covariate V   & 0 & 0.71& 0   & 0.5  & 0.8  & 0   & 0   & 0   & 0   \\
   & & $n_{MS}$ = 2,000 vs $n_{MS}$ = 5,000& 0 & 0.71& 0   & 0.5  & 0.8  & 0   & 0   & 0   & 0   \\
   & &$n_{VS}$ = 150 vs $n_{VS}$ = 400 & 0 & 0.71& 0   & 0.5  & 0.8  & 1   & 1   & 1   & 1   \\
   & DAG 2   & base case  & 0 & 0.71& 0.2 & 0.5  & 0.8  & 0   & 32  & 30  & 2   \\
   & & small $\rho_{x,z|v}$ & 0 & 0.37& 0.2 & 0.5  & 0.8  & 2   & 83  & 77  & 5   \\
   & & small effect $\beta_x$  & 0 & 0.71& 0.2 & 0.1  & 0.8  & 1   & 161 & 156 & 3   \\
   & & large correlation $\rho_{v,z|x}$& 0 & 0.71& 0.97& 0.5  & 0.8  & 0   & 670 & -18 & 838 \\
   & & weak risk factor V, small $\beta_v$ & 0 & 0.71& 0.2 & 0.5  & 0.2  & 0   & 8   & 6   & 2   \\
   & & binary covariate V   & 0 & 0.71& 0.1 & 0.5  & 0.8  & 0   & 7   & 7   & 1   \\
   & & $n_{MS}$ = 2,000 vs $n_{MS}$ = 5,000& 0 & 0.71& 0.2 & 0.5  & 0.8  & 0   & 32  & 30  & 2   \\
   & &$n_{VS}$ = 150 vs $n_{VS}$ = 400 & 0 & 0.71& 0.2 & 0.5  & 0.8  & 1   & 33  & 30  & 3   \\
   & DAG 3   & base case  & 0.37 & 0.71& 0   & 0.5  & 0.8  & 0   & 55  & 67  & -7  \\
   & & small $\rho_{x,z|v}$ & 0.37 & 0.37& 0   & 0.5  & 0.8  & 2   & 57  & 79  & -11 \\
   & & small effect $\beta_x$  & 0.37 & 0.71& 0   & 0.1  & 0.8  & 1   & 276 & 305 & -6  \\
   & & negative $\rho_{v,x}$& -0.37& 0.71& 0   & 0.5  & 0.8  & 0   & -55 & -52 & -6  \\
   & & small $\rho_{v,x}$   & 0.20 & 0.71& 0   & 0.5  & 0.8  & 0   & 31  & 34  & -2  \\
   & & weak risk factor V, small $\beta_v$ & 0.37 & 0.71& 0   & 0.5  & 0.2  & 0   & 14  & 23  & -7  \\
   & & binary covariate V   & 0.19 & 0.71& 0   & 0.5  & 0.8  & 0   & 13  & 15  & -1  \\
   & & $n_{MS}$ = 2,000 vs $n_{MS}$ = 5,000& 0.37 & 0.71& 0   & 0.5  & 0.8  & 0   & 56  & 67  & -7  \\
   & &$n_{VS}$ = 150 vs $n_{VS}$ = 400 & 0.37 & 0.71& 0   & 0.5  & 0.8  & 1   & 56  & 68  & -6  \\
   & DAG 4   & base case  & 0.37 & 0.71& 0.18& 0.5  & 0.8  & 0   & 78  & 87  & -5  \\
   & & small $\rho_{x,z|v}$ & 0.37 & 0.37& 0.18& 0.5  & 0.8  & 2   & 108 & 156 & -17 \\
   & & small effect $\beta_x$  & 0.37 & 0.71& 0.18& 0.1  & 0.8  & 1   & 388 & 413 & -4  \\
   & & negative $\rho_{v,x}$& -0.37& 0.71& 0.18& 0.5  & 0.8  & 0   & -29 & -25 & -5  \\
   & & small $\rho_{v,x}$   & 0.20 & 0.71& 0.19& 0.5  & 0.8  & 0   & 60  & 60  & 0   \\
   & & large $\rho_{v,z|x}$& 0.37 & 0.71& 0.97& 0.5  & 0.8  & 0   & 256 & -8  & 288 \\
   & & weak risk factor V, small $\beta_v$ & 0.37 & 0.71& 0.18& 0.5  & 0.2  & 0   & 20  & 26  & -5  \\
   & & binary covariate V   & 0.19 & 0.71& 0.1 & 0.5  & 0.8  & 0   & 20  & 21  & -1  \\
   & & $n_{MS}$ = 2,000 vs $n_{MS}$ = 5,000& 0.37 & 0.71& 0.18& 0.5  & 0.8  & 0   & 78  & 87  & -5  \\
   & &$n_{VS}$ = 150 vs $n_{VS}$ = 400 & 0.37 & 0.71& 0.18& 0.5  & 0.8  & 1   & 78  & 88  & -4  \\
   & DAG 5   & base case  & 0 & 0.71& 0   & 0.5  & 0 & 0   & 0   & 0   & 0   \\
   & & small $\rho_{x,z|v}$ & 0 & 0.37& 0   & 0.5  & 0 & 2   & 2   & 2   & 2   \\
   & & small effect $\beta_x$  & 0 & 0.71& 0   & 0.1  & 0 & 1   & 1   & 1   & 1   \\
   & & binary covariate V   & 0 & 0.71& 0   & 0.5  & 0 & 0   & 0   & 0   & 0   \\
   & & $n_{MS}$ = 2,000 vs $n_{MS}$ = 5,000& 0 & 0.71& 0   & 0.5  & 0 & 0   & 0   & 0   & 0   \\
   & &$n_{VS}$ = 150 vs $n_{VS}$ = 400 & 0 & 0.71& 0   & 0.5  & 0 & 1   & 1   & 1   & 1   \\
   & DAG 6   & base case  & 0 & 0.71& 0.2 & 0.5  & 0 & 0   & 0   & -2  & 2   \\
   & & small $\rho_{x,z|v}$ & 0 & 0.37& 0.2 & 0.5  & 0 & 2   & 2   & -2  & 5   \\
   & & small effect $\beta_x$  & 0 & 0.71& 0.2 & 0.1  & 0 & 1   & 1   & -1  & 3   \\
   & & large $\rho_{v,z|x}$& 0 & 0.71& 0.97& 0.5  & 0 & 0   & 4   & -89 & 838 \\
   & & binary covariate V   & 0 & 0.71& 0.1 & 0.5  & 0 & 0   & 0   & 0   & 1   \\
   & & $n_{MS}$ = 2,000 vs $n_{MS}$ = 5,000& 0 & 0.71& 0.2 & 0.5  & 0 & 0   & 0   & -2  & 2   \\
   & &$n_{VS}$ = 150 vs $n_{VS}$ = 400 & 0 & 0.71& 0.2 & 0.5  & 0 & 1   & 1   & -1  & 3   \\
   & DAG 7   & base case  & 0.37 & 0.71& 0   & 0.5  & 0 & 0   & 0   & 8   & -7  \\
   & & small $\rho_{x,z|v}$ & 0.37 & 0.37& 0   & 0.5  & 0 & 2   & 1   & 15  & -11 \\
   & & small effect $\beta_x$  & 0.37 & 0.71& 0   & 0.1  & 0 & 1   & 1   & 8   & -6  \\
   & & negative $\rho_{v,x}$& -0.37& 0.71& 0   & 0.5  & 0 & 0   & 1   & 8   & -6  \\
   & & small $\rho_{v,x}$   & 0.20 & 0.71& 0   & 0.5  & 0 & 0   & 0   & 2   & -2  \\
   & & binary covariate V   & 0.19 & 0.71& 0   & 0.5  & 0 & 0   & 0   & 2   & -1  \\
   & & $n_{MS}$ = 2,000 vs $n_{MS}$ = 5,000& 0.37 & 0.71& 0   & 0.5  & 0 & 0   & 0   & 8   & -7  \\
   & &$n_{VS}$ = 150 vs $n_{VS}$ = 400 & 0.37 & 0.71& 0   & 0.5  & 0 & 1   & 0   & 8   & -6  \\
   & DAG 8   & base case  & 0.37 & 0.71& 0.18& 0.5  & 0 & 0   & 0   & 5   & -5  \\
   & & small $\rho_{x,z|v}$ & 0.37 & 0.37& 0.18& 0.5  & 0 & 2   & 1   & 24  & -17 \\
   & & small effect $\beta_x$  & 0.37 & 0.71& 0.18& 0.1  & 0 & 1   & 0   & 6   & -4  \\
   & & negative $\rho_{v,x}$& -0.37& 0.71& 0.18& 0.5  & 0 & 0   & 1   & 6   & -5  \\
   & & small $\rho_{v,x}$   & 0.20 & 0.71& 0.19& 0.5  & 0 & 0   & 0   & 0   & 0   \\
   & & large $\rho_{v,z|x}$& 0.37 & 0.71& 0.97& 0.5  & 0 & 0   & 0   & -74 & 288 \\
   & & binary covariate V   & 0.19 & 0.71& 0.1 & 0.5  & 0 & 0   & 0   & 2   & -1  \\
   & & $n_{MS}$ = 2,000 vs $n_{MS}$ = 5,000& 0.37 & 0.71& 0.18& 0.5  & 0 & 0   & 0   & 5   & -5  \\
   & &$n_{VS}$ = 150 vs $n_{VS}$ = 400 & 0.37 & 0.71& 0.18& 0.5  & 0 & 1   & 0   & 6   & -4  \\
\hline
Binary& DAG 1   & base case  & 0 & 0.71& 0   & 0.5  & 0.8  & 0   & -1  & -1  & 0   \\
   & & small $\rho_{x,z|v}$ & 0 & 0.37& 0   & 0.5  & 0.8  & 1   & 0   & 0   & 1   \\
   & & small effect $\beta_x$  & 0 & 0.71& 0   & 0.1  & 0.8  & -5  & -6  & -6  & -5  \\
   & & weak risk factor V, small $\beta_v$ & 0 & 0.71& 0   & 0.5  & 0.2  & 1   & 1   & 1   & 1   \\
   & & binary covariate V   & 0 & 0.71& 0   & 0.5  & 0.8  & -1  & -1  & -1  & -1  \\
   & & $n_{MS}$ = 5,000 vs $n_{MS}$ = 10,000  & 0 & 0.71& 0   & 0.5  & 0.8  & 0   & -1  & -1  & 0   \\
   & &$n_{VS}$ = 150 vs $n_{VS}$ = 400 & 0 & 0.71& 0   & 0.5  & 0.8  & 0   & -1  & -1  & 0   \\
   & DAG 2   & base case  & 0 & 0.71& 0.2 & 0.5  & 0.8  & 0   & 31  & 28  & 2   \\
   & & small $\rho_{x,z|v}$ & 0 & 0.37& 0.2 & 0.5  & 0.8  & 1   & 81  & 74  & 5   \\
   & & small effect $\beta_x$  & 0 & 0.71& 0.2 & 0.1  & 0.8  & -5  & 154 & 149 & -3  \\
   & & large $\rho_{v,z|x}$& 0 & 0.71& 0.97& 0.5  & 0.8  & 0   & 675 & -17 & 839 \\
   & & weak risk factor V, small $\beta_v$ & 0 & 0.71& 0.2 & 0.5  & 0.2  & 1   & 9   & 7   & 3   \\
   & & binary covariate V   & 0 & 0.71& 0.1 & 0.5  & 0.8  & -1  & 4   & 4   & -1  \\
   & & $n_{MS}$ = 5,000 vs $n_{MS}$ = 10,000  & 0 & 0.71& 0.2 & 0.5  & 0.8  & 0   & 30  & 28  & 2   \\
   & &$n_{VS}$ = 150 vs $n_{VS}$ = 400 & 0 & 0.71& 0.2 & 0.5  & 0.8  & 0   & 31  & 29  & 2   \\
   & DAG 3   & base case  & 0.371391& 0.71& 0   & 0.5  & 0.8  & 0   & 52  & 64  & -7  \\
   & & small $\rho_{x,z|v}$ & 0.371391& 0.37& 0   & 0.5  & 0.8  & 1   & 54  & 74  & -11 \\
   & & small effect $\beta_x$  & 0.371391& 0.71& 0   & 0.1  & 0.8  & -5  & 270 & 298 & -11 \\
   & & negative $\rho_{v,x}$& -0.37139& 0.71& 0   & 0.5  & 0.8  & -1  & -56 & -53 & -7  \\
   & & small $\rho_{v,x}$   & 0.196116& 0.71& 0   & 0.5  & 0.8  & 0   & 29  & 31  & -2  \\
   & & weak risk factor V, small $\beta_v$ & 0.371391& 0.71& 0   & 0.5  & 0.2  & 0   & 14  & 23  & -7  \\
   & & binary covariate V   & 0.192302& 0.71& 0   & 0.5  & 0.8  & -1  & 9   & 11  & -3  \\
   & & $n_{MS}$ = 5,000 vs $n_{MS}$ = 10,000  & 0.371391& 0.71& 0   & 0.5  & 0.8  & -1  & 52  & 63  & -8  \\
   & &$n_{VS}$ = 150 vs $n_{VS}$ = 400 & 0.371391& 0.71& 0   & 0.5  & 0.8  & 0   & 53  & 64  & -7  \\
   & DAG 4   & base case  & 0.371391& 0.71& 0.18& 0.5  & 0.8  & 0   & 75  & 84  & -5  \\
   & & small $\rho_{x,z|v}$ & 0.371391& 0.37& 0.18& 0.5  & 0.8  & 1   & 103 & 150 & -17 \\
   & & small effect $\beta_x$  & 0.371391& 0.71& 0.18& 0.1  & 0.8  & -5  & 382 & 406 & -9  \\
   & & negative $\rho_{v,x}$& -0.37139& 0.71& 0.18& 0.5  & 0.8  & -1  & -31 & -27 & -6  \\
   & & small $\rho_{v,x}$   & 0.196116& 0.71& 0.19& 0.5  & 0.8  & 0   & 57  & 57  & 0   \\
   & & large $\rho_{v,z|x}$& 0.371391& 0.71& 0.97& 0.5  & 0.8  & 0   & 256 & -8  & 287 \\
   & & weak risk factor V, small $\beta_v$ & 0.371391& 0.71& 0.18& 0.5  & 0.2  & 0   & 20  & 26  & -5  \\
   & & binary covariate V   & 0.192302& 0.71& 0.1 & 0.5  & 0.8  & -1  & 14  & 15  & -2  \\
   & & $n_{MS}$ = 5,000 vs $n_{MS}$ = 10,000  & 0.371391& 0.71& 0.18& 0.5  & 0.8  & -1  & 74  & 83  & -6  \\
   & &$n_{VS}$ = 150 vs $n_{VS}$ = 400 & 0.371391& 0.71& 0.18& 0.5  & 0.8  & 0   & 75  & 84  & -5  \\
   & DAG 5   & base case  & 0 & 0.71& 0   & 0.5  & 0 & 1   & 1   & 1   & 1   \\
   & & small $\rho_{x,z|v}$ & 0 & 0.37& 0   & 0.5  & 0 & 4   & 4   & 4   & 4   \\
   & & small effect $\beta_x$  & 0 & 0.71& 0   & 0.1  & 0 & -3  & -3  & -3  & -3  \\
   & & binary covariate V   & 0 & 0.71& 0   & 0.5  & 0 & 1   & 1   & 1   & 1   \\
   & & $n_{MS}$ = 5,000 vs $n_{MS}$ = 10,000  & 0 & 0.71& 0   & 0.5  & 0 & -1  & -1  & -1  & -1  \\
   & &$n_{VS}$ = 150 vs $n_{VS}$ = 400 & 0 & 0.71& 0   & 0.5  & 0 & 1   & 1   & 1   & 1   \\
   & DAG 6   & base case  & 0 & 0.71& 0.2 & 0.5  & 0 & 1   & 1   & -1  & 3   \\
   & & small $\rho_{x,z|v}$ & 0 & 0.37& 0.2 & 0.5  & 0 & 4   & 4   & 1   & 7   \\
   & & small effect $\beta_x$  & 0 & 0.71& 0.2 & 0.1  & 0 & -3  & -1  & -3  & -1  \\
   & & large $\rho_{v,z|x}$& 0 & 0.71& 0.97& 0.5  & 0 & 1   & 12  & -88 & 850 \\
   & & binary covariate V   & 0 & 0.71& 0.1 & 0.5  & 0 & 1   & 1   & 1   & 1   \\
   & & $n_{MS}$ = 5,000 vs $n_{MS}$ = 10,000  & 0 & 0.71& 0.2 & 0.5  & 0 & -1  & -1  & -3  & 1   \\
   & &$n_{VS}$ = 150 vs $n_{VS}$ = 400 & 0 & 0.71& 0.2 & 0.5  & 0 & 1   & 2   & 0   & 4   \\
   & DAG 7   & base case  & 0.371391& 0.71& 0   & 0.5  & 0 & 1   & 2   & 9   & -6  \\
   & & small $\rho_{x,z|v}$ & 0.371391& 0.37& 0   & 0.5  & 0 & 4   & 4   & 18  & -9  \\
   & & small effect $\beta_x$  & 0.371391& 0.71& 0   & 0.1  & 0 & -3  & 0   & 8   & -10 \\
   & & negative $\rho_{v,x}$& -0.37139& 0.71& 0   & 0.5  & 0 & 1   & 0   & 8   & -6  \\
   & & small $\rho_{v,x}$   & 0.196116& 0.71& 0   & 0.5  & 0 & 1   & 1   & 3   & -1  \\
   & & binary covariate V   & 0.192302& 0.71& 0   & 0.5  & 0 & 0   & 0   & 2   & -2  \\
   & & $n_{MS}$ = 5,000 vs $n_{MS}$ = 10,000  & 0.371391& 0.71& 0   & 0.5  & 0 & 0   & 1   & 9   & -7  \\
   & &$n_{VS}$ = 150 vs $n_{VS}$ = 400 & 0.371391& 0.71& 0   & 0.5  & 0 & 2   & 2   & 10  & -5  \\
   & DAG 8   & base case  & 0.371391& 0.71& 0.18& 0.5  & 0 & 1   & 2   & 7   & -4  \\
   & & small $\rho_{x,z|v}$ & 0.371391& 0.37& 0.18& 0.5  & 0 & 4   & 4   & 27  & -16 \\
   & & small effect $\beta_x$  & 0.371391& 0.71& 0.18& 0.1  & 0 & -3  & 2   & 7   & -8  \\
   & & negative $\rho_{v,x}$& -0.37139& 0.71& 0.18& 0.5  & 0 & 1   & 1   & 6   & -5  \\
   & & small $\rho_{v,x}$   & 0.196116& 0.71& 0.19& 0.5  & 0 & 1   & 2   & 2   & 1   \\
   & & large $\rho_{v,z|x}$& 0.371391& 0.71& 0.97& 0.5  & 0 & 1   & 3   & -73 & 293 \\
   & & binary covariate V   & 0.192302& 0.71& 0.1 & 0.5  & 0 & 0   & 0   & 1   & -1  \\
   & & $n_{MS}$ = 5,000 vs $n_{MS}$ = 10,000  & 0.371391& 0.71& 0.18& 0.5  & 0 & 0   & 1   & 7   & -5  \\
   & &$n_{VS}$ = 150 vs $n_{VS}$ = 400 & 0.371391& 0.71& 0.18& 0.5  & 0 & 2   & 2   & 7   & -3  \\
\hline
\end{longtable}

\newpage
\section{All Simulations Results: Analytical and Empirical Asymptotic Relative Efficiency as well as Variance for Valid Estimators}
\begin{longtable}{|| ll| lrrrrr | c c c c ||}
\caption{Analytical ARE ($\widehat{ARE} (\hat{\beta}_{(\cdot)}) = \frac{\widehat{Var}(\hat{\beta}_{(OM)})}{\widehat{Var}(\hat{\beta}_{(\cdot)})}$) and Analytical Variance ($\widehat{Var}(\hat{\beta}_{(\cdot)})$) for Continuous Outcomes in the Simulation Study}\\
\hline
\textbf{Y Type} & \textbf{DAG} & \textbf{Scenario} & $\rho_{v,x}$ & $\rho_{x,z|v}$ & $\rho_{v,z|x}$ & $\beta_x$ & $\beta_v$ & \multicolumn{4}{c||}{ $\widehat{ARE}$ ($\hat{\beta}_{(\cdot)}$) ($\widehat{Var}(\hat{\beta}_{(\cdot)}) \times 10^{3})$} \\
 &  &  &  &  &  &  &  & \textbf{$\hat{\beta}_{(OM)}$} & \textbf{$\hat{\beta}_{(--)}$} & \textbf{$\hat{\beta}_{(-M)}$} & \textbf{$\hat{\beta}_{(O-)}$} \\
 \hline
Cont' & DAG 1  & base case& 0  & 0.71 & 0 & 0.5 & 0.8 & 1 (1.08)& 0.81 (1.33) & 0.81 (1.33) & 1 (1.08) \\
  &   & small $\rho_{x,z|v}$  & 0  & 0.37 & 0 & 0.5 & 0.8 & 1 (5.67)& 0.86 (6.6)& 0.86 (6.6)& 1 (5.67) \\
  &   & small effect $\beta_x$& 0  & 0.71 & 0 & 0.1 & 0.8 & 1 (0.43)& 0.63 (0.68) & 0.63 (0.68) & 1 (0.43) \\
  &   & weak risk factor V, small $\beta_v$& 0  & 0.71 & 0 & 0.5 & 0.2 & 1 (1.08)& 0.99 (1.09) & 0.99 (1.09) & 1 (1.08) \\
  &   & binary covariate V& 0  & 0.71 & 0 & 0.5 & 0.8 & 1 (1.08)& 0.95 (1.14) & 0.95 (1.14) & 1 (1.08) \\
  &   & $n_{MS}$ = 2,000 vs $n_{MS}$=5,000& 0  & 0.71 & 0 & 0.5 & 0.8 & 1 (1.75)& 0.73 (2.39) & 0.73 (2.39) & 1 (1.75) \\
  &   & $n_{VS}$ = 150 vs $n_{VS}$ = 400 & 0  & 0.71 & 0 & 0.5 & 0.8 & 1 (2.12)& 0.89 (2.37) & 0.89 (2.37) & 1 (2.12) \\
  & DAG 5  & base case& 0  & 0.71 & 0 & 0.5 & 0   & 1 (1.08)& 1 (1.08)& 1 (1.08)& 1 (1.08) \\
  &   & small $\rho_{x,z|v}$  & 0  & 0.37 & 0 & 0.5 & 0   & 1 (5.67)& 1 (5.67)& 1 (5.67)& 1 (5.67) \\
  &   & small effect $\beta_x$& 0  & 0.71 & 0 & 0.1 & 0   & 1 (0.43)& 1 (0.43)& 1 (0.43)& 1 (0.43) \\
  &   & binary covariate V& 0  & 0.71 & 0 & 0.5 & 0   & 1 (1.08)& 1 (1.08)& 1 (1.08)& 1 (1.08) \\
  &   & $n_{MS}$ = 2,000 vs $n_{MS}$=5,000& 0  & 0.71 & 0 & 0.5 & 0   & 1 (1.75)& 1 (1.75)& 1 (1.75)& 1 (1.75) \\
  &   & $n_{VS}$ = 150 vs $n_{VS}$ = 400 & 0  & 0.71 & 0 & 0.5 & 0   & 1 (2.12)& 1 (2.12)& 1 (2.12)& 1 (2.12) \\
  & DAG 6   & base case& 0  & 0.71 & 0.2  & 0.5 & 0   & 1 (1.08)& 0.97 (1.11) &&\\
  &  & small $\rho_{x,z|v}$  & 0  & 0.37 & 0.2  & 0.5 & 0   & 1 (5.67)& 0.96 (5.89) &&\\
  &   & small effect $\beta_x$& 0  & 0.71 & 0.2  & 0.1 & 0   & 1 (0.43)& 0.98 (0.44) &&\\
  &   & large $\rho_{v,z|x}$ & 0  & 0.71 & 0.97 & 0.5 & 0   & 1 (1.08)& 0.07 (15.08)&&\\
  &   & binary covariate V& 0  & 0.71 & 0.1  & 0.5 & 0   & 1 (1.08)& 0.99 (1.08) &&\\
  &   & $n_{MS}$ = 2,000 vs $n_{MS}$=5,000& 0  & 0.71 & 0.2  & 0.5 & 0   & 1 (1.75)& 0.97 (1.8)&&\\
  &   & $n_{VS}$ = 150 vs $n_{VS}$ = 400 & 0  & 0.71 & 0.2  & 0.5 & 0   & 1 (2.12)& 0.97 (2.19) &&\\
  & DAG 7  & base case& 0.37   & 0.71 & 0 & 0.5 & 0   & 1 (1.08)& 1.19 (0.9)&&\\
  &   & small $\rho_{v,x}$& 0.20   & 0.71 & 0 & 0.5 & 0   & 1 (1.08)& 1.05 (1.03) &&\\
  &   & negative $\rho_{v,x}$ & -0.37  & 0.71 & 0 & 0.5 & 0   & 1 (1.08)& 1.19 (0.9)&&\\
  &   & small $\rho_{x,z|v}$  & 0.37   & 0.37 & 0 & 0.5 & 0   & 1 (5.67)& 1.2 (4.74)&&\\
  &   & small effect $\beta_x$& 0.37   & 0.71 & 0 & 0.1 & 0   & 1 (0.43)& 1.24 (0.34) &&\\
  &   & binary covariate V& 0.19   & 0.71 & 0 & 0.5 & 0   & 1 (1.08)& 1.05 (1.03) &&\\
  &   & $n_{MS}$ = 2,000 vs $n_{MS}$=5,000& 0.37   & 0.71 & 0 & 0.5 & 0   & 1 (1.75)& 1.21 (1.45) &&\\
  &   & $n_{VS}$ = 150 vs $n_{VS}$ = 400 & 0.37   & 0.71 & 0 & 0.5 & 0   & 1 (2.12)& 1.18 (1.8)&&\\
  & DAG 8  & base case& 0.37   & 0.71 & 0.18 & 0.5 & 0   & 1 (1.08)& 1.29 (0.83) &&\\
  &   & small $\rho_{v,x}$& 0.20   & 0.71 & 0.19 & 0.5 & 0   & 1 (1.08)& 1.08 (1.00) &&\\
  &   & negative $\rho_{v,x}$ & -0.37  & 0.71 & 0.18 & 0.5 & 0   & 1 (1.08)   & 1 (1.04)   &&\\
  &   & small $\rho_{x,z|v}$  & 0.37   & 0.37 & 0.18 & 0.5 & 0   & 1 (5.67)& 1.57 (3.61) &&\\
  &   & small effect $\beta_x$& 0.37   & 0.71 & 0.18 & 0.1 & 0   & 1 (0.43)& 1.3 (0.33)&&\\
  &   & large $\rho_{v,z|x}$ & 0.37   & 0.71 & 0.97 & 0.5 & 0   & 1 (1.07)& 0.52 (2.08) &&\\
  &   & binary covariate V& 0.19   & 0.71 & 0.1  & 0.5 & 0   & 1 (1.08)& 1.07 (1.01) &&\\
  &   & $n_{MS}$ = 2,000 vs $n_{MS}$=5,000& 0.37   & 0.71 & 0.18 & 0.5 & 0   & 1 (1.75)& 1.29 (1.35) &&\\
  &   & $n_{VS}$ = 150 vs $n_{VS}$ = 400 & 0.37   & 0.71 & 0.18 & 0.5 & 0   & 1 (2.12)& 1.29 (1.65) && \\
\hline
\end{longtable}

\newpage
\begin{longtable}{|| ll | lrrrrr | c c c c ||}
\caption{Empirical ARE ($ERE (\hat{\beta}_{(\cdot)}) = \frac{Var(\hat{\beta}_{(OM)})}{Var(\hat{\beta}_{(\cdot)})}$) and Empirical Variance for Continuous Outcomes in the Simulation Study} \\
\hline
\textbf{Y Type} & \textbf{DAG} & \textbf{Scenario} & $\rho_{v,x}$ & $\rho_{x,z|v}$ & $\rho_{v,z|x}$ & $\beta_x$ & $\beta_v$ & \multicolumn{4}{c||}{ $ERE (\hat{\beta}_{(\cdot)}) (Var( \hat{\beta}_{(\cdot)}) \times 10^{3})$}\\
 &  &  &  &  &  &  &  & \textbf{$\hat{\beta}_{(OM)}$} & \textbf{$\hat{\beta}_{(--)}$} & \textbf{$\hat{\beta}_{(-M)}$} & \textbf{$\hat{\beta}_{(O-)}$} \\
\hline
\endhead
Cont' & DAG 1  & base case& 0  & 0.71 & 0 & 0.5 & 0.8 & 1 (1.14)   & 0.79 (1.44)& 0.79 (1.44)& 1 (1.14)   \\
  &   & small $\rho_{x,z|v}$  & 0  & 0.37 & 0 & 0.5 & 0.8 & 1 (6.35)   & 0.87 (7.29)& 0.87 (7.33)& 1.01 (6.32)\\
  &   & small effect $\beta_x$& 0  & 0.71 & 0 & 0.1 & 0.8 & 1 (0.45)   & 0.6 (0.74) & 0.6 (0.74) & 1 (0.45)   \\
  &   & weak risk factor V, small $\beta_v$& 0  & 0.71 & 0 & 0.5 & 0.2 & 1 (1.14)   & 0.98 (1.16)& 0.98 (1.16)& 1 (1.14)   \\
  &   & binary covariate V& 0  & 0.71 & 0 & 0.5 & 0.8 & 1 (1.14)   & 0.96 (1.19)& 0.96 (1.19)& 1 (1.14)   \\
  &   & $n_{MS}$ = 2,000 vs $n_{MS}$=5,000& 0  & 0.71 & 0 & 0.5 & 0.8 & 1 (1.92)   & 0.69 (2.76)& 0.69 (2.77)& 1 (1.91)   \\
  &   & $n_{VS}$ = 150 vs $n_{VS}$ = 400 & 0  & 0.71 & 0 & 0.5 & 0.8 & 1 (2.25)   & 0.9 (2.5)  & 0.89 (2.52)& 1.01 (2.23)\\
  & DAG 5  & base case& 0  & 0.71 & 0 & 0.5 & 0   & 1 (1.14)   & 1 (1.14)   & 1 (1.14)   & 1 (1.14)   \\
  &   & small $\rho_{x,z|v}$  & 0  & 0.37 & 0 & 0.5 & 0   & 1 (6.35)   & 1 (6.32)   & 1 (6.35)   & 1.01 (6.32)\\
  &   & small effect $\beta_x$& 0  & 0.71 & 0 & 0.1 & 0   & 1 (0.45)   & 1 (0.45)   & 1 (0.45)   & 1 (0.45)   \\
  &   & binary covariate V& 0  & 0.71 & 0 & 0.5 & 0   & 1 (1.14)   & 1 (1.14)   & 1 (1.14)   & 1 (1.14)   \\
  &   & $n_{MS}$ = 2,000 vs $n_{MS}$=5,000& 0  & 0.71 & 0 & 0.5 & 0   & 1 (1.92)   & 1.01 (1.91)& 1 (1.92)   & 1 (1.91)   \\
  &   & $n_{VS}$ = 150 vs $n_{VS}$ = 400 & 0  & 0.71 & 0 & 0.5 & 0   & 1 (2.25)   & 1.01 (2.23)& 1 (2.25)   & 1.01 (2.23)\\
  & DAG 6   & base case& 0  & 0.71 & 0.2  & 0.5 & 0   & 1 (1.14)   & \multicolumn{2}{l}{0.97 (1.18)} &\\
  &  & small $\rho_{x,z|v}$  & 0  & 0.37 & 0.2  & 0.5 & 0   & 1 (6.35)   & \multicolumn{2}{l}{0.97 (6.53)} &\\
  &   & small effect $\beta_x$& 0  & 0.71 & 0.2  & 0.1 & 0   & 1 (0.45)   & \multicolumn{2}{l}{0.98 (0.46)} &\\
  &   & large $\rho_{v,z|x}$ & 0  & 0.71 & 0.97 & 0.5 & 0   & 1 (1.14)   & \multicolumn{2}{l}{0.05 (23.68)}&\\
  &   & binary covariate V& 0  & 0.71 & 0.1  & 0.5 & 0   & 1 (1.14)   & \multicolumn{2}{l}{0.99 (1.15)} &\\
  &   & $n_{MS}$ = 2,000 vs $n_{MS}$=5,000& 0  & 0.71 & 0.2  & 0.5 & 0   & 1 (1.92)   & \multicolumn{2}{l}{0.98 (1.97)} &\\
  &   & $n_{VS}$ = 150 vs $n_{VS}$ = 400 & 0  & 0.71 & 0.2  & 0.5 & 0   & 1 (2.25)   & \multicolumn{2}{l}{0.97 (2.31)} &\\
  & DAG 7  & base case& 0.37   & 0.71 & 0 & 0.5 & 0   & 1 (1.14)   & \multicolumn{2}{l}{1.19 (0.96)} &\\
  &   & small $\rho_{v,x}$& 0.20   & 0.71 & 0 & 0.5 & 0   & 1 (1.14)   & \multicolumn{2}{l}{1.05 (1.09)} &\\
  &   & negative $\rho_{v,x}$ & -0.37  & 0.71 & 0 & 0.5 & 0   & 1 (1.14)   & \multicolumn{2}{l}{1.16 (0.99)} &\\
  &   & small $\rho_{x,z|v}$  & 0.37   & 0.37 & 0 & 0.5 & 0   & 1 (6.35)   & 1.25 (5.1) &&\\
  &   & small effect $\beta_x$& 0.37   & 0.71 & 0 & 0.1 & 0   & 1 (0.45)   & \multicolumn{2}{l}{1.22 (0.37)} &\\
  &   & binary covariate V& 0.19   & 0.71 & 0 & 0.5 & 0   & 1 (1.14)   & \multicolumn{2}{l}{1.04 (1.09)} &\\
  &   & $n_{MS}$ = 2,000 vs $n_{MS}$=5,000& 0.37   & 0.71 & 0 & 0.5 & 0   & 1 (1.92)   & \multicolumn{2}{l}{1.21 (1.59)} &\\
  &   & $n_{VS}$ = 150 vs $n_{VS}$ = 400 & 0.37   & 0.71 & 0 & 0.5 & 0   & 1 (2.25)   & \multicolumn{2}{l}{1.23 (1.82)} &\\
  & DAG 8  & base case& 0.37   & 0.71 & 0.18 & 0.5 & 0   & 1 (1.14)   & \multicolumn{2}{l}{1.28 (0.89)} &\\
  &   & small $\rho_{v,x}$& 0.20   & 0.71 & 0.19 & 0.5 & 0   & 1 (1.14)   & \multicolumn{2}{l}{1.08 (1.05)} &\\
  &   & negative $\rho_{v,x}$ & -0.37  & 0.71 & 0.18 & 0.5 & 0   & 1 (1.14)   & 1 (1.14)   &&\\
  &   & small $\rho_{x,z|v}$  & 0.37   & 0.37 & 0.18 & 0.5 & 0   & 1 (6.35)   & \multicolumn{2}{l}{1.66 (3.83)} &\\
  &   & small effect $\beta_x$& 0.37   & 0.71 & 0.18 & 0.1 & 0   & 1 (0.45)   & \multicolumn{2}{l}{1.27 (0.35)} &\\
  &   & large $\rho_{v,z|x}$ & 0.37   & 0.71 & 0.97 & 0.5 & 0   & 1 (1.14)   & \multicolumn{2}{l}{0.46 (2.45)} &\\
  &   & binary covariate V& 0.19   & 0.71 & 0.1  & 0.5 & 0   & 1 (1.14)   & \multicolumn{2}{l}{1.06 (1.08)} &\\
  &   & $n_{MS}$ = 2,000 vs $n_{MS}$=5,000& 0.37   & 0.71 & 0.18 & 0.5 & 0   & 1 (1.92)   & 1.28 (1.5) &&\\
  &   & $n_{VS}$ = 150 vs $n_{VS}$ = 400 & 0.37   & 0.71 & 0.18 & 0.5 & 0   & 1 (2.25)   & \multicolumn{2}{l}{1.35 (1.67)} &\\
  \hline
\end{longtable}
\newpage
\begin{longtable}{|| ll| lrrrrr | c c c c ||}
\caption{Empirical ARE ($ERE (\hat{\beta}_{(\cdot)}) = \frac{Var(\hat{\beta}_{(OM)})}{Var(\hat{\beta}_{(\cdot)})}$) and Empirical Variance for Binary Outcomes in the Simulation Study}\\
\hline
  \textbf{Y Type} & \textbf{DAG} & \textbf{Scenario} & $\rho_{v,x}$ & $\rho_{x,z|v}$ & $\rho_{v,z|x}$ & $\beta_x$ & $\beta_v$ & \multicolumn{4}{c||}{ $ERE (\hat{\beta}_{(\cdot)}) (Var( \hat{\beta}_{(\cdot)}) \times 10^{3})$}\\
 &  &  &  &  &  &  &  & \textbf{$\hat{\beta}_{(OM)}$} & \textbf{$\hat{\beta}_{(--)}$} & \textbf{$\hat{\beta}_{(-M)}$} & \textbf{$\hat{\beta}_{(O-)}$} \\
  \hline
  \endhead
Binary & DAG 1& base case & 0& 0.71 & 0& 0.5& 0.8& 1 (2.11) & 1.01 (2.09)& 1.01 (2.09)& 1 (2.11) \\
& & small $\rho_{x,z|v}$& 0& 0.37 & 0& 0.5& 0.8& 1 (8.22) & 1.01 (8.1) & 1.02 (8.1) & 1 (8.22) \\
& & small effect $\beta_x$& 0& 0.71 & 0& 0.1& 0.8& 1 (2.19) & 1.01 (2.18)& 1.01 (2.18)& 1 (2.19) \\
& & weak risk factor V, small $\beta_v$ & 0& 0.71 & 0& 0.5& 0.2& 1 (2.84) & 1 (2.85) & 1 (2.84) & 1 (2.85) \\
& & binary covariate V& 0& 0.71 & 0& 0.5& 0.8& 1 (1.65) & 1 (1.65) & 1 (1.65) & 1 (1.66) \\
& & $n_{MS}$ = 5,000 vs $n_{MS}$=10,000 & 0& 0.71 & 0& 0.5& 0.8& 1 (4.58) & 1.02 (4.5) & 1.02 (4.5) & 1 (4.58) \\
& & $n_{VS}$ = 150 vs $n_{VS}$ = 400& 0& 0.71 & 0& 0.5& 0.8& 1 (2.16) & 1.01 (2.15)& 1.01 (2.14)& 1 (2.16) \\
& DAG 5& base case & 0& 0.71 & 0& 0.5& 0 & 1 (2.9)& 1 (2.91) & 1 (2.9)& 1 (2.91) \\
& & small $\rho_{x,z|v}$& 0& 0.37 & 0& 0.5& 0 & 1 (11.14)& 1 (11.14)& 1 (11.14)& 1 (11.14)\\
& & small effect $\beta_x$& 0& 0.71 & 0& 0.1& 0 & 1 (3.23) & 1 (3.23) & 1 (3.23) & 1 (3.23) \\
& & binary covariate V& 0& 0.71 & 0& 0.5& 0 & 1 (2.91) & 1 (2.91) & 1 (2.9)& 1 (2.91) \\
& & $n_{MS}$ = 5,000 vs $n_{MS}$=10,000 & 0& 0.71 & 0& 0.5& 0 & 1 (6.26) & 1 (6.26) & 1 (6.26) & 1 (6.26) \\
& & $n_{VS}$ = 150 vs $n_{VS}$ = 400& 0& 0.71 & 0& 0.5& 0 & 1 (3)& 1 (3)& 1 (3)& 1 (3)\\
& DAG 6& base case & 0& 0.71 & 0.2& 0.5& 0 & 1 (2.9)& 0.97 (2.99)& & \\
& & small $\rho_{x,z|v}$& 0& 0.37 & 0.2& 0.5& 0 & 1 (11.14)& 0.96 (11.62) & & \\
& & small effect $\beta_x$& 0& 0.71 & 0.2& 0.1& 0 & 1 (3.23) & 0.98 (3.29)& & \\
& & large $\rho_{v,z|x}$ & 0& 0.71 & 0.97 & 0.5& 0 & 1 (2.9)& 0.09 (32.25) & & \\
& & binary covariate V& 0& 0.71 & 0.1& 0.5& 0 & 1 (2.91) & 1 (2.92) & & \\
& & $n_{MS}$ = 5,000 vs $n_{MS}$=10,000 & 0& 0.71 & 0.2& 0.5& 0 & 1 (6.26) & 0.97 (6.45)& & \\
& & $n_{VS}$ = 150 vs $n_{VS}$ = 400& 0& 0.71 & 0.2& 0.5& 0 & 1 (3)& 0.97 (3.09)& & \\
& DAG 7& base case & 0.37 & 0.71 & 0& 0.5& 0 & 1 (2.84) & 1.25 (2.28)& & \\
& & small $\rho_{x,z|v}$& 0.37 & 0.37 & 0& 0.5& 0 & 1 (10.76)& 1.33 (8.09)& & \\
& & small effect $\beta_x$& 0.37 & 0.71 & 0& 0.1& 0 & 1 (3.21) & 1.25 (2.56)& & \\
& & negative $\rho_{v,x}$ & -0.37& 0.71 & 0& 0.5& 0 & 1 (2.78) & 1.24 (2.24)& & \\
& & small $\rho_{v,x}$& 0.20 & 0.71 & 0& 0.5& 0 & 1 (2.89) & 1.06 (2.74)& & \\
& & binary covariate V& 0.19 & 0.71 & 0& 0.5& 0 & 1 (2.63) & 1.05 (2.52)& & \\
& & $n_{MS}$ = 5,000 vs $n_{MS}$=10,000 & 0.37 & 0.71 & 0& 0.5& 0 & 1 (6.26) & 1.22 (5.13)& & \\
& & $n_{VS}$ = 150 vs $n_{VS}$ = 400& 0.37 & 0.71 & 0& 0.5& 0 & 1 (2.89) & 1.24 (2.34)& & \\
& DAG 8& base case & 0.37 & 0.71 & 0.18 & 0.5& 0 & 1 (2.84) & 1.31 (2.17)& & \\
& & small $\rho_{x,z|v}$& 0.37 & 0.37 & 0.18 & 0.5& 0 & 1 (10.76)& 1.71 (6.29)& & \\
& & small effect $\beta_x$& 0.37 & 0.71 & 0.18 & 0.1& 0 & 1 (3.21) & 1.32 (2.44)& & \\
& & negative $\rho_{v,x}$ & -0.37& 0.71 & 0.18 & 0.5& 0 & 1 (2.78) & 1.14 (2.44)& & \\
& & small $\rho_{v,x}$& 0.20 & 0.71 & 0.19 & 0.5& 0 & 1 (2.89) & 1.07 (2.69)& & \\
& & large $\rho_{v,z|x}$  & 0.37 & 0.71 & 0.97 & 0.5& 0 & 1 (2.84) & 0.71 (4) & & \\
& & binary covariate V& 0.19 & 0.71 & 0.1& 0.5& 0 & 1 (2.63) & 1.06 (2.49)& & \\
& & $n_{MS}$ = 5,000 vs $n_{MS}$=10,000 & 0.37 & 0.71 & 0.18 & 0.5& 0 & 1 (6.26) & 1.27 (4.93)& & \\
& & $n_{VS}$ = 150 vs $n_{VS}$ = 400& 0.37 & 0.71 & 0.18 & 0.5& 0 & 1 (2.89) & 1.3 (2.22) && \\
\hline

\end{longtable}

\newpage
\section{Additional Description of Real Data Example}
For this example, we created a cohort consisting of participants who completed the sleep duration question in 1987 and sunscreen use question in 1992, as these two questions were asked among a subset of HPFS cohort along with the 4-year interval waves of FFQs. Among 22,577 eligible cohort participants, 198 were missing marital status, smoking status, BMI or physical activity. Since the percentage of missingness was so small, we conducted the complete case analysis assuming missing at random\cite{rubin1976inference,tsiatis2006semiparametric}.

In figure 2, we additionally note that sunscreen use here likely does not directly cause changes in fiber intake. Thus,  rather than drawing a  direct arrow in DAG 7, we illustrate the backdoor path through unmeasured covariates, $U$, (e.g. healthy lifestyle) to connect $V_{7(X-{}-)}$ and $X$. Nevertheless, $V_{7(X-{}-)}$ represented in this somewhat more complex manner still has the required feature that it is not a direct cause of cardiovascular disease and that it is independent of measurement error, conditional on the covariates that cause measurement error. Furthermore, by considering sleep duration and depression as $V_{2(-ZY)}$, we are assuming that conditional on $V_{3(X-Y)}$ and $V_{4(XZY)}$, these variables are independent of the unmeasured covariate $U$, $V_{7(X-{}-)}$, and the true exposure $X$.   

We also empirically assessed how each covariate, conditional on the other covariates, was associated with validated fiber intake $X$, measurement error thus mismeasured FFQ fiber intake $Z$ (conditional on true intake) and CVD outcome $Y$, to ensure that the associations are in line with our a priori knowledge. However, we note that such evaluation might be underpowered in validation study and is only suggestive evidence of presence or absence of arrows for the DAG. The result of the assessment is presented in the table below. 

\begin{threeparttable}
\caption{Partial Correlation Coefficient, Odds Ratio (OR) and P values in HPFS}
\begin{tabular}{|l l | c c c |}
\hline
$V_j$ & Covariate $V=\overset{\cdot}{\cup} V_i$ \tnote{\textdagger} & $\rho_{_{X,V_i| V \setminus V_i}}$ (p value) \tnote{\textdaggerdbl} & $\rho_{_{Z,V_i|X, V \setminus V_i}}$ (p value) \tnote{\textdaggerdbl} &  Y $\sim$V: OR(p value) \\
\hline
\textbf{$V_{_{1(-{}-{}Y)}}$} & No Family history of MI vs Yes  & -0.069(0.161) & 0.023(0.645)  & 0.792(0.001) \\
\hline
$V_{_{3(X{}-{}Y)}}$ & Energy intake & 0.053(0.281)  & 0.023(0.647)  & 1(0.122)  \\
 & Cholesterol& -0.033(0.507) & 0.042(0.4) & 0.853(0.002) \\
 & Diabetes& -0.009(0.85)  & 0.037(0.456)  & 0.582(0)  \\
 & No high blood pressure & 0.028(0.576)  & 0.037(0.457)  & 0.753(0)  \\
 & Former vs current smoker  & -0.009(0.854) & -0.086(0.083) & 0.789(0.009) \\
 & Never vs current smoker& 0.054(0.273)  & 0.115(0.02)& 0.727(0.001) \\
 & Not married vs married & -0.001(0.977) & 0.04(0.415)& 0.968(0.709) \\
 \hline
$V_{_{4(X{}Z{}Y)}}$ & Metabolic equivalent hours& 0.204(0)& 0.232(0)& 0.998(0.09)  \\
 & Baseline age  & 0.031(0.534)  & 0.131(0.008)  & 1.054(0)  \\
 & BMI  & -0.159(0.001) & -0.096(0.053) & 1.054(0)  \\
 & Taking multivitamin vs not& -0.001(0.976) & 0.086(0.083)  & 0.975(0.599) \\
 & Alcohol intake $\ge$ 45 vs 0-5 g/d & 0.045(0.361)  & -0.088(0.075) & 0.979(0.865) \\
 & Alcohol intake 5 - 45 vs 0-5 g/d& -0.153(0.002) & -0.163(0.001) & 0.877(0.008) \\
 & sleep duration $\le$ 6 vs 6-8 hrs  & 0.01(0.842)& 0.087(0.08)& 0.913(0.122) \\
 & sleep duration $\ge$ 8 vs 6-8 hrs  & -0.017(0.729) & -0.009(0.85)  & 0.955(0.657) \\
 & Not depressed vs depressed& -0.06(0.229)  & -0.067(0.179) & 0.866(0.395) \\
 \hline
$V_{_{7(X{}-{}-)}}$ & Sunscreen use 100\%  vs 0-25\%  & 0.062(0.21)& 0.04(0.414)& 0.938(0.289) \\
 & Sunscreen use 25-50\%  vs 0-25\%& -0.092(0.062) & -0.072(0.146) & 0.901(0.115) \\
 & Sunscreen use Not in the sun vs 0-25\%& 0(0.997)& 0.089(0.072)  & 0.992(0.912)\\
 \hline
\end{tabular}
    \begin{tablenotes}
        \item[\textdagger] $V_i$ is each of the covariate in this column under consideration and belongs to one of the four covariate sets $V_{_{1(-{}-{}Y)}}$, $V_{_{3(X{}-{}Y)}}$,  $V_{_{4(X{}Z{}Y)}}$ and $V_{_{7(-{}-{}Y)}}$. 
        \item[\textdaggerdbl] $\rho_{\cdot}$ is the partial correlation coefficient where $V \setminus V_i$ indicates the covaraite sets that are conditioned on, i.e. the full set of covariate $V$ minus each covaraite $V_i$ under consideration. 
    \end{tablenotes}
\end{threeparttable}

\newpage
\section{Effect Modification by Covariate $V$} \label{C}
\normalsize	

\subsection{Correct Modeling of Outcome Model under Simple $X-V$ Interaction}
Suppose we have the following true model for the observed outcome and true exposure:$E[Y|X,V] = \beta_0 + \beta_x X + \beta_v V + \beta_{xv} XV$ and the true model for the observed true exposure and surrogate exposure $E[X|Z,V] = \alpha_0 + \alpha_{z}Z + \alpha_{v}V$. Then the true model relating observed outcome to the surrogate exposure is obtained as:

\begin{equation} \tag{C2} \label{model_interaction}
    \begin{aligned}
    E[Y|Z,V] & = E_{X|Z,V}[E[Y|Z,X,V]] & \\
             & = E_{X|Z,V}[E[Y|X,V]]  \text{    (surrogacy assumption)} & \\
             & = E_{X|Z,V}[\beta_0 + \beta_x X + \beta_v V + \beta_{xv} XV] & \\
             & = \beta_0 + \beta_x E[X|Z,V] + \beta_v V + \beta_{xv} E[X|Z,V] V& \\
             & = (\beta_0 + \beta_x \alpha_0) + \beta_x \alpha_z Z + (\beta_x \alpha_v + \beta_{xv} \alpha_0 + \beta_v) V + \beta_{xv} \alpha_z ZV + \beta_{xv}  \alpha_v V^2 & \\
             & = \beta_0^* + \beta_z^* Z + \beta_v^* V + \beta_{zv}^* ZV + \beta_{v^2}^* V^2 & \\
    \end{aligned}
\end{equation}
This result suggests that for a simple $X-V$ interaction on the additive scale, our conditional effect $\beta(V)$ can only be consistently estimated \textit{if} we include both the $Z-V$ product term as well as the $V^2$ term into the outcome model of the RSW estimator $E[Y|Z,V]$, which gives $\hat{\beta}_{_{(OM)}} (V) = \frac{\hat{\beta}_z^* + \hat{\beta}_{zv}^* V}{\hat{\alpha}_z} \overset{p}{\to} \beta_{_{(OM)}} (V) = \beta (V) = \frac{\beta_z^* + \beta_{zv}^* V}{\alpha_z}$. 

\subsection{Validity of General Formula of RSW Estimators under Effect Modification by $V$}

To understand whether the validity results in Table 1 of the manuscript under the four different covariate adjustment strategies still apply in the presence of effect modification, we first redefine the four RSW estimators now in the following general forms: 

\begin{align*}
\hat{\beta}_{(OM)} (V)& = \frac{\hat{E}[Y|Z=z,V] - \hat{E}[Y|Z=z' , V]}{\hat{E}[X|Z=z,V] - \hat{E}[X|Z=z' , V]}\\
\hat{\beta}_{(--)} & = \frac{\hat{E}[Y|Z=z] - \hat{E}[Y|Z=z' ]}{\hat{E}[X|Z=z] - \hat{E}[X|Z=z']} \text{ (note this estimator cannot depend on covariate value $V$)}\\
\hat{\beta}_{(-M)} (V)& = \frac{\hat{E}[Y|Z=z] - \hat{E}[Y|Z=z']}{\hat{E}[X|Z=z,V] - \hat{E}[X|Z=z' , V]}\\
\hat{\beta}_{(O-)} (V)& = \frac{\hat{E}[Y|Z=z,V] - \hat{E}[Y|Z=z' , V]}{\hat{E}[X|Z=z] - \hat{E}[X|Z=z' ]},
\end{align*} where $\hat{E}[Y|Z=z, V], \hat{E}[Y|Z=z],\hat{E}[X|Z=z, V]$ and $\hat{E}[X|Z=z]$ can be parametrically, semiparametrically estimated or are just plug-in estimators (i.e. empirical expectation) and are potentially function of $V$. But all estimators will be restricted in the sense that they cannot depend on the value of $Z$; this can be achieved, for example for $\hat{\beta}_{(OM)} (V)$, by modeling $E[Y|Z,V] = Z \phi_1 (V)$ and  $E[X|Z,V] = Z \phi_2 (V)$, where $\phi_1 (V)$ and $\phi_2 (V)$ are different flexible functions of $V$. 

If $\hat{E}[\cdot]$ is estimated using plug-in estimator, then under conditions such as prescribed in the Gilvenko-Centelli theorem, the empirical distribution (of random variables $Y$, $X$ for any fixed $Z$ and $V$) converge almost surely to the true distribution and $\hat{E}[Y|Z=z, V] \overset{p}{\to}E[Y|Z=z, V] , \hat{E}[Y|Z=z]\overset{p}{\to} E[Y|Z=z] ,\hat{E}[X|Z=z, V]\overset{p}{\to} E[X|Z=z, V] $ and $\hat{E}[X|Z=z]\overset{p}{\to} E[X|Z=z]$, where mean is a smooth function (see Vaart et al (1996)) \cite{vaart1996weak}. These consistency results still hold when additional modeling assumptions are used to estimate $E[\cdot]$. Thus under Slutsky theorem, $$\hat{\beta}_{_{(OM)}} (V) \overset{p}{\to} \frac{E[Y|Z=z,V] - E[Y|Z=z' , V]}{E[X|Z=z,V] - E[X|Z=z' , V]} =  \beta_{_{(OM)}} (V) = \beta (V),$$ and 
$$\hat{\beta}_{_{(--)}} \overset{p}{\to} \frac{E[Y|Z=z] - E[Y|Z=z^{'}]}{E[X|Z=z] - E[X|Z=z^{'}]} = \beta_{_{(--)}}$$, where $\beta_{_{(--)}}$ cannot identify conditional average treatment effect when covariate $V$ is indeed an effect modification. 

We note that $V_{5(-{}-{}-)},V_{6(-{}Z{}-)},V_{7(X{}-{}-)},V_{8(X{}Z{}-)}$ cannot be effect modifiers\cite{webster2021directed} and that $V_{2(-{}Z{}Y)}$, $V_{3(X{}-{}Y)}$ and $V_{4(X{}Z{}Y)}$ can only be included in both outcome model and MEM thus only $\hat{\beta}_{_{(OM)}} (V)$ can be used to estiamte the conditional average treatment effect $\beta (V)$ under DAG 2, 3 and 4. 

Under DAG 1, $X\perp V|Z$ and thus
$$\hat{\beta}_{_{(O-)}} (V) \overset{p}{\to} \frac{E[Y|Z=z, V] - E[Y|Z=z^{'},V]}{E[X|Z=z] - E[X|Z=z^{'}]} = \frac{E[Y|Z=z, V] - E[Y|Z=z^{'},V]}{E[X|Z=z, V] - E[X|Z=z^{'},V]} =  \beta (V), $$
 and
$$\hat{\beta}_{_{(-M)}} (V) \overset{p}{\to} \frac{E[Y|Z=z] - E[Y|Z=z^{'}]}{E[X|Z=z,V] - E[X|Z=z^{'},V]} = \frac{E[Y|Z=z] - E[Y|Z=z^{'}]}{E[X|Z=z] - E[X|Z=z^{'}]} =  \beta_{_{(--)}}.$$ 

Therefore when $V_{1(-{}-{}Y)}$ is an effect modifier, only $\hat{\beta}_{_{(OM)}}$ and $\hat{\beta}_{_{(O-)}}$ are valid estimators for conditional average treatment effect under DAG 1.

\newpage
\section{Bibliography}
\bibliography{sample.bib}


